\documentclass[final,onefignum,onetabnum]{siamonline171218}

\graphicspath{{figures/}}

\usepackage[utf8]{inputenc} 
\usepackage[T1]{fontenc}
\usepackage[english]{babel}
\usepackage{algorithmic}      
\usepackage[algo2e,lined,ruled,linesnumbered]{algorithm2e}
\SetAlFnt{\small\sffamily}
\DontPrintSemicolon
\usepackage{array}   
\usepackage{amsmath,amssymb,amsfonts}
\usepackage{amsopn}
\allowdisplaybreaks
\usepackage{dsfont,bm,bbm} 
\usepackage{mathtools}
\usepackage{xfrac}
\usepackage{mathrsfs}
\usepackage{booktabs}
\usepackage[margin=1in]{geometry}
\usepackage{graphicx} 
\usepackage[caption=false]{subfig}
\usepackage{textcomp}
\usepackage{epstopdf}
\usepackage{lipsum}
\usepackage{multirow}
\usepackage{enumitem}
\usepackage{nicefrac}
\usepackage[colorinlistoftodos,prependcaption,textsize=small,textwidth=20mm]{todonotes}
\usepackage[thinlines]{easytable}
\usepackage[export]{adjustbox}

\setlist[enumerate]{leftmargin=.5in}
\setlist[itemize]{leftmargin=.5in}
\ifpdf
\DeclareGraphicsExtensions{.eps,.pdf,.png,.jpg}
\else
\DeclareGraphicsExtensions{.eps}
\fi

\newcommand{\given}{\;\ifnum\currentgrouptype=16 \middle\fi|\;}
\newcommand{\suchthat}{\;\ifnum\currentgrouptype=16 \middle\fi|\;}
\newcommand{\norm}[1]{\left\Vert#1\right\Vert}

\newcommand{\dd}{\mathrm{d}}
\newcommand{\tran}{\mathsf{T}}

\newcommand{\pr}[1]{\mathds{P}\!\left[#1\right]}
\newcommand{\pf}{p_\mathcal{F}}
\newcommand{\pfe}{\widehat{p}_\mathcal{F}}

\newcommand{\I}[2]{{\mathds{1}_{#1}\!\left(#2\right)}}
\newcommand{\E}[2]{\mathds{E}_{#1}\!\left[{#2}\right]}
\newcommand{\V}[2]{\mathds{V}_{#1}\!\left[{#2}\right]}
\newcommand{\ve}[1]{\bm{#1}}
\newcommand{\mat}[1]{\mathbf{#1}}
\newcommand{\cov}[1]{\mathds{C}\mathrm{ov}\left[#1\right]}
\newcommand{\cv}[1]{\widehat{\mathrm{cv}}\left(#1\right)} 
\newcommand{\cve}{\widehat{\mathrm{cv}}}

\newcommand{\Dkl}[2]{{D}_{\text{KL}}\left(#1||#2\right)}
\newcommand{\Dklb}[2]{D_{\text{KL}}(#1||#2)}

\DeclareMathOperator*{\argmax}{arg\,max}
\DeclareMathOperator*{\argmin}{arg\,min}

\DeclareMathOperator{\supp}{supp} 

\newsiamremark{remark}{Remark}
\newsiamremark{hypothesis}{Hypothesis}
\crefname{hypothesis}{Hypothesis}{Hypotheses}
\crefname{remark}{remark}{remark}
\newsiamthm{claim}{Claim}

\Crefname{ALC@unique}{Line}{Lines}

\headers{Cross-entropy method with dimension reduction}{F. Uribe et al.}

\title{Cross-entropy-based importance sampling with failure-informed dimension reduction for rare event simulation \thanks{Submitted to the editors \today.\funding{This research has been supported by the \emph{Deutsche Forschungsgemeinschaft} (DFG) through the TUM International Graduate School of Science and Engineering (IGSSE) within the project 10.02 BAYES.}}}

\author{
	Felipe Uribe\thanks{Engineering Risk Analysis Group, Technische Universität München. Arcisstraße 21, 80333 Munich, Germany (\email{felipe.uribe@tum.de}, \email{iason.papaioannou@tum.de}, \email{straub@tum.de}).}
	\and Iason Papaioannou\footnotemark[2] 	
	\and Youssef M. Marzouk\thanks{Department of Aeronautics and Astronautics, Massachusetts Institute of Technology. 77 Massachusetts Avenue, Cambridge, MA 02139 (\email{ymarz@mit.edu}).}
	\and Daniel Straub\footnotemark[2] 
}

\ifpdf
\hypersetup{
	pdftitle={Failure-informed cross-entropy method},
	pdfauthor={F. Uribe et al.}
}
\fi

\begin{document}
	
\maketitle

\begin{abstract}
The estimation of rare event or failure probabilities in high dimensions is of interest in many areas of science and technology. We consider problems where the rare event is expressed in terms of a computationally costly numerical model. Importance sampling with the cross-entropy method offers an efficient way to address such problems provided that a suitable parametric family of biasing densities is employed. Although some existing parametric distribution families are designed to perform efficiently in high dimensions, their applicability within the cross-entropy method is limited to problems with dimension of $\mathcal{O}(10^2)$. In this work, rather than directly building sampling densities in high dimensions, we focus on identifying the intrinsic low-dimensional structure of the rare event simulation problem. To this end, we exploit a connection between rare event simulation and Bayesian inverse problems. This allows us to adapt dimension reduction techniques from Bayesian inference to construct new, effectively low-dimensional, biasing distributions within the cross-entropy method. In particular, we employ the approach in \cite{zahm_et_al_2018a}, as it enables control of the error in the approximation of the optimal biasing distribution. We illustrate our method using two standard high-dimensional reliability benchmark problems and one structural mechanics application involving random fields.
\end{abstract}

\begin{keywords}
rare event simulation, reliability analysis, likelihood-informed subspace, importance sampling, cross-entropy method, random fields.
\end{keywords}

\begin{AMS}
60G60, 62L12, 65C05, 65C60, 65F15. 
\end{AMS}

\section{Introduction}
Computational models of physical systems in engineering and science are controlled by inputs and parameters whose values are random or uncertain. The treatment and modeling of this uncertainty is fundamental to the analysis and design of physical systems. When satisfactory or safe operation of the system under consideration is a main concern, system performance may be evaluated in terms of the probability of undesirable events---e.g., system failure. These events correspond to the system response exceeding predefined bounds, where the response is described by a forward model with uncertain inputs (typically a partial differential equation). A special challenge in reliability and rare event simulation involves the analysis of failure events whose probabilities are very small, and for which the associated dimension of the input parameter space is very large \cite{bucklew_2004}.

The estimation of failure probabilities involves the exploration of tails of probability distributions. One practical way to approach this problem is via approximation methods such as the first-order \cite{rackwitz_and_fiessler_1978} and second-order \cite{fiessler_et_al_1979} reliability methods. Both approaches require an optimization task to find the point of minimum distance from the failure hypersurface to the origin; the failure surface is then approximated through a first- or second-order Taylor series expansion at this point. A drawback of these methods is that their accuracy decreases with increasing dimension of the parameter space and with the nonlinearity of the failure hypersurface \cite{hohenbichler_and_rackwitz_1988, valdebenito_et_al_2010}. Monte Carlo (MC) methods provide another way to solve the rare event simulation problem. They involve statistical estimation of the averages of response quantities, which are formulated as probability-weighted integrals over the parameter space \cite{owen_2013,rubinstein_and_kroese_2017}. Standard MC simulation can become intractable if the underlying mathematical model is expensive to evaluate and/or if the failure probability is small. Essentially, the sample size required to obtain an estimate of a fixed relative error is inversely proportional to the failure probability, which limits the application of simple MC methods to relatively frequent events, or computationally inexpensive system models. The number of samples can be reduced by concentrating the sampling on the region of the parameter space that contributes most to failure. This is the idea of importance sampling (IS) \cite{kahn_and_marshall_1953, shinozuka_1983}, where a biasing distribution is employed to draw rare failure samples more frequently. In principle, it is possible to derive an optimal biasing distribution leading to a zero-variance estimator. Such a construction is infeasible in practice since one requires \emph{a priori} knowledge of the failure probability. However, it still provides an indication of how to build effective biasing distributions. 

Large deviation techniques can be employed to derive asymptotic approximations of the optimal biasing distribution, such that they describe the most probable path leading to the rare event (see, e.g., \cite{asmussen_2002, bucklew_2004}). Alternatively, the cross-entropy (CE) method constructs an approximation of the optimal biasing distribution by minimizing the Kullback--Leibler (KL) divergence to a given parametric family of distributions, usually selected from the exponential family. A study on the performance of single Gaussian and Gaussian mixture distributions in the CE method is carried out in \cite{geyer_et_al_2019}. Moreover, the generalized CE method \cite{botev_et_al_2007} uses kernel mixture distributions as nonparametric models to approximate the optimal biasing distribution. In \cite{peherstorfer_et_al_2018}, Gaussian distributions are employed in combination with a multifidelity approach that uses low-cost surrogates to efficiently build a sequence of biasing distributions. These approaches are able to estimate small failure probabilities; however, their application has been limited to low-dimensional parameter spaces.

Different parametric families have been proposed to extend the applicability of the CE method to higher dimensions. For instance, \cite{wang_and_song_2016} exploits the geometry of the standard Gaussian space in high dimensions and employs a von Mises--Fisher mixture model which is optimal for sampling on the surface of a hypersphere. Although the method is applied to high-dimensional problems, it requires the generation of a large number of samples; moreover, its performance degrades in low-dimensional problems. Hence, the more flexible von Mises--Fisher--Nakagami mixture distribution is proposed in the improved CE method \cite{papaioannou_et_al_2019} to extend the applicability to low and moderate dimensions. \cite{papaioannou_et_al_2019} also proposes a smooth approximation of the optimal biasing distribution that allows information from all the samples to be used in fitting the parametric distribution.

While the CE method provides a flexible way to construct good biasing distributions for IS, an accurate solution of the underlying optimization task (i.e., fitting the parametric biasing densities) requires a large number of samples for problems with high-dimensional input parameter spaces \cite{papaioannou_et_al_2019}. Moreover, the likelihood ratios or weights used within the IS framework often degenerate in high dimensions \cite{rubinstein_and_glynn_2009}. As a result, the application of the CE method remains largely limited to low and moderate dimensional spaces. 

In this paper, we introduce an approach for IS with the CE method that is able to exploit the intrinsic low-dimensional structure of rare event simulation problems. A main source of such a structure is the smoothing effect of the forward operator defining the system response. A consequence is that this response might vary predominantly along a few directions of the input parameter space, while being essentially constant in the remaining directions. Successful strategies for identifying such directions have been developed in the context of Bayesian inverse problems, and include the likelihood-informed subspace method \cite{cui_et_al_2014, spantini_et_al_2015}, the active subspace method \cite{constantine_et_al_2014, constantine_2015}, and the certified dimension reduction approach \cite{zahm_et_al_2018a}. In this contribution, we exploit the fact that the rare event simulation problem can be expressed as a Bayesian inverse problem, where the failure indicator and the probability of failure are equivalent to the likelihood function and the Bayesian model evidence, respectively. The resulting posterior distribution coincides with the optimal zero-variance biasing distribution of IS. This connection allows us to adapt dimension reduction techniques from Bayesian inversion to construct effective biasing distribution models within the CE method that operate on a low-dimensional subspace. By analogy to the Bayesian inversion context, we refer to this subspace as the \emph{failure-informed} subspace (FIS). We remark that the method proposed in \cite{wahal_and_biros_2019a, wahal_and_biros_2019b} also relies on the link between rare event simulation and Bayesian inference to build biasing distributions within IS; however, the accuracy of the method still deteriorates in high dimensions.

In order to identify and construct the FIS, we build on the ideas of the certified dimension reduction approach \cite{zahm_et_al_2018a}, which is applicable to nonlinear problems and provides a way to regulate the error in the approximation of the optimal biasing distribution on the FIS. In principle, the method requires the computation of a conditional expectation and the second moment matrix of the gradient of the log-failure indicator function. However, we show that when adapting certified dimension reduction for the CE method: (i) it is no longer necessary to compute the conditional expectation explicitly, and (ii) the failure indicator function needs to be approximated by a smooth function to ensure sufficient regularity. Therefore, instead of employing the standard CE method, we utilize the improved CE method \cite{papaioannou_et_al_2019} in which the failure indicator is approximated by a suitable smooth function. We term the resulting improved CE method with failure-informed dimension reduction, iCEred. %

Since the FIS is effectively low-dimensional, the optimization problem within iCEred is only solved along the failure-informed directions. Hence, a Gaussian parametric family of biasing densities is in general sufficient. This makes the approach very efficient for high-dimensional rare event simulation problems that are equipped with a low-dimensional structure. We also discuss a refinement step that can be applied at the end of the iCEred algorithm. The idea is to further reduce the coefficient of variation of the failure probability estimate based on a user-defined threshold. This extra step requires additional limit-state function evaluations, but no additional gradient computations. We test the proposed approach on both linear and nonlinear reliability problems. These include two algebraic problems where the reference failure probability is easy to compute, and a high-dimensional structural mechanics application where the Young's modulus is spatially variable and modeled by a random field. 

The organization of the paper is as follows: in \cref{sec:IS_CE}, we introduce the connection between rare event simulation and Bayesian inversion, and describe IS with the standard and improved CE methods. In \cref{sec:dimred}, we adapt the approach in \cite{zahm_et_al_2018a} to the rare event simulation context. The major contribution of this work is presented in \cref{sec:iCEred}, where we combine the FIS with the improved CE method. \Cref{sec:numexp} presents three application examples. The paper ends with a summary of the work in \cref{sec:conclusions}.

\section{Mathematical and computational framework}\label{sec:IS_CE}
We first introduce the reliability problem related to the task of estimating rare event probabilities, and discuss the fundamentals of importance sampling together with the standard and improved cross-entropy methodologies.

\subsection{Rare event simulation}\label{subsec:rare_event}
Consider the canonical probability space $(\Omega,\mathscr{F},\mathds{P})= (\mathbbm{R}^d, \mathscr{B}(\mathbbm{R}^d),\allowbreak \mathds{P})$, with $\mathscr{B}(\mathbbm{R}^d)$ denoting Borel sets on $\mathbbm{R}^d$ and $\mathds{P}$ a probability measure \cite{ash_and_doleansdade_2000}. The uncertain parameter vector $\ve{\theta}$ is modeled as a random vector taking values on $\mathbbm{R}^d$, such that $\ve{\theta}(\omega) = \omega$. The distribution of $\ve{\theta}$ is assumed to have a density $ \pi_{\text{pr}}(\ve{\theta}) = \dd \mathds{P}/ \dd\lambda$ with respect to the Lebesgue measure $\lambda$ on $\mathbbm{R}^d$; we call this the \emph{prior} or nominal probability density.

Different modes of failure can be grouped in a so-called \emph{limit-state function} (LSF) $g: \mat{\Theta}\to\mathbbm{R}$, usually defined as $g(\ve{\theta}) = \beta - \mathcal{Q}(\ve{\theta})$. Here, $\beta$ is a predefined maximum allowed threshold, and $\mathcal{Q}(\ve{\theta}): \mat{\Theta}\to\mathbbm{R}$ is a forward response operator that maps the parameter $\ve{\theta}$ to a quantity of interest (QoI) characterizing the performance of the system. The failure hypersurface defined by $g(\ve{\theta})=0$ splits the parameter space into two subsets, namely the \emph{safe set} $\mathcal{S}=\{\ve{\theta}: g(\ve{\theta}) > 0 \}$ and the \emph{failure set} $\mathcal{F} = \{\ve{\theta}:g(\ve{\theta}) \leq 0 \}$. The probability of $\mathcal{F}$ under the prior distribution, also known as the \emph{probability of failure} $p_\mathcal{F}$, is defined as 
\begin{equation}\label{eq:pf}
\pf = \pr{\mathcal{F}}  = \int_{\ve{\Theta}} \I{\mathcal{F}}{\ve{\theta}}\pi_{\text{pr}}(\ve{\theta}) ~\dd\ve{\theta}  = \E{\pi_{\text{pr}}}{\I{\mathcal{F}}{\ve{\theta}}}
\end{equation}
where $\mathds{1}_\mathcal{F}:\mathbbm{R}^d\rightarrow \{0,1\}$ stands for the indicator function, taking values $\I{\mathcal{F}}{\ve{\theta}}=1$ when $\ve{\theta}\in \mathcal{F}$, and $\I{\mathcal{F}}{\ve{\theta}}=0$ otherwise. In rare event simulation, the probability of failure \cref{eq:pf} represents the evaluation of a potentially high-dimensional integral for which $\pf$ is very small (typically in the range $10^{-3}-10^{-10}$). Specialized Monte Carlo algorithms are used in those cases, for instance, directional and line sampling \cite{ditlevsen_et_al_1990, koutsourelakis_et_al_2004}, importance sampling schemes \cite{shinozuka_1983, engelund_and_rackwitz_1993, papaioannou_et_al_2016} including the cross-entropy method \cite{rubinstein_and_kroese_2017}, and multilevel splitting methods \cite{cerou_et_al_2012, botev_and_kroese_2012, ullmann_and_papaioannou_2015} including subset simulation \cite{au_and_beck_2001}. 

From \cref{eq:pf}, one sees that the probability of failure is obtained by integrating the product of the indicator function and the prior density over the input parameter space $\ve{\Theta}$. This is analogous to the {model evidence} $Z$ in the context of Bayesian inference \cite{stuart_2010}. $Z$ is computed similarly by integrating the product of a likelihood function and the prior density. We can exploit this connection to formulate the rare event simulation problem as the Bayesian inference task: 
\begin{equation}\label{eq:post_F}
\pi_\mathcal{F}(\ve{\theta}) = \dfrac{1}{\pf} \I{\mathcal{F}}{\ve{\theta}} \pi_{\text{pr}}(\ve{\theta}),
\end{equation}
where the indicator function acts as a likelihood function, and $\pi_\mathcal{F}$ can be interpreted as a `posterior-failure' density of the parameters given the occurrence of the failure event $\mathcal{F}$. Note that $\pi_\mathcal{F}$ is not a posterior density in the Bayesian inference sense, since there is no data entering into the likelihood function. In this setting, $\pi_\mathcal{F}$ can be seen as a density conditional on the failure domain with its normalizing constant equal to the target probability of failure (see, \cite{rubinstein_and_kroese_2017}). We remark that the formulation \cref{eq:post_F} is valid if the integral \cref{eq:pf} is finite and $\I{\mathcal{F}}{\ve{\theta}}$ is measurable (which is true if and only if the failure set $\mathcal{F}$ is measurable, see, e.g., \cite{schilling_2005}). 


\begin{remark}\label{rem:N01}
Classical approaches for estimating \eqref{eq:pf} operate in the independent standard Gaussian space---meaning that $\mathds{P}$ is assumed to be a standard Gaussian measure on $\mathbbm{R}^d$. Several isoprobabilistic transformations, such as the Knothe--Rosenblatt and Nataf constructions, exist to perform this `whitening' task (see, e.g., \cite[Ch.4]{lemaire_et_al_2009}). Therefore, we assume that the uncertain parameters are distributed as $\ve{\theta}\sim \pi_{\text{pr}}=\mathcal{N}(\ve{\mu}_{\text{pr}}, \mat{\Sigma}_{\text{pr}})$, with $\ve{\mu}_{\text{pr}}=\ve{0}$ and $\mat{\Sigma}_{\text{pr}}=\mat{I}_d$, where $\mat{I}_d\in \mathbbm{R}^{d\times d}$ denotes the identity matrix. 
\end{remark}

\subsection{Importance sampling}\label{subsec:IS}
Standard Monte Carlo simulation of \cref{eq:pf} requires a large number of samples from the prior distribution to achieve a suitable accuracy on $\pf$. The idea of \emph{importance sampling} (IS) \cite{kahn_and_marshall_1953, shinozuka_1983} is to employ an auxiliary distribution that concentrates the samples in the failure region. Consider the following modified version of \cref{eq:pf} 
\begin{equation}\label{eq:pfis}
p_\mathcal{F} = \int_{\mat{\Theta}}  \dfrac{\I{\mathcal{F}}{\ve{\theta}} \pi_{\text{pr}}(\ve{\theta})}{\pi_{\text{bias}}(\ve{\theta})} \pi_{\text{bias}}(\ve{\theta}) \dd\ve{\theta} = \E{\pi_{\text{bias}}}{\dfrac{\I{\mathcal{F}}{\ve{\theta}} \pi_{\text{pr}}(\ve{\theta})}{\pi_{\text{bias}}(\ve{\theta})}}, 
\end{equation}
where $\pi_{\text{bias}}$ is the importance or \emph{biasing density}, satisfying the relation $\supp(\I{\mathcal{F}}{\ve{\theta}} \allowbreak\pi_{\text{pr}}(\ve{\theta}) ) \allowbreak \subseteq \supp(\I{\mathcal{F}}{\ve{\theta}}\pi_{\text{bias}}(\ve{\theta}))$. The purpose of the biasing density is to make the occurrence of the rare event $\mathcal{F}$ more likely. Based on \cref{eq:pfis}, the IS estimate of the probability of failure \cref{eq:pf} is \cite{owen_2013}
\begin{equation}\label{eq:IS}
\widehat{p}_\mathcal{F}^{~\mathrm{IS}} = \dfrac{1}{N}\sum_{i=1}^{N} \I{\mathcal{F}}{\ve{\theta}_i}w(\ve{\theta}_i)\qquad\text{with }\quad w(\ve{\theta}_i) =\frac{\pi_{\text{pr}}(\ve{\theta}_i)}{{\pi}_{\text{bias}}(\ve{\theta}_i)},
\end{equation}
where $\{\ve{\theta}_i\}_{i=1}^{N} \overset{\text{i.i.d.}}{\sim} {\pi}_{\text{bias}}$, and each value $w(\ve{\theta}_i)$ represents a \emph{weight} that corrects for the use of the biasing density and ensures that the IS estimator remains unbiased, $\E{{\pi}_{\text{bias}}}{\widehat{p}_\mathcal{F}^{~\mathrm{IS}}}=\pf$. Moreover, the variance of the IS estimator is
\begin{equation}\label{eq:ISvar}
 \V{{\pi}_{\text{bias}}}{\widehat{p}_\mathcal{F}^{~\mathrm{IS}}}=\frac{1}{N}\V{{\pi}_{\text{bias}}}{\pf} = \frac{1}{N} \left( \E{\pi_{\text{bias}}}{(\I{\mathcal{F}}{\ve{\theta}} w(\ve{\theta}))^2} - {p}_\mathcal{F}^2\right). 
\end{equation}

To reduce the variance \cref{eq:ISvar}, one aims at selecting the biasing density $\pi_{\text{bias}}$ that minimizes the term $\E{\pi_{\text{bias}}}{(\I{\mathcal{F}}{\ve{\theta}} w(\ve{\theta}))^2}$. The resulting \emph{optimal biasing} density $\pi_{\text{bias}}^\star$, generating a zero-variance IS estimator, is given by \cite{bucklew_2004,owen_2013}
\begin{equation}\label{eq:post_F_CE}
\pi_{\text{bias}}^\star(\ve{\theta}) = \dfrac{\I{\mathcal{F}}{\ve{\theta}}\pi_{\text{pr}}(\ve{\theta})}{\int_{\mat{\Theta}} \I{\mathcal{F}}{\ve{\theta}} \pi_{\text{pr}}(\ve{\theta}) \dd\ve{\theta}}= \dfrac{1}{p_\mathcal{F}} \I{\mathcal{F}}{\ve{\theta}} \pi_{\text{pr}}(\ve{\theta})=\pi_\mathcal{F}(\ve{\theta}).
\end{equation}

The optimal biasing density \cref{eq:post_F_CE} is equal to the posterior-failure density defined in \cref{eq:post_F}, and it is not available without knowing the target failure probability in advance. Although $\pi_{\text{bias}}^\star$ is inaccessible in practice, it still provides a guideline on how to build useful IS schemes. This is exploited in the cross-entropy method.

\subsection{Cross-entropy method} \label{subsec:CEmethod}
The standard \emph{cross-entropy} (CE) method \cite{rubinstein_1997} approximates $\pi_{\text{bias}}^\star$ by a \emph{parametric biasing} density ${\pi}_{\text{bias}}(\ve{\theta}; \ve{\upsilon})$, with {reference parameters} $\ve{\upsilon}$. The approximation is selected from a family of densities $\Pi= \{{\pi}_{\text{bias}}(\ve{\theta}; \ve{\upsilon})\given\allowbreak \ve{\upsilon}\in\Upsilon\}$ designed to be of simpler form than $\pi_{\text{bias}}^\star$. Thereafter, the objective is to find $\ve{\upsilon}^\star\in\Upsilon$ such that the distance between the optimal and approximated biasing densities is minimal. The dissimilarity between these distributions is measured by the cross-entropy or Kullback--Leibler (KL) divergence
\begin{align}\nonumber
\Dkl{\pi_{\text{bias}}^\star}{{\pi}_{\text{bias}}} &= \int_{\mat{\Theta}} \ln\left(\dfrac{\pi_{\text{bias}}^\star(\ve{\theta})}{{\pi}_{\text{bias}}(\ve{\theta};\ve{\upsilon})}\right)  \pi_{\text{bias}}^\star(\ve{\theta}) \dd\ve{\theta}\\ \label{eq:KL2}
&=  \int_{\mat{\Theta}} \ln\pi_{\text{bias}}^\star(\ve{\theta})~ \pi_{\text{bias}}^\star(\ve{\theta}) \dd\ve{\theta} - \int_{\mat{\Theta}} \ln{\pi}_{\text{bias}}(\ve{\theta};\ve{\upsilon})~ \pi_{\text{bias}}^\star(\ve{\theta}) \dd \ve{\theta}.
\end{align}

The first term in \cref{eq:KL2} is invariant with respect to any choice of ${\pi}_{\text{bias}}$ and the problem reduces to the optimization task:
\begin{equation}\label{eq:CE_obj}
\ve{\upsilon}^\star = \argmax_{\ve{\upsilon}\in \Upsilon}\E{\pi^\star_{\text{bias}}}{\ln{\pi}_{\text{bias}}(\ve{\theta};\ve{\upsilon})},
\end{equation}
where $\ve{\upsilon}^\star$ denotes the optimal reference parameters. We can substitute the optimal biasing density from \cref{eq:post_F_CE} into \cref{eq:CE_obj} to express the optimization program as
\begin{equation}\label{eq:CE1}
\ve{\upsilon}^\star = \argmax_{\ve{\upsilon}\in \Upsilon} \E{\pi_{\text{pr}}}{\ln{\pi}_{\text{bias}}(\ve{\theta};\ve{\upsilon})~ \I{\mathcal{F}}{\ve{\theta}} }. 
\end{equation}

The expectation \cref{eq:CE1} can be estimated by Monte Carlo using samples from the prior distribution. However, this is impractical if $\mathcal{F}$ defines a rare event. In order to efficiently evaluate \cref{eq:CE1}, we apply IS with biasing distribution ${\pi}_{\text{bias}}(\ve{\theta}; \ve{\upsilon}')\in\Pi$ (for reference parameters $\ve{\upsilon}'\in \Upsilon$):
\begin{equation}\label{eq:CE2}
\ve{\upsilon}^\star = \argmax_{\ve{\upsilon}\in \Upsilon} \E{{\pi}_{\text{bias}}(\cdot; \ve{\upsilon}')}{\ln{\pi}_{\text{bias}}(\ve{\theta};\ve{\upsilon})~\I{\mathcal{F}}{\ve{\theta}}~ w(\ve{\theta};\ve{\upsilon}') } \qquad\text{with }\quad w(\ve{\theta};\ve{\upsilon}') =\frac{\pi_{\text{pr}}(\ve{\theta})}{{\pi}_{\text{bias}}(\ve{\theta};\ve{\upsilon}')}.
\end{equation}

We can further employ the IS estimator of the expectation \cref{eq:CE2} to define the stochastic optimization problem:
\begin{equation}\label{eq:IS_CE}
{\ve{\upsilon}}^\star\approx \widehat{\ve{\upsilon}}^\star = \argmax_{\ve{\upsilon}\in \Upsilon} {\mathcal{J}}(\ve{\upsilon}) \qquad \text{with} \qquad {\mathcal{J}}(\ve{\upsilon}) =  \dfrac{1}{N}\sum_{i=1}^{N} \ln{\pi}_{\text{bias}}(\ve{\theta}_i;\ve{\upsilon})\I{\mathcal{F}}{\ve{\theta}_i} w(\ve{\theta}_i;\ve{\upsilon}') 
\end{equation}
where $\{\ve{\theta}_i\}_{i=1}^{N} \overset{\text{i.i.d.}}{\sim} {\pi}_{\text{bias}}(\cdot;\ve{\upsilon}')$. If ${\mathcal{J}}(\ve{\upsilon})$ is convex and differentiable with respect to $\ve{\upsilon}$, the solution to \cref{eq:IS_CE} can be computed by $\nabla_{\ve{\upsilon}}{\mathcal{J}}(\ve{\upsilon}) = 0$ \cite{rubinstein_1997}. Moreover, if the biasing distribution belongs to the natural exponential family, the solution of the stochastic optimization problem can be computed analytically. For instance, if $\Pi$ is a collection of Gaussian densities, the parameter $\ve{\upsilon}$ is selected from the space $\Upsilon$ containing mean vectors and covariance matrices. In this case, the reference parameter estimator $\widehat{\ve{\upsilon}}^\star$ has an explicit updating rule (see, e.g., \cite{geyer_et_al_2019}).

In principle, estimating the optimal reference parameters using ${\pi}_{\text{bias}}(\ve{\theta}; \ve{\upsilon}')$ in \cref{eq:CE2} yields better efficiency than using  $\pi_{\text{pr}}(\ve{\theta})$ in \cref{eq:CE1}. However, one still requires a good initial choice of $\ve{\upsilon}'$, such that a substantial number of samples from ${\pi}_{\text{bias}}(\ve{\theta}; \ve{\upsilon}')$ lie in the failure domain. This is addressed in the CE method by gradually approaching the target failure event. The idea is to construct a sequence of intermediate sets $\mathcal{F}_j = \{\ve{\theta}\in \mat{\Theta}: g(\ve{\theta})\leq \gamma_j\}$, with intermediate thresholds $\gamma_j\geq 0$. 

The CE optimization task \cref{eq:IS_CE} is now solved at each level with respect to an intermediate optimal biasing density ${\pi}_{\text{bias},j}^\star(\ve{\theta}) \propto  \I{\mathcal{F}_j}{\ve{\theta}}\pi_{\text{pr}}(\ve{\theta})$ associated to a failure threshold $\gamma_j$. Starting from an initial reference parameter estimate $\widehat{\ve{\upsilon}}_{0}$, the sequential stochastic CE program reads
\begin{equation}\label{eq:IS_CE_levels}
\widehat{\ve{\upsilon}}_{j+1} = \argmax_{\ve{\upsilon}\in\Upsilon}\dfrac{1}{N}\sum_{i=1}^{N} \ln{\pi}_{\text{bias}}(\ve{\theta}_i;\ve{\upsilon}) \widetilde{w}_i^{(j)} \quad \text{with }\quad \widetilde{w}_i^{(j)} =\I{\mathcal{F}_j}{\ve{\theta}_i}\frac{\pi_{\text{pr}}(\ve{\theta}_i)}{{\pi}_{\text{bias}}(\ve{\theta}_i;\widehat{\ve{\upsilon}}_{j})},
\end{equation}
where $\{\ve{\theta}_i\}_{i=1}^{N} \overset{\text{i.i.d.}}{\sim} {\pi}_{\text{bias}}(\cdot;\widehat{\ve{\upsilon}}_{j})$, and the vector $\widetilde{\ve{w}}^{(j)}$ contains the $j$-th level weights computed with respect to the failure indicator function. Each failure threshold $\gamma_j$ is defined as the $N\cdot(1-\rho)$-th order statistic ($\rho$-quantile) of the sequence of LSF values $\{g_i=g(\ve{\theta}_i)\}_{i=1}^{N}$. The quantile value $\rho$ is chosen to ensure that a good portion of the samples from ${\pi}_{\text{bias}}(\cdot;\widehat{\ve{\upsilon}}_{j})$ fall in the failure set $\mathcal{F}_j$, usually $\rho\in[0.01,0.1]$ \cite{rubinstein_and_kroese_2017}.

The CE algorithm proceeds until an intermediate threshold is such that $\gamma_j\leq0$, for which at least $N\cdot \rho$ samples lie in the target failure set $\mathcal{F}$. These final `elite' samples are used to estimate the reference parameter of the approximated optimal biasing density, i.e., $\widehat{\ve{\upsilon}}_{n_{\text{lv}}}= \widehat{\ve{\upsilon}}^\star$, where $n_{\text{lv}}$ denotes the total number of levels or iterations. Samples from the latter are then used to compute the probability of failure via \cref{eq:IS}. We remark that if the prior and the biasing densities belong to the same parametric family, the initial estimate of the reference parameters is typically selected as the parameters defining the prior (e.g.,  $\widehat{\ve{\upsilon}}_{0}=[\ve{\mu}_{\text{pr}}, \mat{\Sigma}_{\text{pr}}]$).

\subsection{Improved cross-entropy method}\label{subsec:iCEmethod}
In the standard CE method, only $N\cdot\rho$ samples drawn from the intermediate biasing densities contribute to the estimation of each $\widehat{\ve{\upsilon}}_{j+1}$ in \cref{eq:IS_CE_levels}. This is due to the selection of the failure thresholds $\gamma_j$ and the shape of the intermediate optimal biasing densities. These are conditional distributions that are defined at each intermediate failure event using an indicator function. Since the remaining $N\cdot(1-\rho)$ samples can potentially support the estimation of each reference parameter $\widehat{\ve{\upsilon}}_{j+1}$, it would be better to use all samples. 

The \emph{improved cross-entropy} (iCE) method \cite{papaioannou_et_al_2019} re-defines the intermediate optimal biasing densities using an approximation of the indicator function that guarantees a smooth transition towards the optimal biasing density. This modification allows one to employ all the $N$ samples in the estimation of the reference parameters at each intermediate level. Several smooth approximations of the indicator function $f(\ve{\theta}; s) \approx \I{\mathcal{F}}{\ve{\theta}}$ are available. For instance, we consider the approximations:
\begin{equation}\label{eq:ind}
f^{\mathrm{log}}(\ve{\theta};s) =
\dfrac{1}{2} \left[1 +  \text{tanh}\left(-\dfrac{g(\ve{\theta})}{s}\right)\right] \qquad \mathrm{and} \qquad f^{\mathrm{erf}}(\ve{\theta};s) =
\Phi\left(-\dfrac{g(\ve{\theta})}{s}\right)
\end{equation}
which correspond to the standard logistic function and the standard Gaussian CDF, respectively. Note that in the limit, when the \emph{smoothing parameter} $s\to 0$, both functions converge to the indicator function. A comparison of different approximations is reported in \cite{lacaze_et_al_2015}, in the context of sensitivity analysis.

After using any of the smooth approximations \cref{eq:ind}, the sequential CE optimization problem \cref{eq:IS_CE_levels} is now defined by the sequence $\{s_j>0\}_{j=0}^{n_{\text{lv}}}$, instead of the failure thresholds $\gamma_j$. The new intermediate optimal biasing density associated with a smoothing parameter $s_j$ is defined as ${\pi}_{\text{bias},j}^\star(\ve{\theta};s_j) = \nicefrac{1}{p_j}(f(\ve{\theta};s_j) \pi_{\text{pr}}(\ve{\theta}))$, where $p_j$ is the normalizing constant. Therefore, we can re-formulate the sequential CE stochastic optimization problem \cref{eq:IS_CE_levels} as
\begin{equation}\label{eq:IS_iCE_levels}
\widehat{\ve{\upsilon}}_{j+1} = \argmax_{\ve{\upsilon}\in \Upsilon}\dfrac{1}{N}\sum_{i=1}^{N} \ln{\pi}_{\text{bias}}(\ve{\theta}_i;\ve{\upsilon}) \widetilde{w}_i^{(j+1)} \quad \text{with }\quad \widetilde{w}_i^{(j+1)} = f(\ve{\theta}_i;s_{j+1})\frac{\pi_{\text{pr}}(\ve{\theta}_i)}{{\pi}_{\text{bias}}(\ve{\theta}_i;\widehat{\ve{\upsilon}}_{j})},
\end{equation}
where $\{\ve{\theta}_i\}_{i=1}^{N} \overset{\text{i.i.d.}}{\sim} {\pi}_{\text{bias}}(\cdot;\widehat{\ve{\upsilon}}_{j})$, and the vector $\widetilde{\ve{w}}^{(j+1)}$ contains the $(j+1)$-th level weights computed with respect to the smooth approximation of the indicator function at $s_{j+1}$. Note that the objective function in \cref{eq:IS_iCE_levels} is analogous to the CE stochastic optimization \cref{eq:IS_CE_levels}, and thus both problems are solved identically. 

In order to ensure that each consecutive pair of intermediate optimal biasing distributions do not differ significantly from one another, the smoothing parameters are chosen adaptively. The idea is to match the effective sampling size to a target predefined value, as it is typically carried out in sequential MC approaches \cite{delmoral_et_al_2006}. Such task can be equivalently performed by requiring that the sample coefficient of variation ($\cve$) of the weights at each level is equal to a target value $\delta$ \cite{papaioannou_et_al_2016}. Therefore, the smoothing parameters are estimated via the optimization problem 
\begin{equation}\label{eq:s_adapt}
s_{j+1} = \argmin_{s\in(0,s_{j})} \left( \cv{f\!\left(\ve{\theta}; s\right) \ve{w}^{(j)}}-\delta \right)^2
\end{equation}
where ${\ve{w}}^{(j)} =\{\pi_{\text{pr}}(\ve{\theta}_i)/{\pi}_{\text{bias}}(\ve{\theta}_i;\widehat{\ve{\upsilon}}_{j})\}_{i=1}^{N}$. Notice that \cref{eq:s_adapt} is solved without further evaluations of the LSF. The iCE algorithm proceeds until the $\widehat{\mathrm{cv}}$ of the ratio between the indicator function and its smooth approximation is smaller than the target $\delta$. The final $N$ samples are used to approximate the probability of failure with the IS estimator \cref{eq:IS}.sing a closed-form update if available.

\begin{figure}[!ht]
\centering
\hspace{0.5cm}\raisebox{-0.5cm}{\small \rotatebox{0}{\textbf{CE method}}}\hspace{6cm} \raisebox{-0.5cm}{\small \rotatebox{0}{\textbf{iCE method}}}\\
\includegraphics[width=0.24\textwidth]{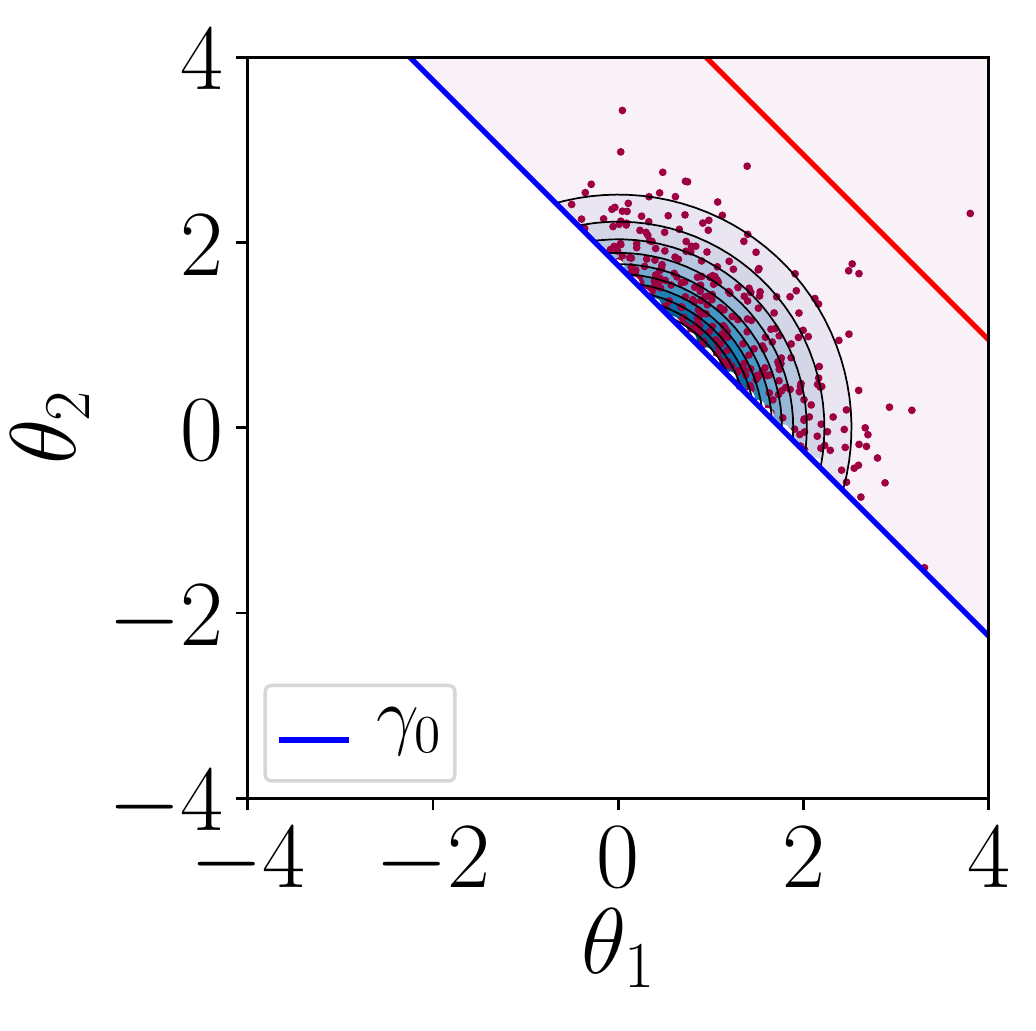}
\includegraphics[width=0.24\textwidth]{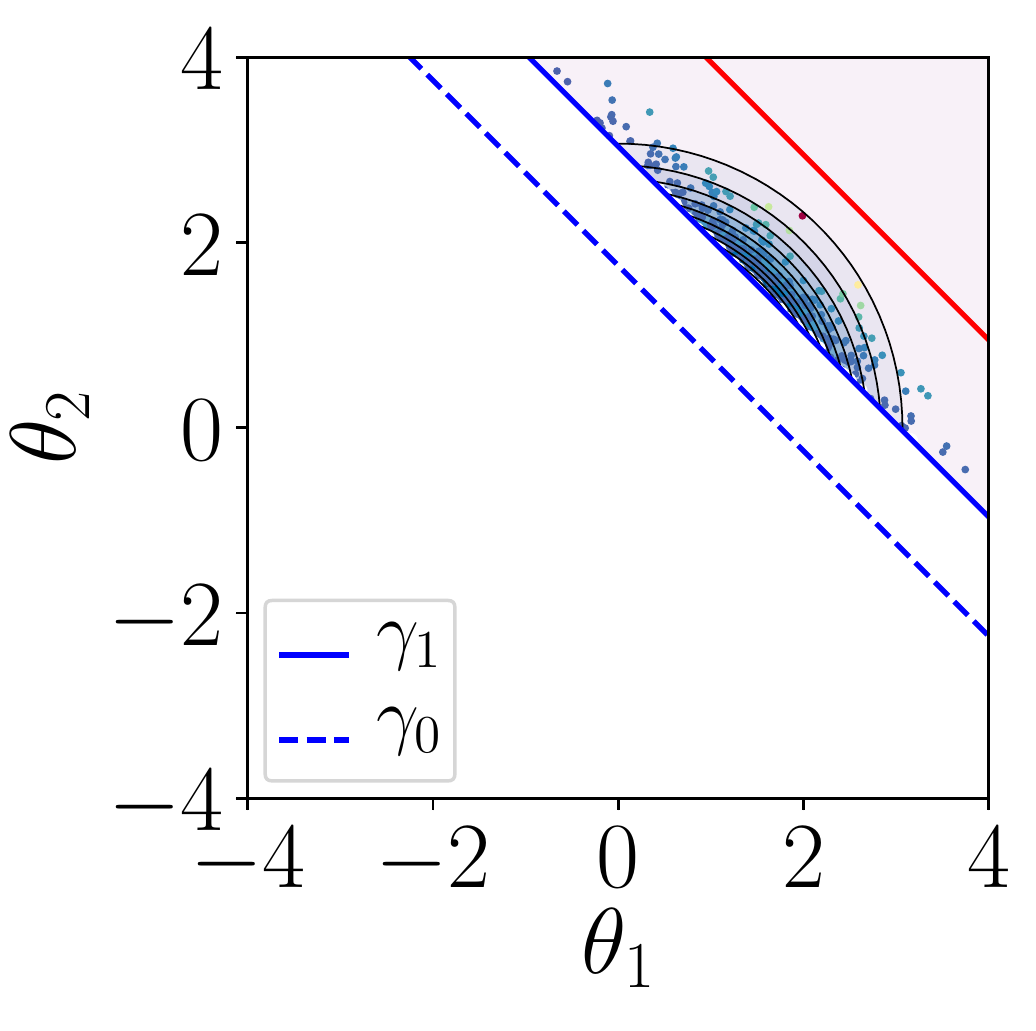}
\includegraphics[width=0.24\textwidth]{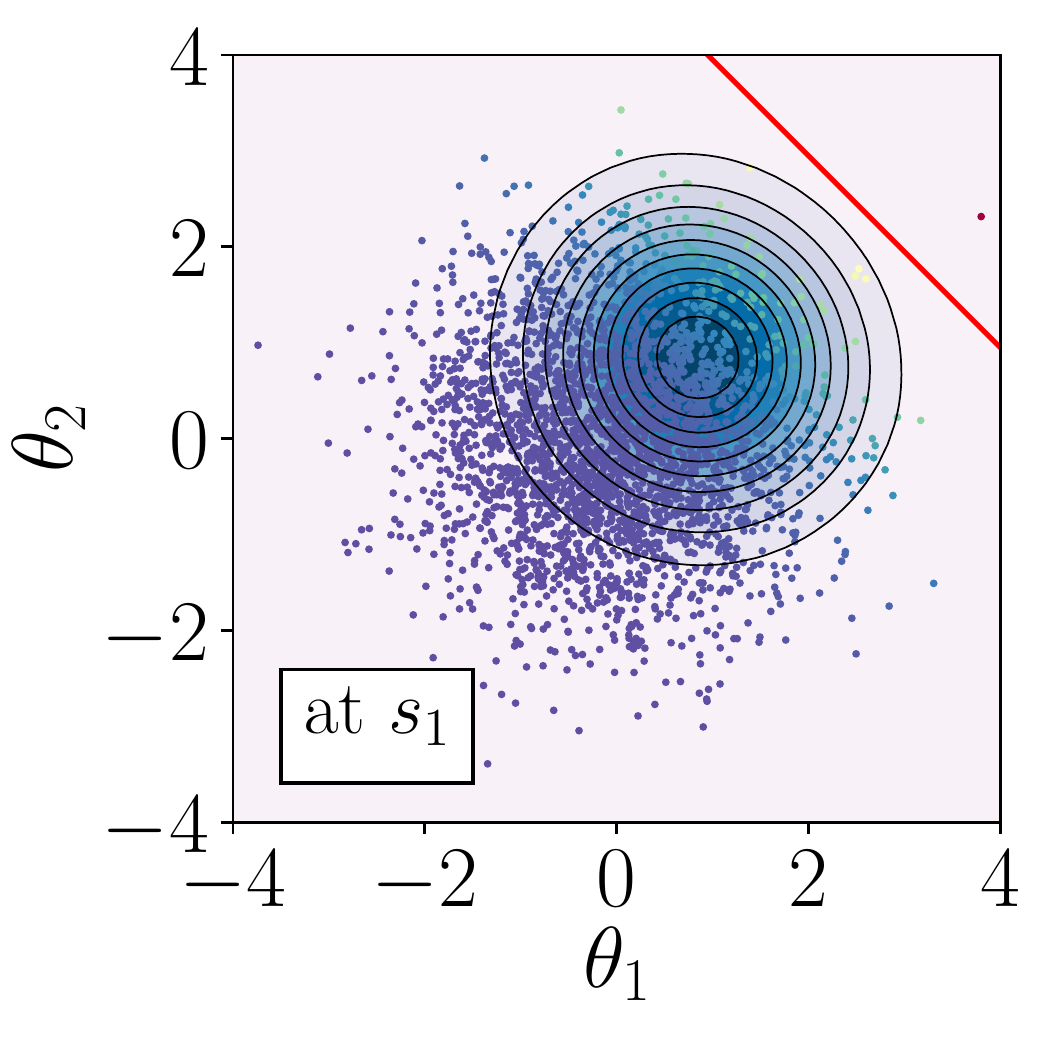}
\includegraphics[width=0.24\textwidth]{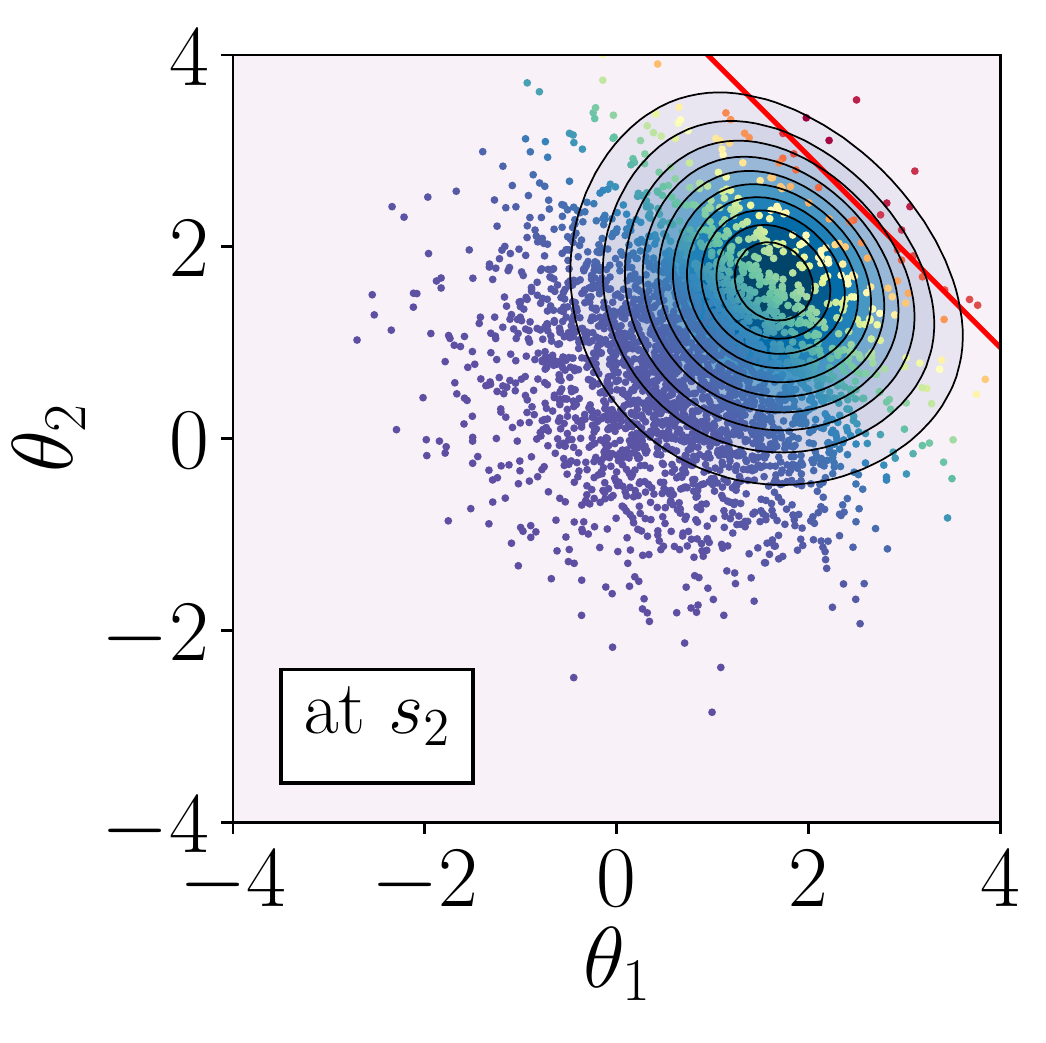}
\caption{Samples and contours of the intermediate optimal biasing densities at two initial simulation levels ($d=2$). The standard CE method uses only $N\cdot\rho$ elite weighted samples. The iCE method uses all $N$ weighted samples. The red solid line marks the target failure level and the color of the samples represents their weights.}	
\label{fig:optbias}
\end{figure}

A comparison between the intermediate optimal biasing densities in CE and iCE is shown in \cref{fig:optbias}.

\section{Construction of the failure-informed subspace}\label{sec:dimred}
As described in \cref{subsec:rare_event}, the probability of failure $p_\mathcal{F}$ and indicator $\I{\mathcal{F}}{\ve{\theta}}$ are analogous to the model evidence and likelihood function in Bayesian inference; thus, rare event simulation can be interpreted as a Bayesian inverse problem. Several dimension reduction techniques are available for nonlinear Bayesian inference \cite{constantine_et_al_2014, cui_et_al_2014, zahm_et_al_2018a}. The idea is to identify a certain low-dimensional structure in the parameter space, corresponding to the parameter directions along which the posterior differs most strongly from the prior. Particularly, the certified dimension reduction (CDR) approach \cite{zahm_et_al_2018a} can be applied to any type of measurable likelihood function, as long as the gradient of its natural logarithm is square integrable. CDR provides low-dimensional approximations that are equipped with certified error bounds (in the sense of the KL divergence). As a result, the approximation can be controlled by some user-defined threshold bounding the KL divergence from the exact to the approximated posterior. We build on the ideas of \cite{zahm_et_al_2018a} to construct low-dimensional approximations of the posterior-failure distribution. 

The indicator function $\I{\mathcal{F}}{\ve{\theta}}$ is upper semi-continuous and does not provide sufficient regularity to construct the certified approximation ($\nabla \ln\I{\mathcal{F}}{\ve{\theta}}$ is not square-integrable). Hence, we use one of the smooth representations $f({\ve{\theta};s})$ in \cref{eq:ind}, to express the posterior-failure density as $\pi_{\mathcal{F}}\propto f({\ve{\theta};s}) \pi_{\text{pr}}(\ve{\theta})$, with $s\to 0$. The idea is to find a low-dimensional approximation ${\pi}_{\mathcal{F}}^{(r)}$ such that,
\begin{equation}\label{eq:FIS}
\pi_{\mathcal{F}}(\ve{\theta}) \approx \pi_{\mathcal{F}}^{(r)}(\ve{\theta}) \propto (h \circ \mat{P}_r)(\ve{\theta})\pi_{\text{pr}}(\ve{\theta}),
\end{equation}
where $h:\mathbbm{R}^{d}\to \mathbbm{R}_{>0}$ is a \emph{profile function} and the \emph{projector} $\mat{P}_r\in \mathbbm{R}^{d\times d}$ is a linear transformation such that $\mat{P}_r^2=\mat{P}_r$, but not necessarily $\mat{P}_r^\tran=\mat{P}_r$; we can also define the \emph{complementary projector} to $\mat{P}_r$ as the matrix $\mat{P}_\perp = \mat{I}_d-{\mat{P}_r}$, which satisfies $\mathrm{Im}(\mat{P}_\perp) = \mathrm{Ker}({\mat{P}_r})$. 

The profile function in \cref{eq:FIS} depends only on $\mat{P}_r\ve{\theta}=\ve{\theta}_r$, which is defined on the \emph{failure-informed subspace} (FIS) $\mat{\Theta}_r=\mathrm{Im}({\mat{P}_r})$, and it is essentially constant along the \emph{complementary subspace} (CS) $\mat{\Theta}_\perp=\mathrm{Ker}({\mat{P}_r})$. As a result, if $r\ll d$, the goal of the approximation \cref{eq:FIS} is to replace the high-dimensional smooth indicator by a function of fewer variables. In the following, we employ the ideas of \cite{zahm_et_al_2018a} to describe how the profile function and projector are obtained.

\subsection{Optimal profile function}
For a fixed projector $\mat{P}_r$, $f(\ve{\theta};s)$ can be approximated by its average over all values of $\ve{\theta}$ that map to $\ve{\theta}_r=\mat{P}_r\ve{\theta}$ \cite{rosenthal_2006}. This is the \emph{conditional expectation} of the smooth indicator given the projector under the prior distribution \cite{zahm_et_al_2018a}
\begin{equation}\label{eq:cond_expec}
\E{{\pi}_{\text{pr}}}{f(\ve{\theta};s)\given \mat{P}_r\ve{\theta}} = \int_{\widetilde{\mat{\Theta}}_\perp}  f(\mat{P}_r\ve{\theta}+\mat{\Phi}_\perp\ve{\xi}_\perp; s)~ \underbrace{\dfrac{{\pi}_{\text{pr}}(\mat{P}_r\ve{\theta}+\mat{\Phi}_\perp\ve{\xi}_\perp)}{\int_{\widetilde{\mat{\Theta}}_\perp} {\pi}_{\text{pr}}(\mat{P}_r\ve{\theta}+\mat{\Phi}_\perp\ve{\xi}_\perp')\dd \ve{\xi}_\perp'}}_{\pi(\ve{\xi}_\perp\given\mat{P}_r\ve{\theta} )} ~\dd \ve{\xi}_\perp,
\end{equation}
where $\ve{\xi}_\perp\in\widetilde{\mat{\Theta}}_\perp\subseteq \mathbbm{R}^{d-r}$, and the columns of $\mat{\Phi}_\perp\in\mathbbm{R}^{d\times d-r}$ form a basis for $\mathrm{Ker}({\mat{P}_r})$. 

The optimal profile function $h^\star$ is obtained by minimizing the KL divergence between the exact ${\pi}_{\mathcal{F}}$ and approximated $\pi_{\mathcal{F}}^{(r)}$ densities. We can use the conditional expectation \cref{eq:cond_expec} to define:
\begin{equation}\label{eq:opt_approx}
\pi_{\mathcal{F}}^{(r)\star}\propto \underbrace{ \E{{\pi}_{\text{pr}}}{f(\ve{\theta};s)\given \mat{P}_r\ve{\theta}}}_{(h^\star\circ\mat{P}_r)(\ve{\theta})} \pi_{\text{pr}}(\ve{\theta}),
\end{equation}
for which the relation, $\Dklb{{\pi}_{\mathcal{F}}}{\pi_{\mathcal{F}}^{(r)}}-\Dklb{{\pi}_{\mathcal{F}}}{\pi_{\mathcal{F}}^{(r)\star}}=\Dklb{\pi_{\mathcal{F}}^{(r)\star}}{\pi_{\mathcal{F}}^{(r)}}\geq 0$ holds (see, \cite{zahm_et_al_2018a}). In particular, $\Dklb{\pi_{\mathcal{F}}}{\pi_{\mathcal{F}}^{(r)}} \geq \Dklb{{\pi}_{\mathcal{F}}}{\pi_{\mathcal{F}}^{(r)\star}}$, and hence a profile function of the form $(h^\star\circ\mat{P}_r)(\ve{\theta})=\E{{\pi}_{\text{pr}}}{f(\ve{\theta};s)\given \mat{P}_r\ve{\theta}}$ is a minimizer of $\Dklb{{\pi}_{\mathcal{F}}}{\pi_{\mathcal{F}}^{(r)}}$.

\subsection{Optimal projector}
Consider a Gaussian prior density $\pi_{\mathrm{pr}}(\ve{\theta})\propto \exp(-V(\ve{\theta}))$, where the function $V(\ve{\theta})\allowbreak= \frac{1}{2}(\ve{\theta}-\ve{\mu}_{\text{pr}})^\tran\mat{\Sigma}_{\text{pr}}^{-1}(\ve{\theta}-\ve{\mu}_{\text{pr}})$ is twice differentiable and satisfies $\nabla^2 V(\ve{\theta})=\mat{\Sigma}_{\text{pr}}^{-1}\succeq c\cdot\mat{I}_d$, with $c> 0$. Theorem 1 in \cite{zahm_et_al_2018a} shows that, for any continuously differentiable function $h:\mathbbm{R}^d\to\mathbbm{R}$, such that $\mathds{E}_{\pi_{\mathrm{pr}}}[\norm{\nabla h(\ve{\theta}) }_{\mat{\Sigma}_{\text{pr}}}^2] <\infty$, the following \emph{subspace logarithmic Sobolev inequality} holds 
\begin{align}\label{eq:subspace_sobolev}
\E{\pi_{\mathrm{pr}}}{h^2(\ve{\theta}) \ln\left(\frac{ h^2(\ve{\theta})}{\E{\pi_{\mathrm{pr}}}{h^2(\ve{\theta})\given\mat{P}_r\ve{\theta}}}\right) } &\leq 2\E{\pi_{\mathrm{pr}}}{ \norm{(\mat{I}_d-\mat{P}_r^\tran)\nabla h(\ve{\theta}) }_{\mat{\Sigma}_{\text{pr}}}^2},
\end{align}
with the notation $\norm{\nabla h(\ve{\theta}) }_{\mat{\Sigma}_{\text{pr}}}^2 = \nabla h(\ve{\theta})^\tran\mat{\Sigma}_{\text{pr}}\nabla h(\ve{\theta})$. The inequality \cref{eq:subspace_sobolev} allows one to bound $\Dklb{{\pi}_{\mathcal{F}}}{\pi_{\mathcal{F}}^{(r)\star}}$, which in turn provides a way to characterize the optimal projector in the approximation $\pi_{\mathcal{F}}^{(r)\star}$ of \cref{eq:opt_approx}. 

Let $h^2(\ve{\theta})$ be equal to the normalized smooth indicator function $f(\ve{\theta};s)/p$, such that $\nabla h(\ve{\theta}) = \frac{1}{2} (f(\ve{\theta};s)/p)^{\nicefrac{1}{2}} \nabla \ln f(\ve{\theta};s)$; then \cref{eq:subspace_sobolev} becomes
\begin{align}\label{eq:logsobsub}
\E{\pi_{\mathrm{pr}}}{\frac{f(\ve{\theta};s)}{p} \ln\left(\frac{ f(\ve{\theta};s)/p}{\E{\pi_{\mathrm{pr}}}{f(\ve{\theta};s)\given\mat{P}_r\ve{\theta}}/\pf}\right)  } &\leq \frac{1}{2}~ \E{\pi_{\mathrm{pr}}}{\frac{f(\ve{\theta};s)}{p} \norm{(\mat{I}_d-\mat{P}_r^\tran)\nabla \ln f(\ve{\theta};s) }_{\mat{\Sigma}_{\text{pr}}}^2 }.
\end{align}

The left-hand side in \cref{eq:logsobsub} is equal to $\Dklb{{\pi}_{\mathcal{F}}}{\pi_{\mathcal{F}}^{(r)\star}}$; hence we obtain the bound
\begin{align}\nonumber
\Dkl{{\pi}_{\mathcal{F}}}{\pi_{\mathcal{F}}^{(r)\star}} &\leq \frac{1}{2}~\E{\pi_{\mathcal{F}}}{ \norm{(\mat{I}_d-\mat{P}_r^\tran)\nabla \ln f(\ve{\theta};s) }_{\mat{\Sigma}_{\text{pr}}}^2 }
\\\nonumber
&= \frac{1}{2}~\E{\pi_{\mathcal{F}}}{ \mathrm{tr}\left(   \mat{\Sigma}_{\text{pr}} (\mat{I}_d-\mat{P}_r^\tran) \nabla \ln f(\ve{\theta};s) \nabla \ln f(\ve{\theta};s)^\tran (\mat{I}_d-\mat{P}_r)  \right) }\\ \label{eq:dKL_Pr_bound}
&= \frac{1}{2}~\mathrm{tr}\left(   \mat{\Sigma}_{\text{pr}} (\mat{I}_d-\mat{P}_r^\tran) ~\mat{H}~ (\mat{I}_d-\mat{P}_r)  \right) \eqqcolon \frac{1}{2}~\mathcal{R}(\mat{P}_r, \mat{H}),
\end{align}
where $\mathcal{R}(\mat{P}_r, \mat{H})$ is the mean-squared error of the approximation of $\nabla \ln f(\ve{\theta};s)$ by $\mat{P}_r^\tran \nabla \ln f(\ve{\theta};s)$ (with the parameter $\ve{\theta}\sim \pi_{\mathcal{F}}$), and $\mat{H}\in \mathbbm{R}^{d\times d}$ is the second moment matrix of the gradient of the log-smooth indicator function \cite{zahm_et_al_2018a}
\begin{equation}\label{eq:smooth_H}
\mat{H} = \int_{\mat{\Theta}} \nabla \ln f(\ve{\theta};s) \nabla \ln f(\ve{\theta};s)^\tran {\pi}_{\mathcal{F}}(\ve{\theta}) \dd\ve{\theta}= \E{{\pi}_{\mathcal{F}}}{\nabla \ln f(\ve{\theta};s) \nabla \ln f(\ve{\theta};s)^\tran }.
\end{equation}

Note from \cref{eq:dKL_Pr_bound} that the mean-squared error is quadratic in $\mat{P}_r$. Therefore, one can minimize the upper bound in \cref{eq:dKL_Pr_bound} over the collection of $r$-rank projectors to find an optimal projector:
\begin{equation}\label{eq:min_reconstruct}
\mat{P}_r = \argmin_{\mat{P}_r'\in\mathbbm{R}^{d\times d}} \mathcal{R}(\mat{P}_r',\mat{H}). 
\end{equation}

As shown by Proposition 2 in \cite{zahm_et_al_2018a}, a solution to \cref{eq:min_reconstruct} is
\begin{equation}\label{eq:optimal_project}
\mat{P}_r = \left(\sum_{i=1}^r \ve{\phi}_i \ve{\phi}_i^\tran\right) \mat{\Sigma}_{\text{pr}}^{-1} \qquad \mathrm{and} \qquad \min \mathcal{R}(\mat{P}_r,\mat{H})= \sum_{i=r+1}^d \lambda_i,
\end{equation}
where the eigenpairs $(\lambda_i,\ve{\phi}_i)$ correspond to the solution of the generalized eigenvalue problem $\mat{H}\ve{\phi}_i =\lambda_i\mat{\Sigma}_{\text{pr}}^{-1}\ve{\phi}_i,~i=1,\ldots, d$. Hence, if $\pi_{\mathrm{pr}}$ satisfies the inequality \cref{eq:subspace_sobolev} and $\mat{P}_r$ is defined as in \cref{eq:optimal_project}, the approximation $\pi_{\mathcal{F}}^{(r)\star}$ in \cref{eq:opt_approx} can be controlled by a desired tolerance $\varepsilon\geq0$ as 
\begin{equation}\label{eq:KL_opt}
\Dkl{{\pi}_{\mathcal{F}}}{\pi_{\mathcal{F}}^{(r)\star}} \leq \frac{1}{2}\sum_{i=r+1}^d \lambda_i \leq \varepsilon.
\end{equation}

In this case, bounding $\Dklb{{\pi}_{\mathcal{F}}}{\pi_{\mathcal{F}}^{(r)\star}}$ amounts to  selecting the rank $r$ as the smallest integer such that the left-hand side of \cref{eq:KL_opt} is below a prescribed $\varepsilon$, i.e., $r = \min\left\{r': \frac{1}{2}\sum_{i=r'+1}^d \lambda_i \leq \varepsilon\right\}$. Note that, along with the prior, the second-moment matrix of the log-smooth indicator function $\mat{H}$ reveals the effective low dimension of the posterior-failure distribution. A sharp decay in the generalized spectrum of $(\mat{H},\mat{\Sigma}_{\text{pr}}^{-1})$ guarantees the existence of a low-rank approximation $r\ll d$.

\section{Failure-informed cross-entropy-based importance sampling}\label{sec:iCEred} 
As shown in \cref{subsec:IS}, the optimal IS biasing density $\pi^\star_{\text{bias}}$ is equivalent to the posterior-failure density $\pi_{\mathcal{F}}$. Hence, we can employ the theory of \cref{sec:dimred} to derive low-dimensional biasing distributions that extend the application of the iCE method to high-dimensional parameter spaces that have an intrinsic low-dimensional structure. Recall that the goal is to represent $\pi_{\mathcal{F}}(\ve{\theta})= {\pi}_{\text{bias}}^\star(\ve{\theta}) \propto f(\ve{\theta};s) \pi_{\text{pr}}(\ve{\theta})$ (as $s\to 0$) on the FIS using the certified approximation \cref{eq:opt_approx}.

\subsection{Formulation}
Based on \cref{rem:N01}, we can impose that the prior $\pi_{\text{pr}}(\ve{\theta})$ is standard Gaussian. The solution of the generalized eigenvalue problem $(\mat{H},\mat{\Sigma}_{\mathrm{pr}}^{-1})$ required for the construction of $\mat{P}_r$ in \cref{eq:optimal_project}, reduces to a standard eigenvalue problem for $\mat{H}$. As a result, the projector can be written as $\mat{P}_r= \mat{\Phi}_r\mat{\Phi}_r^\tran$, where $\mat{\Phi}_r\in \mathbbm{R}^{d\times r}$ contains the first $r$ eigenvectors defining a basis for the FIS. Similarly, the complementary projector can be obtained as $\mat{P}_\perp= \mat{\Phi}_\perp\mat{\Phi}_\perp^\tran$, where $\mat{\Phi}_\perp\in \mathbbm{R}^{d\times d-r}$ is defined by the remaining eigenvectors, generating a basis for the CS.

\begin{figure}[!ht]
\centering
\includegraphics[width=0.9\textwidth]{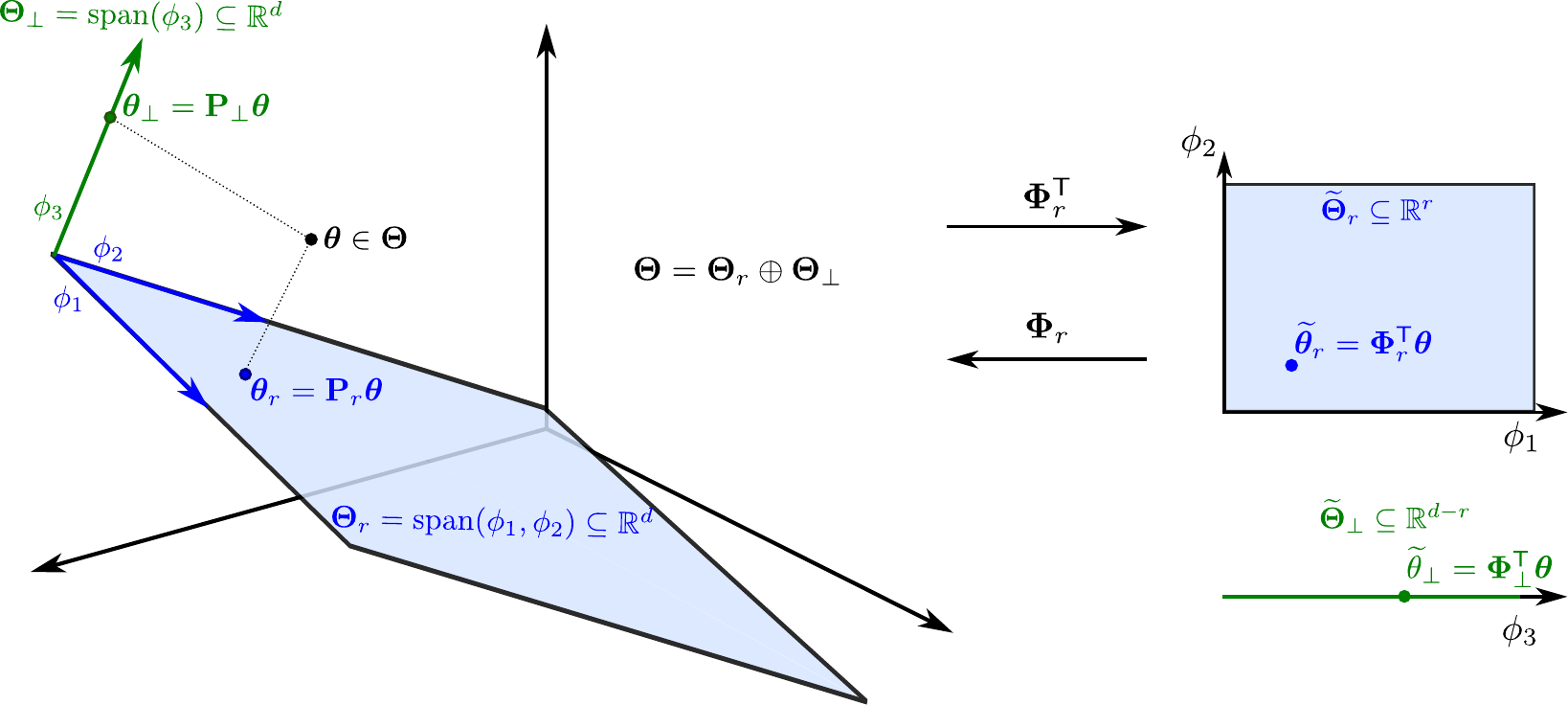}
\caption{Schematic representation of a parameter $\ve{\theta}\in\mathbbm{R}^3$ projected onto the FIS with $r=2$ (in blue): global ${\ve{\theta}}_r\in\mathbbm{R}^d$, local $\widetilde{\ve{\theta}}_r\in\mathbbm{R}^r$ (adapted from \cite{cui_et_al_2014}).}
\label{fig:fis}
\end{figure}

The projector $\mat{P}_r$ is orthogonal with respect to the prior precision matrix $\mat{\Sigma}_{\mathrm{pr}}^{-1} = \mat{I}_d$. This induces a decomposition of the parameter space as the direct sum ${\ve{\Theta}} = {\ve{\Theta}}_r \oplus {\ve{\Theta}}_\perp$, where any element of ${\ve{\Theta}}$ can be uniquely represented as $\ve{\theta}= \ve{\theta}_r + \ve{\theta}_{\perp}$, with $\ve{\theta}_r = \mat{P}_r\ve{\theta}\in \ve{\Theta}_r$ and $\ve{\theta}_{\perp} = \mat{P}_\perp\ve{\theta}\in \ve{\Theta}_\perp$. Moreover, we can use the operators $\mat{\Phi}_r$ and $\mat{\Phi}_\perp$ to map the parameters onto a local FIS $\widetilde{\mat{\Theta}}_r\subseteq\mathbbm{R}^{r}$ and local CS $\widetilde{\mat{\Theta}}_\perp\subseteq\mathbbm{R}^{d-r}$, respectively. In this case, the parameter vector can be written as $\ve{\theta}=\mat{\Phi}_r\widetilde{\ve{\theta}}_r + \mat{\Phi}_\perp\widetilde{\ve{\theta}}_\perp$, where $\widetilde{\ve{\theta}}=[\widetilde{\ve{\theta}}_r,\widetilde{\ve{\theta}}_\perp]$ has components $\widetilde{\ve{\theta}}_r = \mat{\Phi}_r^\tran{\ve{\theta}}\in \widetilde{\mat{\Theta}}_r$ and $\widetilde{\ve{\theta}}_\perp = \mat{\Phi}_\perp^\tran{\ve{\theta}}\in \widetilde{\mat{\Theta}}_\perp$. \Cref{fig:fis} illustrates the action of the projectors and basis operators.

From the discussion above, we see that the prior distribution can be factorized as $\pi_{\mathrm{pr}}(\widetilde{\ve{\theta}})= \pi_{\text{pr}}^{(r)}(\widetilde{\ve{\theta}}_r){\pi}_{\text{pr}}^{(\perp)}(\widetilde{\ve{\theta}}_\perp)$, where $\pi_{\text{pr}}^{(r)}(\widetilde{\ve{\theta}}_r)$ and ${\pi}_{\text{pr}}^{(\perp)}(\widetilde{\ve{\theta}}_\perp)$ are densities on the local FIS and CS, respectively \cite{cui_et_al_2014}. This choice, together with the certified approximation in \cref{eq:opt_approx}, allows us to define the optimal low-dimensional biasing density:
\begin{equation}\label{eq:IS_iCE_red_loc}
{\pi}_{\mathcal{F}}^{(r)\star}(\widetilde{\ve{\theta}})= {\pi}_{\text{bias}}^{\star}(\widetilde{\ve{\theta}}) \propto\underbrace{ \E{{\pi}_{\text{pr}}}{f(\ve{\theta};s)\given\ve{\Phi}_r\widetilde{\ve{\theta}}_r}~{\pi}_{\text{pr}}^{(r)}(\widetilde{\ve{\theta}}_r)}_{\text{reduced optimal biasing}} ~\underbrace{{\pi}_{\text{pr}}^{(\perp)}(\widetilde{\ve{\theta}}_\perp). }_{\text{complementary prior}}
\end{equation}

We select the parametric biasing distribution from a Gaussian family that decomposes as follows:
\begin{equation}\label{eq:near_opt_bias_red}
{\pi}_{\text{bias}}(\widetilde{\ve{\theta}}; \ve{\upsilon}_r) = {\pi}_{\text{bias}}^{(r)}(\widetilde{\ve{\theta}}_r;\ve{\upsilon}_r)~{\pi}_{\text{pr}}^{(\perp)}(\widetilde{\ve{\theta}}_\perp);
\end{equation}
this family has a general Gaussian form on the local FIS, where $\ve{\upsilon}_r\in\Upsilon_r$ represents the mean and covariance matrix of the $r$-dimensional reduced Gaussian density, and is equal to the standard Gaussian prior in the complementary directions.

Aiming at the optimal biasing density \cref{eq:IS_iCE_red_loc} using biasing densities of the form \cref{eq:near_opt_bias_red}, the CE optimization problem can be re-defined on a low-dimensional space. The KL divergence between \cref{eq:IS_iCE_red_loc} and \cref{eq:near_opt_bias_red} reads 
\begin{equation}
\Dkl{\pi_{\text{bias}}^\star}{{\pi}_{\text{bias}}}  = \E{\pi_{\text{bias}}^\star}{\ln\left(\dfrac{\pf^{-1}\E{{\pi}_{\text{pr}}}{f(\ve{\theta};s)\given\mat{\Phi}_r\widetilde{\ve{\theta}}_r}~{\pi}_{\text{pr}}^{(r)}(\widetilde{\ve{\theta}}_r)  }{{\pi}_{\text{bias}}^{(r)}(\widetilde{\ve{\theta}}_r;\ve{\upsilon}_r) }\right)}.
\end{equation}

As in \cref{subsec:CEmethod}, minimizing $\Dklb{\pi_{\text{bias}}^\star}{{\pi}_{\text{bias}}}$ is equivalent to the following maximization:
\begin{align}\nonumber
\ve{\upsilon}^\star_r &= \argmax_{\ve{\upsilon}_r\in\Upsilon_r} \E{\pi_{\text{bias}}^\star}{\ln{\pi}_{\text{bias}}^{(r)}(\widetilde{\ve{\theta}}_r; \ve{\upsilon}_r)} \\\nonumber
&= \argmax_{\ve{\upsilon}_r\in\Upsilon_r} \int_{\widetilde{\mat{\Theta}}_r}\int_{\widetilde{\mat{\Theta}}_\perp} \ln{\pi}_{\text{bias}}^{(r)}(\widetilde{\ve{\theta}}_r; \ve{\upsilon}_r) \E{{\pi}_{\text{pr}}}{f(\ve{\theta};s)\given\ve{\Phi}_r\widetilde{\ve{\theta}}_r}~{\pi}_{\text{pr}}^{(r)}(\widetilde{\ve{\theta}}_r){\pi}_{\text{pr}}^{(\perp)}(\widetilde{\ve{\theta}}_\perp)\dd\widetilde{\ve{\theta}}_{\perp}\dd\widetilde{\ve{\theta}}_{r}\\\nonumber
&= \argmax_{\ve{\upsilon}_r\in\Upsilon_r} \int_{\widetilde{\mat{\Theta}}_r}\left[\int_{\widetilde{\mat{\Theta}}_\perp}  f(\mat{\Phi}_r\widetilde{\ve{\theta}}_r+\mat{\Phi}_\perp\widetilde{\ve{\xi}}_{\perp}; s)~{\pi}_{\text{pr}}^{(\perp)}(\widetilde{\ve{\xi}}_\perp) \dd \widetilde{\ve{\xi}}_\perp\right]  \ln{\pi}_{\text{bias}}^{(r)}(\widetilde{\ve{\theta}}_r;\ve{\upsilon}_r)\pi_{\text{pr}}^{(r)}(\widetilde{\ve{\theta}}_r) \dd \widetilde{\ve{\theta}}_r\\ \label{eq:iCEred_ini}
&=  \argmax_{\ve{\upsilon}_r\in\Upsilon_r} \E{\pi_{\text{pr}}}{ f(\mat{\Phi}_r\widetilde{\ve{\theta}}_r + \mat{\Phi}_\perp\widetilde{\ve{\theta}}_\perp;s) \ln{\pi}_{\text{bias}}^{(r)}(\widetilde{\ve{\theta}}_r; \ve{\upsilon}_r)}.
\end{align}

Note that in this CE formulation there is no need for evaluating the conditional expectation in \cref{eq:IS_iCE_red_loc}, as it is normally the case when applying sequential IS in the Bayesian inference setting where Markov chain Monte Carlo techniques are required to draw samples from \cref{eq:IS_iCE_red_loc} (see, e.g., \cite{cui_et_al_2014,zahm_et_al_2018a}). Indeed, only an expectation with respect to the full prior density is required, thus avoiding the computational demands associated with double-loop procedures. 

Instead of directly taking the expectation with respect to the prior in \cref{eq:iCEred_ini}, we apply IS with biasing distribution ${\pi}_{\text{bias}}(\widetilde{\ve{\theta}}; \ve{\upsilon}'_r)$ in \cref{eq:near_opt_bias_red}:
\begin{equation}\label{eq:iCEred}
\ve{\upsilon}^\star_r = \argmax_{\ve{\upsilon}_r\in\Upsilon_r} \E{{\pi}_{\text{bias}}(\cdot; \ve{\upsilon}_r')}{f(\mat{\Phi}_r\widetilde{\ve{\theta}}_r + \mat{\Phi}_\perp\widetilde{\ve{\theta}}_\perp;s)  \ln{\pi}_{\text{bias}}^{(r)}(\widetilde{\ve{\theta}}_r;\ve{\upsilon}_r)~\dfrac{\pi_{\text{pr}}(\widetilde{\ve{\theta}})}{{\pi}_{\mathrm{bias}}(\widetilde{\ve{\theta}};\ve{\upsilon}_r')}}
\end{equation}
for reference parameters $\ve{\upsilon}'_r$. We apply the IS estimator of the expectation \cref{eq:iCEred} to define the stochastic optimization problem:
\begin{equation}\label{eq:IS_iCEred}
{\ve{\upsilon}}_r^\star\approx \widehat{\ve{\upsilon}}_r^\star = \argmax_{\ve{\upsilon}_r\in \Upsilon_r} \dfrac{1}{N}\sum_{i=1}^{N} f\left(\mat{\Phi}_r\widetilde{\ve{\theta}}_{r,i}+\mat{\Phi}_\perp\widetilde{\ve{\theta}}_{\perp,i}; s\right) \ln{\pi}_{\text{bias}}^{(r)}\left(\widetilde{\ve{\theta}}_{r,i};\ve{\upsilon}_r\right)  w\left(\widetilde{\ve{\theta}}_{i};\ve{\upsilon}'_r\right) 
\end{equation}
where $\{\widetilde{\ve{\theta}}_i=[\widetilde{\ve{\theta}}_{r,i},\widetilde{\ve{\theta}}_{\perp,i}]\}_{i=1}^{N} \overset{\text{i.i.d.}}{\sim} {\pi}_{\text{bias}}(\cdot;\ve{\upsilon}'_r)$ and $w(\widetilde{\ve{\theta}}_{i};\ve{\upsilon}'_r) = \pi_{\text{pr}}(\widetilde{\ve{\theta}})/{\pi}_{\mathrm{bias}}(\widetilde{\ve{\theta}};\ve{\upsilon}_r')$. As in the iCE method, the optimization \cref{eq:IS_iCEred} is performed sequentially by introducing a set of smoothing parameters $\{s_j>0\}_{j=0}^{n_{\text{lv}}}$. 

Starting in the full-dimensional space at level $j=0$, the iCE method applied to the FIS requires two main steps: (i) finding the projector and associated basis operators, and (ii) updating the reduced reference parameters $\widehat{\ve{\upsilon}}_{r}^{(j+1)}$ of the next intermediate biasing density. To achieve both steps, we use samples from the current biasing density ${{\pi}_{\text{bias}}(\cdot; \widehat{\ve{\upsilon}}_{r}^{(j)})}$. In the following, we refer to $\ve{\theta}^{(j)}\in\mathbbm{R}^{d\times N}$ as the matrix containing the full set of samples, such that $\ve{\theta}^{(j)}=\{\ve{\theta}_i\}_{i=1}^{N}\sim {{\pi}_{\text{bias}}(\cdot; \widehat{\ve{\upsilon}}_{r}^{(j)})}$ for each sample $\ve{\theta}_i\in\mathbbm{R}^{d}$.

For the first step, we approximate the matrix $\mat{H}$ in \cref{eq:smooth_H} at each level by the self-normalized IS estimator:
\begin{align}\label{eq:H_MCS}
{\mat{H}}^{(j+1)} \approx \widehat{\mat{H}}^{(j+1)} &= \dfrac{1}{\widetilde{W}^{(j+1)} } \sum_{i=1}^{N}\widetilde{w}_i^{(j+1)}\left[\nabla \ln f(\ve{\theta}_i; s_{j+1})\right] \left[\nabla \ln f(\ve{\theta}_i; s_{j+1})\right]^\tran 
\end{align}
where $\widetilde{\ve{w}}^{(j+1)} = f(\ve{\theta}^{(j)}; s_{j+1}){\ve{w}}^{(j)}$ represents the weight vector computed with respect to the smooth indicator and $\widetilde{W}^{(j+1)}=\sum_{i=1}^{N}\widetilde{w}_i^{(j)}$ is the sum of the weights. Thereafter, we compute the eigenpairs of the estimator \cref{eq:H_MCS} to construct  $\mat{\Phi}_r^{(j+1)},\mat{\Phi}_{\perp}^{(j+1)}$ and each basis is used to project the full set of samples ${\ve{\theta}}^{(j)}$ onto the local FIS and CS as $\widetilde{\ve{\theta}}^{(j+1)} = [\widetilde{\ve{\theta}}_r^{(j+1)},\widetilde{\ve{\theta}}_\perp^{(j+1)}] = [\mat{\Phi}_{r}^{(j+1),\tran},\mat{\Phi}_{\perp}^{(j+1),\tran}]\cdot\ve{\theta}^{(j)}$, with the notation $\widetilde{\ve{\theta}}_r^{(j+1)}\in\mathbbm{R}^{r\times N}$ and $\widetilde{\ve{\theta}}_\perp^{(j+1)}\in\mathbbm{R}^{d-r\times N}$.

For the second step, we update the reference parameters at each level solving the iCE stochastic optimization problem \cref{eq:IS_iCEred} in the reduced coordinate system:
\begin{equation}\label{eq:IS_iCEred_levels}
\widehat{\ve{\upsilon}}_{r}^{(j+1)} = \argmax_{\ve{\upsilon}_r\in \Upsilon_r} \dfrac{1}{N}\sum_{i=1}^{N} \ln{\pi}_{\text{bias}}^{(r)}\left(\widetilde{\ve{\theta}}_{r,i}^{(j+1)};\ve{\upsilon}_r\right) \overline{w}_i^{(j+1)};\quad \overline{w}_i^{(j+1)} = f(\ve{\theta}^{(j)}_{i}; s_{j+1}) \dfrac{\pi_{\text{pr}}(\widetilde{\ve{\theta}}^{(j+1)}_{i})}{{\pi}_{\text{bias}}(\widetilde{\ve{\theta}}^{(j+1)}_{i}; \overline{\ve{\upsilon}}^{(j+1)})},
\end{equation} 
where $\overline{\ve{w}}^{(j+1)}=\{\overline{w}_i^{(j+1)}\}_{i=1}^N$ denotes adjusted weights. This correction is required because the parametric biasing density is evaluated on samples that belong to the new basis, but it is defined with reference parameters estimated at the previous level. Therefore, the reference parameters $\widehat{\ve{\upsilon}}_{r}^{(j)}$ (estimated with the basis $\mat{\Phi}_{r}^{(j)}$) need to be expressed in the coordinate system induced by $\mat{\Phi}_{r}^{(j+1)}, \mat{\Phi}_{\perp}^{(j+1)}$. Since we employ the Gaussian parametric family and the coordinate transformation is linear, the corresponding mean and covariance are computed as
\begin{subequations}\label{eq:IS_iCEred_rottrans_params}
{\small
\begin{align}
\E{}{\widetilde{\ve{\theta}}^{(j+1)}} &=\left[\mat{\Phi}_{r}^{(j+1),\tran},~ \mat{\Phi}_{\perp}^{(j+1),\tran}\right]\cdot \ve{\mu}\\ 
\cov{\widetilde{\ve{\theta}}^{(j+1)},\widetilde{\ve{\theta}}^{(j+1)}} &=
\begin{bmatrix}
\mat{\Phi}_{r}^{(j+1),\tran}\mat{\Sigma}_1\mat{\Phi}_{r}^{(j+1)}+\mat{\Phi}_{r}^{(j+1),\tran}\mat{\Sigma}_2\mat{\Phi}_{r}^{(j+1)} &  \mat{\Phi}_{r}^{(j+1),\tran}\mat{\Sigma}_1\mat{\Phi}_{\perp}^{(j+1)}+\mat{\Phi}_{r}^{(j+1),\tran}\mat{\Sigma}_2\mat{\Phi}_{\perp}^{(j+1)}\\
\mat{\Phi}_{\perp}^{(j+1),\tran}\mat{\Sigma}_1\mat{\Phi}_{r}^{(j+1)}+\mat{\Phi}_{\perp}^{(j+1),\tran}\mat{\Sigma}_2\mat{\Phi}_{r}^{(j+1)} &  \mat{\Phi}_{\perp}^{(j+1),\tran}\mat{\Sigma}_1\mat{\Phi}_{\perp}^{(j+1)}+\mat{\Phi}_{\perp}^{(j+1),\tran}\mat{\Sigma}_2\mat{\Phi}_{\perp}^{(j+1)}
\end{bmatrix},
\end{align} }%
\end{subequations}
where $\ve{\mu}=\mat{\Phi}_{r}^{(j)} \ve{\mu}_r^{(j)}$, $\mat{\Sigma}_1=\mat{\Phi}_{r}^{(j)} \mat{\Sigma}_r^{(j)} \mat{\Phi}_{r}^{(j),\tran}$, $\mat{\Sigma}_2=\mat{\Phi}_{\perp}^{(j)}  \mat{\Phi}_{\perp}^{(j),\tran}$, and the parameters $\ve{\mu}_r^{(j)}$ and $\mat{\Sigma}_r^{(j)}$ constitute the reference parameter $\widehat{\ve{\upsilon}}_r^{(j)}$ estimated at level $j$. The components of the adjusted reference parameter $\overline{\ve{\upsilon}}^{(j+1)}$ used for the computation of the adjusted weights are then given by \cref{eq:IS_iCEred_rottrans_params}. 

The complete procedure of the improved cross-entropy method with failure-informed dimension reduction is summarized in \cref{alg:iCE_red}.
\begin{remark}
	The number of samples required to obtain an accurate estimate $\widehat{\mat{H}}$ in \cref{eq:H_MCS} is related to the effective rank of $\mat{H}$. The heuristic rule $N=\alpha\cdot r\ln(d)$ provides some intuition about the required sample size; here $\alpha\in[2,10]$ is an oversampling factor (see, e.g., \cite[p.35]{constantine_2015}). As a result, we can modify \cref{alg:iCE_red} such that we employ two types of sample sizes, one $N_{\mathrm{LSF}}$ associated to LSF evaluations and another $N_{\mathrm{grad}}$ related to the gradient evaluations. The value of $N_{\mathrm{grad}}$ can be adapted using the heuristic rule at each intermediate level of iCEred. This suggested adaption strategy is left for future study. 
\end{remark}

\IncMargin{1em}
\begin{algorithm}[!ht]
	\small
	\caption{iCEred: improved cross-entropy method with failure-informed dimension reduction.}
	\label{alg:iCE_red}
	\SetKwInOut{Input}{Input}
	\SetKwInOut{Output}{Output}
	\Input{dimension of the parameter space $d$, number of samples per level $N$, LSF $g(\ve{\theta})$, smooth indicator function $f(\ve{\theta};s)$, tolerance in the approximation $\varepsilon$, target coefficient of variation $\delta$, maximum iterations $t_{\mathrm{max}}$}
	\BlankLine
	
	Set the standard Gaussian prior density ${\pi}_{\text{pr}}(\ve{\theta})$\;
	
	Set the parametric family of standard Gaussian biasing densities ${\pi}_{\text{bias}}(\ve{\theta}; \ve{\upsilon})$\;
	
	Set initial reference parameters $\widehat{\ve{\upsilon}}^{(0)}$ from the prior parameters, and the smoothing parameter $s_0\leftarrow \infty$\; 
	
	Initial level $j\leftarrow0$\;
	
	\While{\texttt{True}}{			
		\uIf{$j=0$}{
			Generate $N$ samples from the biasing distribution $\ve{\theta}^{(j)}\sim {\pi}_{\text{bias}}(\cdot; \widehat{\ve{\upsilon}}^{(j)})$\;	
			
			Since $\widehat{\ve{\upsilon}}^{(0)}$ are selected from the prior, set weights $\ve{w}^{(j)}$ equal to one \;
		}
		\Else{
			Generate $N$ samples from the reduced biasing distribution $\widetilde{\ve{\theta}}_{r}^{(j)}\sim {\pi}_{\text{bias}}(\cdot; \widehat{\ve{\upsilon}}_r^{(j)})$\;
			
			Compute the weights $\ve{w}^{(j)}\leftarrow\exp\left( \ln\pi_{\text{pr}}^{(r)}(\widetilde{\ve{\theta}}_r^{(j)})-\ln {\pi}_{\text{bias}}^{(r)}(\widetilde{\ve{\theta}}_r^{(j)};\widehat{\ve{\upsilon}}_r^{(j)})\right)$\;
			
			Contruct the full set of parameters $\ve{\theta}^{(j)} \leftarrow \mat{\Phi}_r^{(j)} \widetilde{\ve{\theta}}^{(j)}_r + \mat{\Phi}_\perp^{(j)} \widetilde{\ve{\theta}}_\perp^{(j)}$\;
		}
		
		Evaluate LSF $\ve{g}_{\text{eval}}\leftarrow g(\ve{\theta}^{(j)})$ and indicator function $\ve{d}_{\text{eval}} \leftarrow \ve{g}_{\text{eval}}\leq 0  = \I{\mathcal{F}}{\ve{\theta}^{(j)}}$\;
		
		Evaluate smooth indicator function $\ve{f}_{\text{eval}} \leftarrow f\!\left(\ve{\theta}^{(j)};s_j\right)$\;
		
		Compute the coefficient of variation of the ratio between the indicator function and its smooth approximation \vspace*{-.3cm}$$\widehat{\mathrm{cv}}^{(j)} = \dfrac{\sqrt{\widehat{\mathds{V}}\left[\ve{d}_{\text{eval}}/\ve{f}_{\text{eval}}\right]}}{{\widehat{\mathds{E}}[\ve{d}_{\text{eval}}/\ve{f}_{\text{eval}}]}}$$\; \vspace*{-.3cm}		
		
		\If{$(\widehat{\mathrm{cv}}^{(j)} \leq \delta)$ \emph{or} $(j\geq t_{\mathrm{max}})$}{\texttt{Break}}
		
		Use the samples $\ve{\theta}^{(j)}$ and the smooth indicator function to compute $s_{j+1}$ as per \cref{eq:s_adapt}\;
		
		Update the smooth indicator values  $\ve{f}_{\mathrm{eval}}\leftarrow f\!\left(\ve{\theta}^{(j)}; s_{j+1}\right)$\;
		
		Compute the weights associated to the smooth indicator function $\widetilde{\ve{w}}^{(j+1)}\leftarrow \ve{w}^{(j)} \cdot \ve{f}_{\mathrm{eval}}$ and its sum $\widetilde{W}$\;
		
		Compute the gradient of the log-smooth indicator function, $\nabla \ln f(\ve{\theta}^{(j)};s_{j+1})$\;
		
		Use $\nabla \ln f(\ve{\theta}^{(j)};s_{j+1})$ to compute the estimator $\widehat{\mat{H}}$ in \cref{eq:H_MCS} and solve the eigenvalue problem $\widehat{\mat{H}}\mat{\Phi}=\mat{\Lambda}\mat{\Phi}$\;
		
		Use the tolerance $\varepsilon$ to find the rank $r$ based on \cref{eq:KL_opt}\;
		
		Construct the FIS basis $\mat{\Phi}_r^{(j+1)} \leftarrow \ve{\Phi}_{1:r}$, and the CS basis $\mat{\Phi}_\perp^{(j+1)} \leftarrow \ve{\Phi}_{r+1:d}$\;
		
		Project the samples onto the local FIS and CS, $\widetilde{\ve{\theta}}_r\leftarrow\mat{\Phi}_r^{(j+1),\tran}\ve{\theta}^{(j)},\quad \widetilde{\ve{\theta}}_\perp\leftarrow\mat{\Phi}_\perp^{(j+1),\tran}\ve{\theta}^{(j)}$
		
		\uIf{$j>0$}{
			
			Compute the adjusted reference parameters $\overline{\ve{\upsilon}}$ using the mean and covariance in \cref{eq:IS_iCEred_rottrans_params}\;
			
			Compute the corrected weights $\overline{\ve{w}}\leftarrow \ve{f}_{\mathrm{eval}}\cdot\exp\left( \ln\pi_{\text{pr}}(\widetilde{\ve{\theta}})-\ln {\pi}_{\text{bias}}(\widetilde{\ve{\theta}};\overline{\ve{\upsilon}})\right)$\;
		}		
		
		Find the next reference parameters $\widehat{\ve{\upsilon}}^{(j+1)}_r$ by solving the stochastic optimization problem \cref{eq:IS_iCEred_levels} using samples $\widetilde{\ve{\theta}}_r$ and weights $\overline{\ve{w}}$.
		
		$j\leftarrow j+1$
	}
	Compute the IS estimator $\widehat{p}_{\mathcal{F}}\leftarrow \frac{1}{N}\sum_{i=1}^{N} {d}_{\text{eval}}(i)\cdot w_i^{(j)}$\;
	\BlankLine
	\Output{$\widehat{p}_{\mathcal{F}}$}
\end{algorithm}
\DecMargin{1em}

\begin{figure}[!ht]
	\centering
	\includegraphics[width=0.24\textwidth]{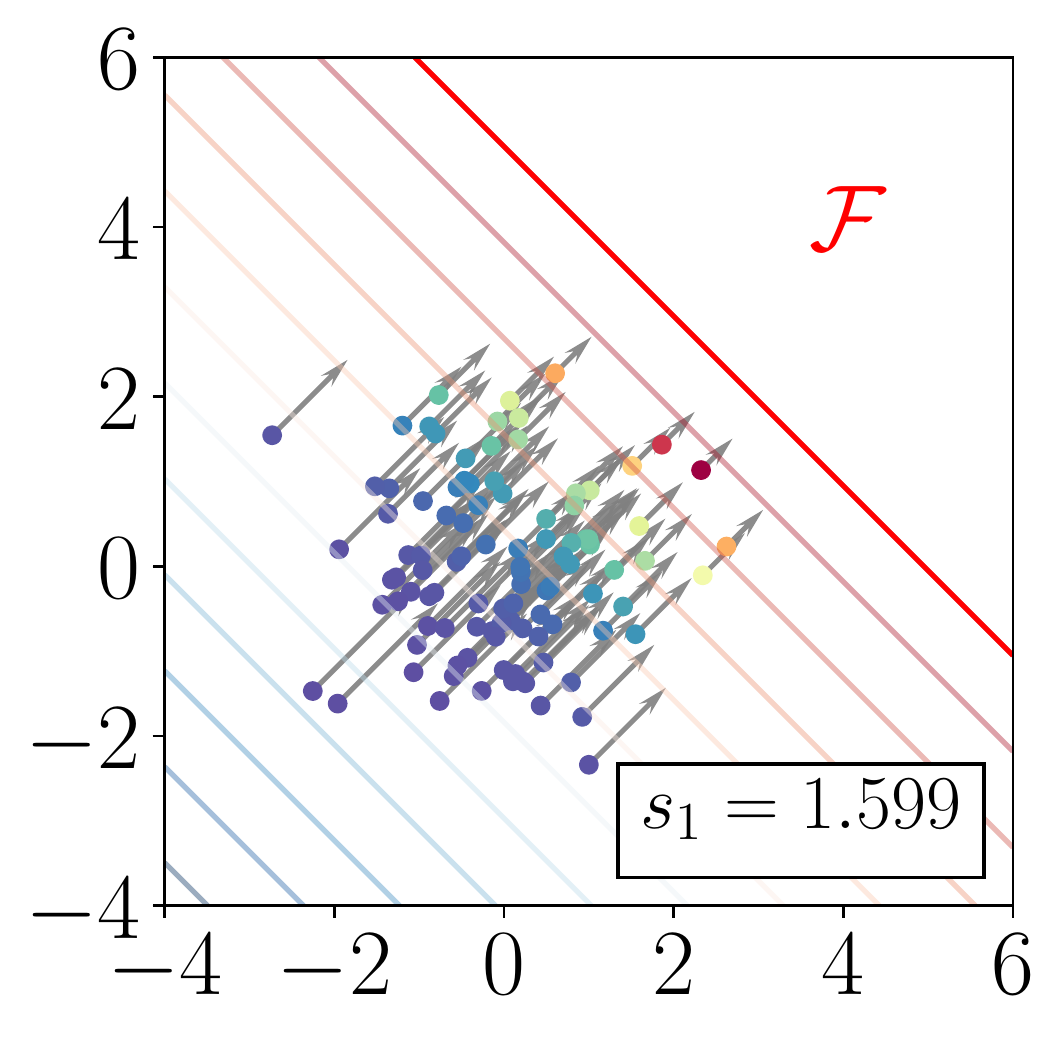}
	\includegraphics[width=0.24\textwidth]{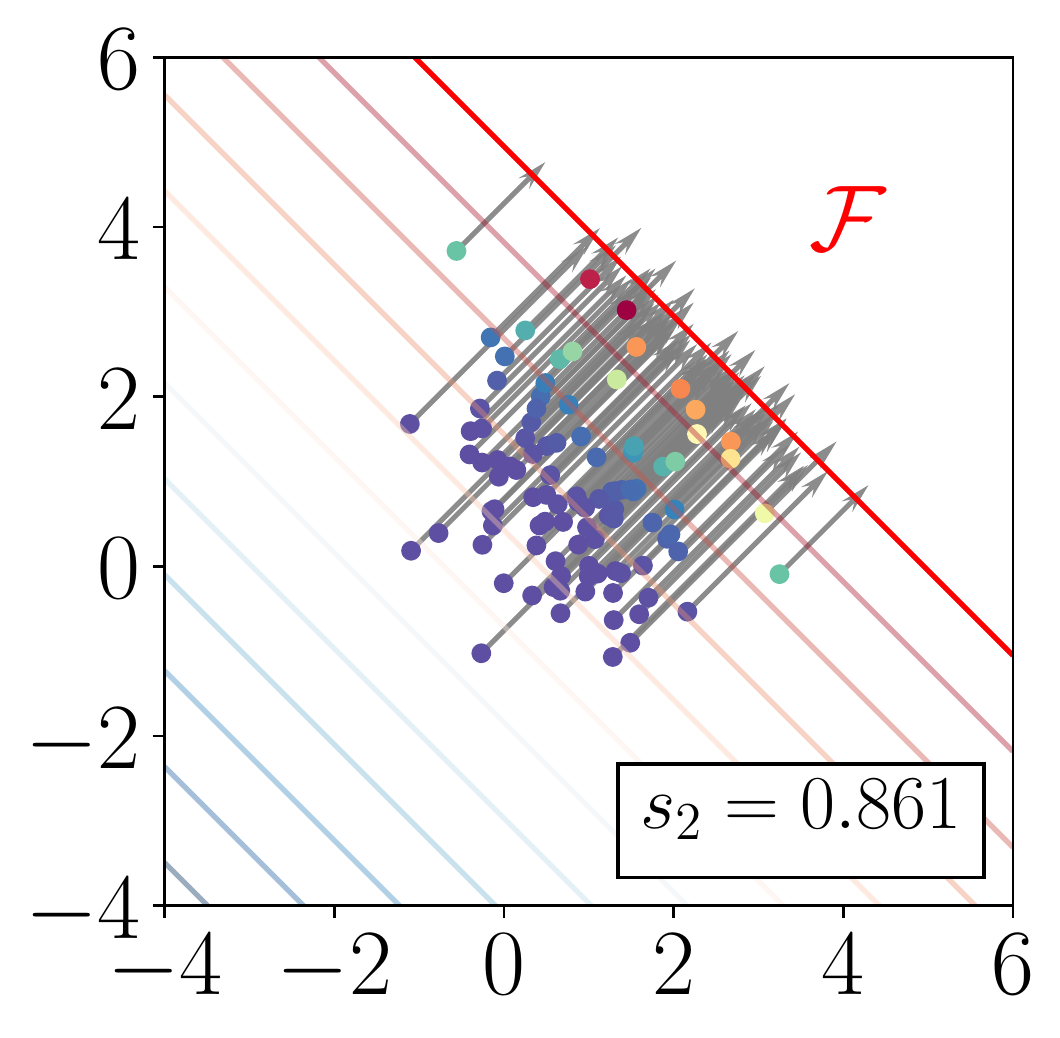}
	\includegraphics[width=0.24\textwidth]{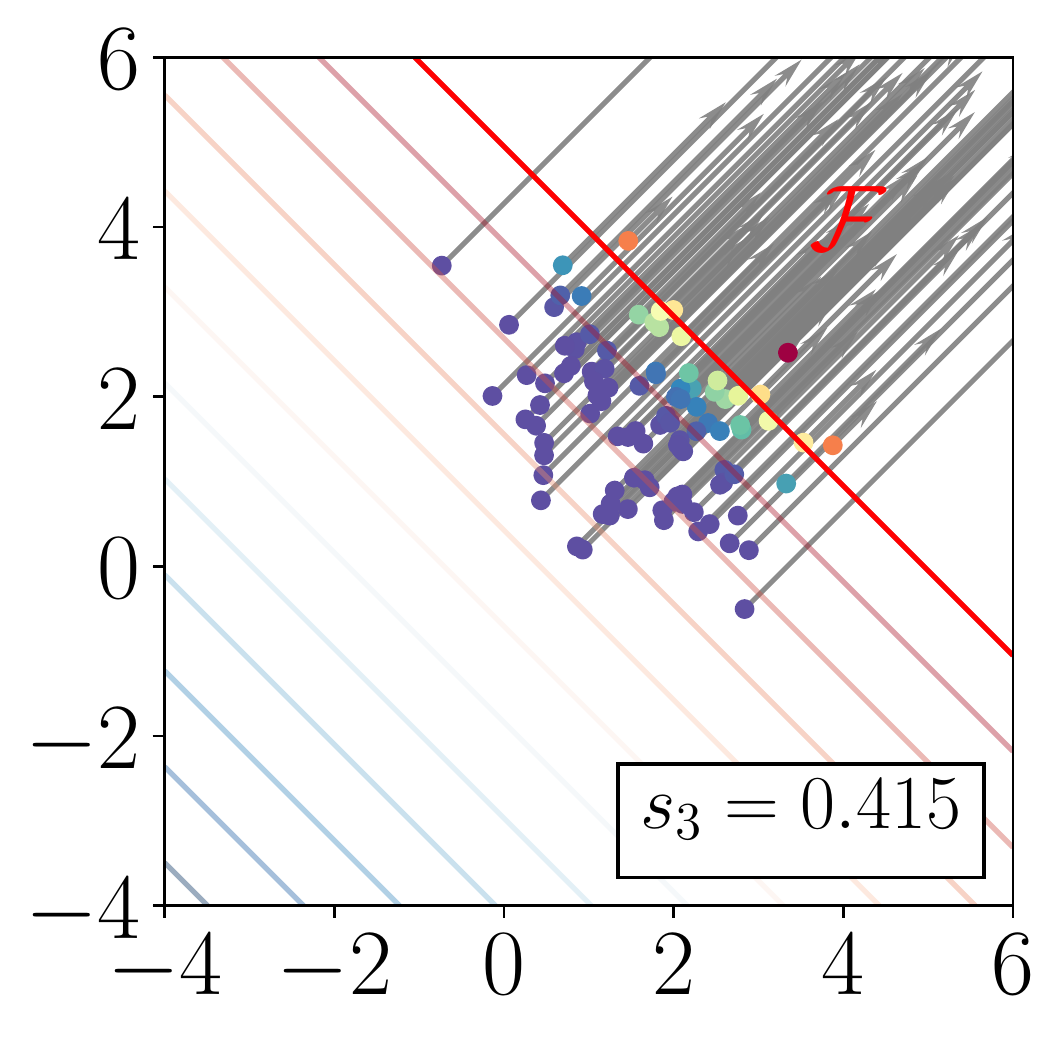}
	\includegraphics[width=0.24\textwidth]{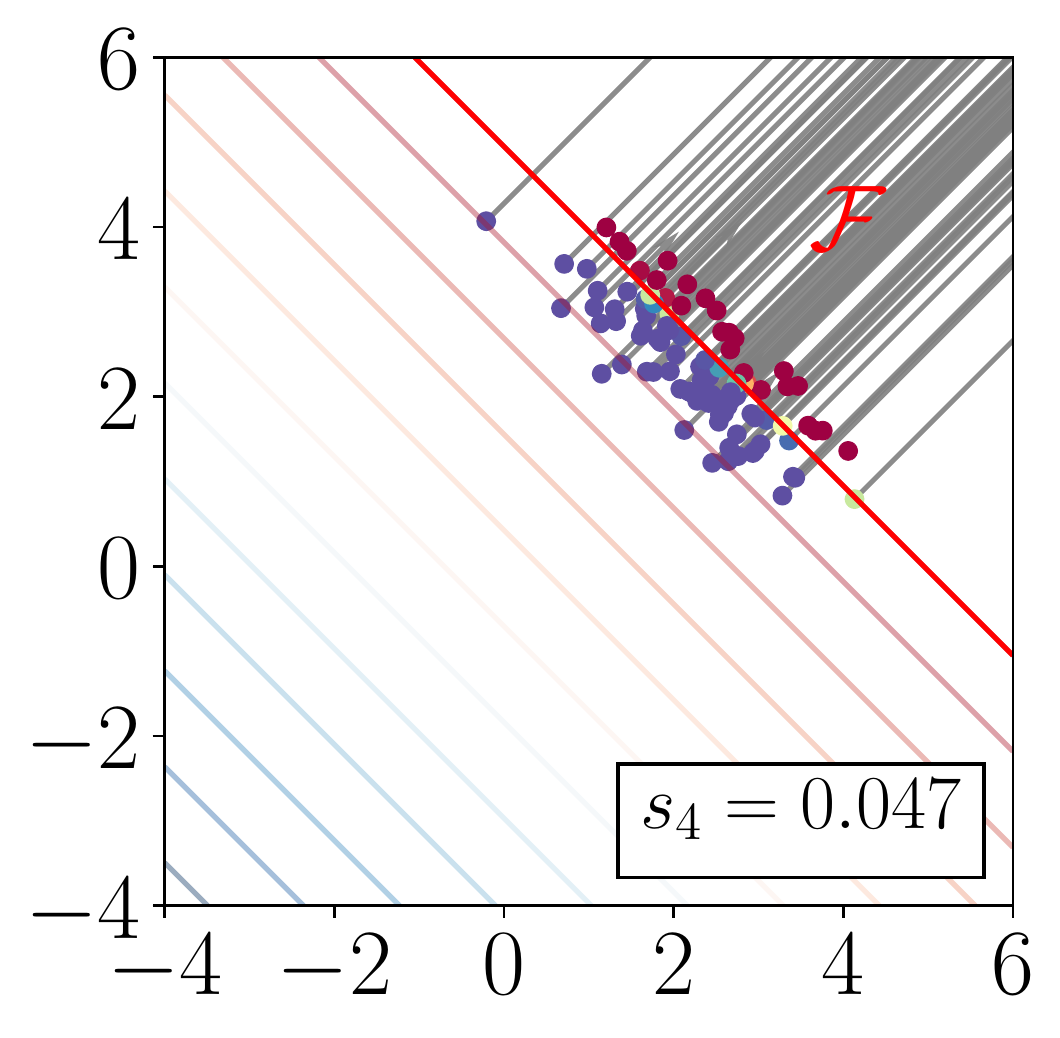}\\
	\includegraphics[width=0.24\textwidth]{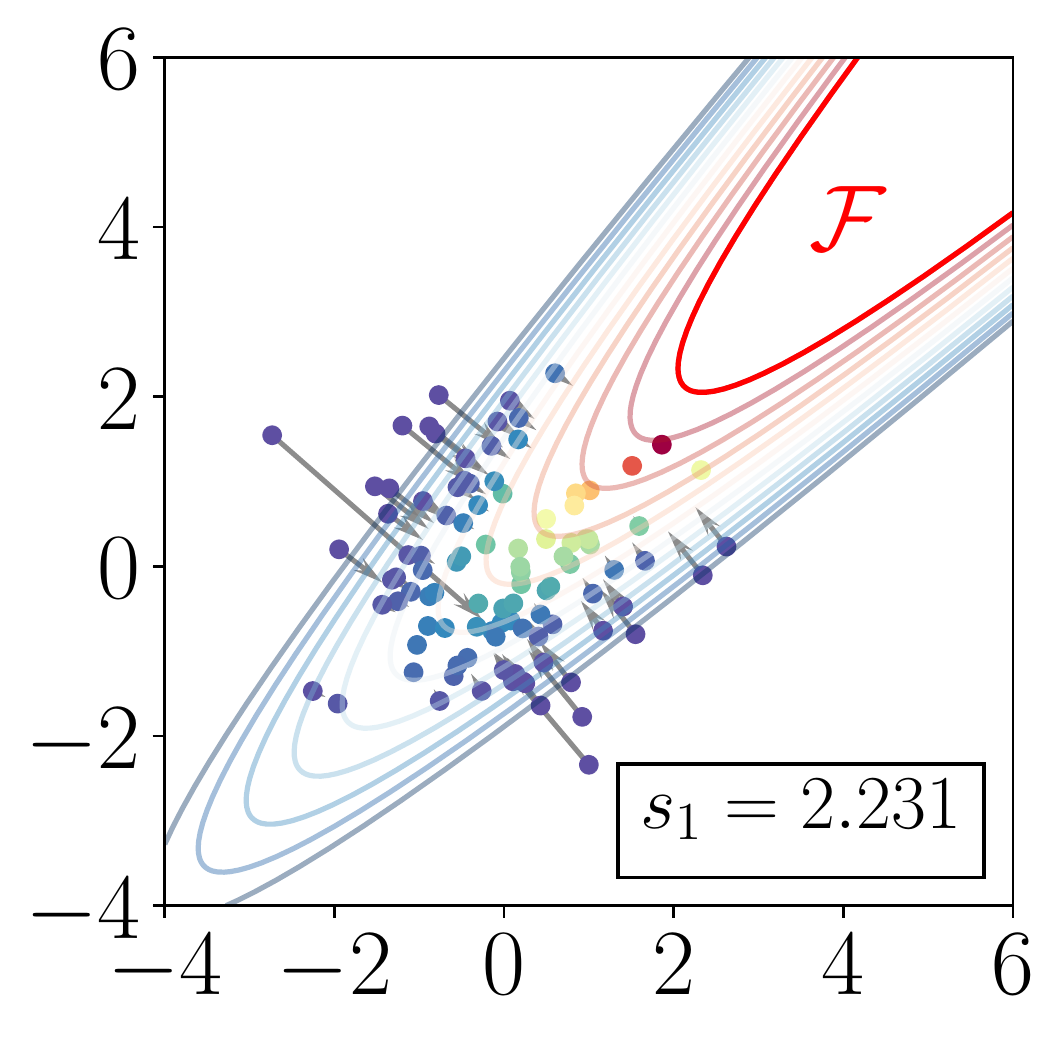}
	\includegraphics[width=0.24\textwidth]{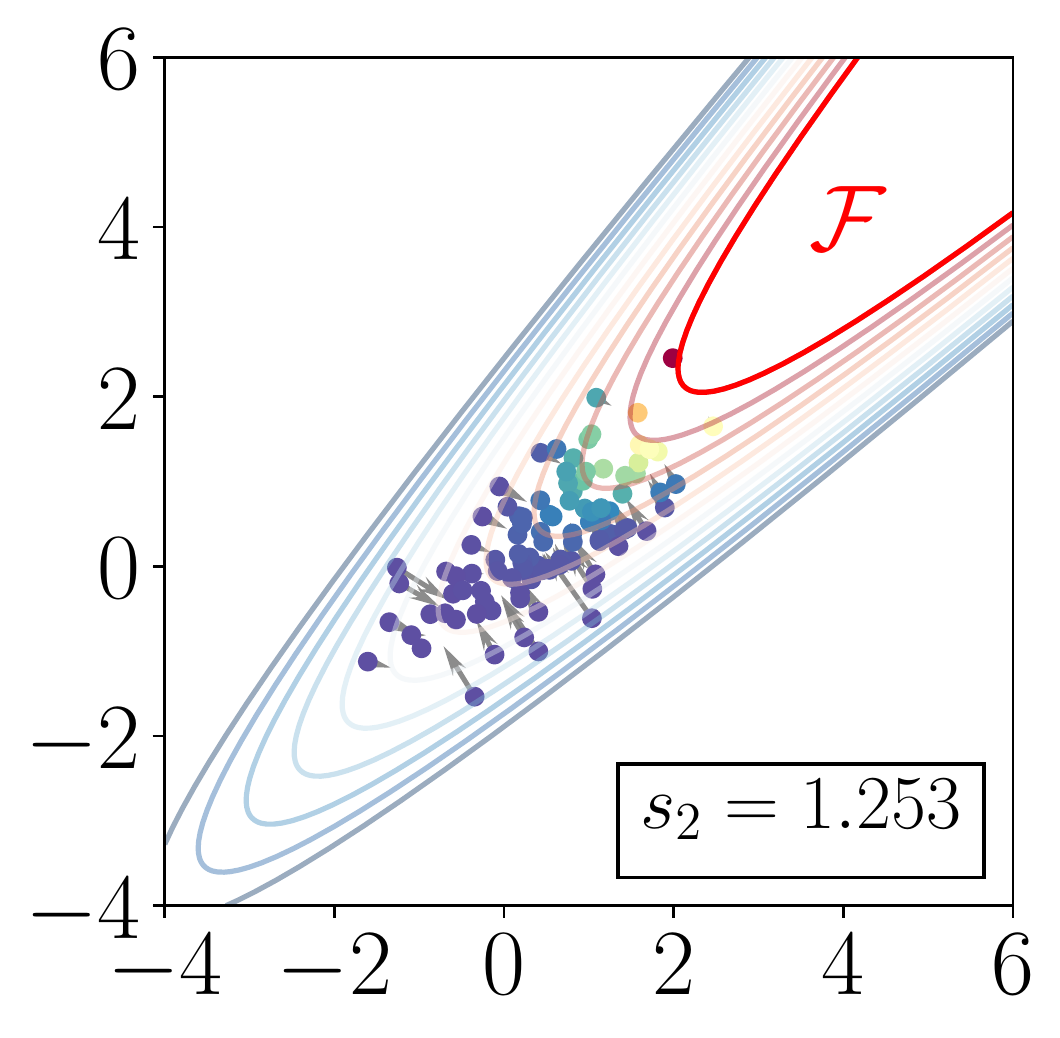}
	\includegraphics[width=0.24\textwidth]{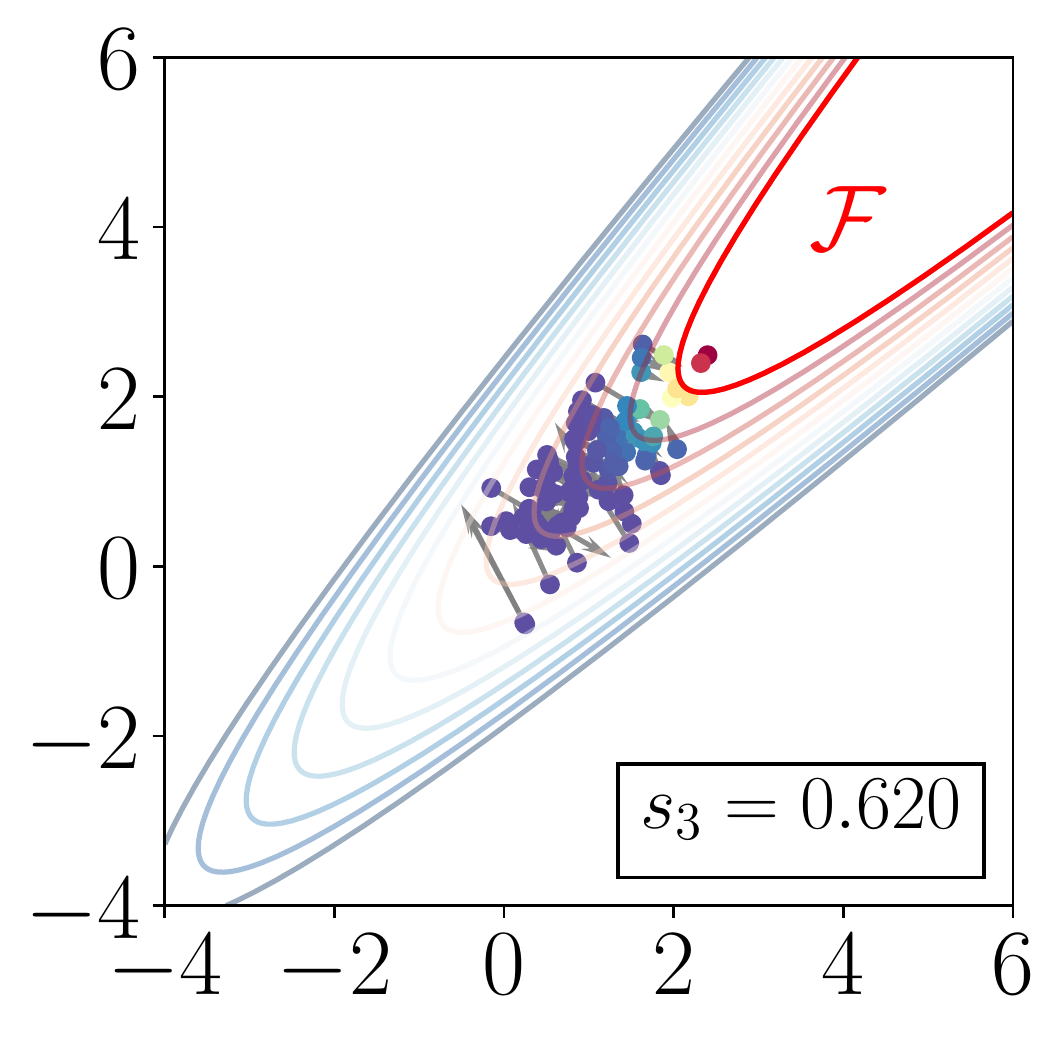}
	\includegraphics[width=0.24\textwidth]{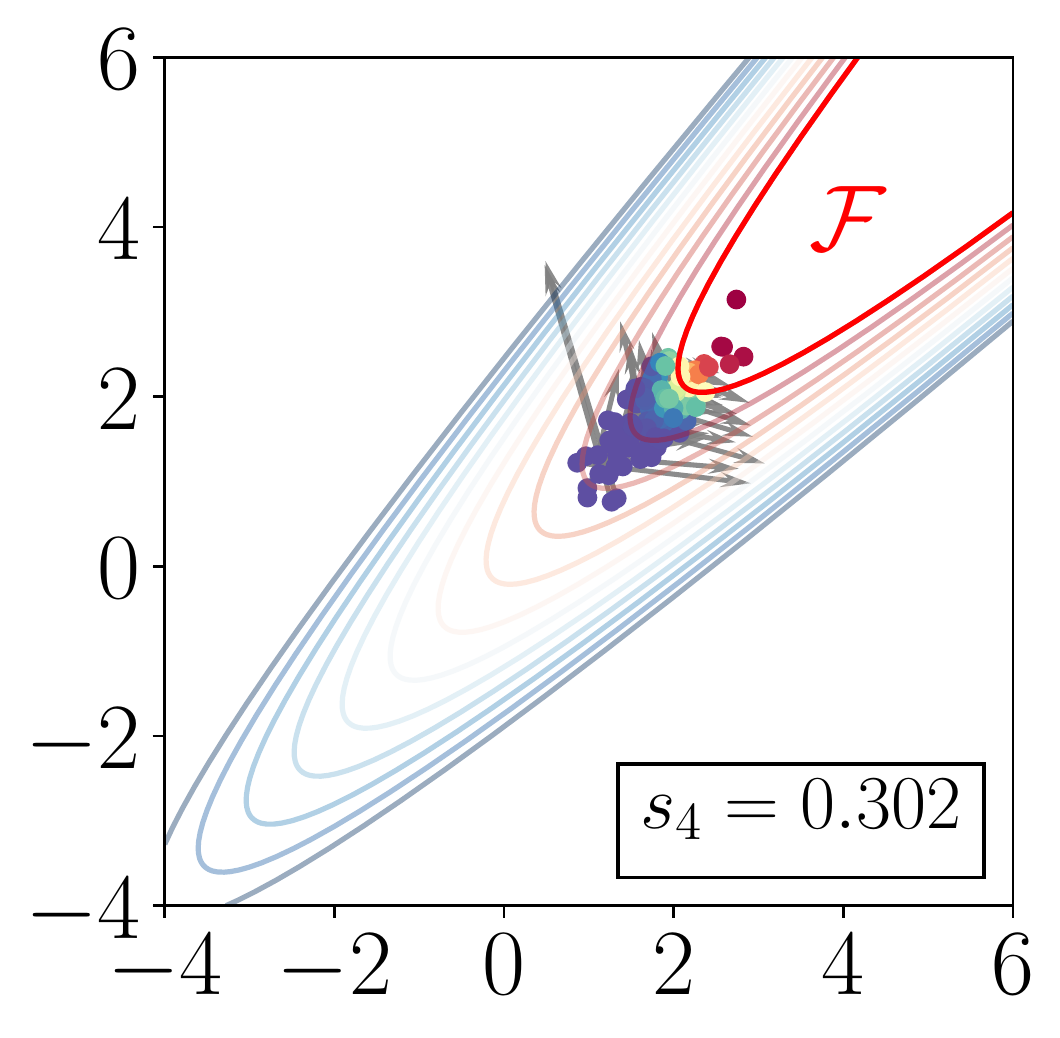}
	\caption{Evolution of the iCEred method for problems \emph{with} (top $r=1<d$) and \emph{without} (bottom $r=2=d$) intrinsic low-dimensional structure of the parameter space. The arrows show the values of $\nabla\ln f(\ve{\theta};s_j)$ and the color in the samples correspond to the magnitude of the smooth indicator function $f(\ve{\theta};s_j)$.}
	\label{fig:EX1EX2_gradplot}
\end{figure}

We conclude this subsection by showing the evolution of the iCEred method for two basic applications. The purpose here is to illustrate the method, and the details on these examples are discussed in \cref{sec:numexp}. For both problems, the dimension of the parameter space is $d=2$ and we select the approximation tolerance as $\varepsilon=0.01$. The first row of \Cref{fig:EX1EX2_gradplot} shows the results for a linear LSF problem that has a clear direction in which the parameter space varies the most; the rank of the projector is found to be $r=1<d$. The second row of \Cref{fig:EX1EX2_gradplot} plots the solution for a quadratic LSF in which the parameter space does not have a low-dimensional structure; the rank of the projector is $r=2=d$ and the iCEred algorithm reduces to the standard iCE method. 

\subsection{Refinement step}
Computational approaches such as adjoint methods \cite{arora_and_haug_1979, constantine_et_al_2014} and automatic differentiation \cite{margossian_2019} can be used to calculate efficiently the gradient of the log-smooth indicator function in \cref{eq:H_MCS}. However, when these tools are not applicable, gradient evaluation is a computationally demanding task. 

This motivates the definition of a refinement step for the iCEred method that allows improving the failure probability estimate, keeping the number of gradient evaluation fixed. The idea is to employ the reference parameters of the low-dimensional biasing density at the last level, to draw additional samples from the fitted parametric density. Such strategy has been applied to the standard CE method in \cite{wang_and_song_2016}. This complementary step to iCEred adaptively increases the LSF evaluations but it does not involve additional gradient computations. The sample size is augmented at each iteration to obtain new LSF evaluations that will improve the IS estimate. The coefficient of variation of the updated failure probability estimate $\cv{\pfe}$ is then monitored until it reaches a predefined threshold $\overline{\delta}$. Since the value of $\cv{\pfe}$ at each iteration is typically noisy, we control its mean from a set of previous iteration values. \cref{alg:iCE_red_adapt} details this `post-processing' step.

\IncMargin{1em}
\begin{algorithm}[!ht]
	\small	
	\caption{Refinement step for iCEred.}
	\label{alg:iCE_red_adapt}
	\setcounter{AlgoLine}{0}
	\SetNoFillComment
	\SetKwInOut{Input}{Input}
	\SetKwInOut{Output}{Output}
	\Input{target coefficient of variation of the failure probability estimate $\overline{\delta}$, number of iterations after the coefficient of variation is checked $m$, sample size increment $M$, last iCEred iteration values of: reference parameters $\widehat{\ve{\upsilon}}_r$, indicator function $\ve{d}_{\mathrm{eval}}$, IS weights $\ve{w}$, and bases $\mat{\Phi}_r,\mat{\Phi}_\perp$}
	\vspace*{0.1cm}
	
	$k=1$\;
	
	\While{\texttt{True}}{	
		Compute IS estimator $\widehat{p}_{\mathcal{F}}\leftarrow \frac{1}{N}\sum_{i=1}^{N} {d}_{\text{eval}}(i)\cdot w_i$\;
		
		Estimate the variance of the IS estimator
		$\widehat{\mathds{V}}[\widehat{p}_{\mathcal{F}}]\leftarrow \frac{1}{N-1}\left[\frac{1}{N}\sum_{i=1}^{N}  {d}_{\text{eval}}(i)\cdot w_i^2 - \widehat{p}_{\mathcal{F}}^2\right]$	\;
		
		Estimate the coefficient of variation $\cve(k)\leftarrow \sqrt{\widehat{\mathds{V}}[\widehat{p}_{\mathcal{F}}]}/\widehat{p}_{\mathcal{F}}$\;
		
		\uIf{$k~\mathrm{mod}~m = 0$}{
			Compute the mean of the previous $m$ values stored in $\cve$ $\rightarrow \mu_{\cve}$\;
			
			\If{$\mu_{\cve} \leq \overline{\delta}$}{
				\texttt{Break}}
		}
		\Else{
			Generate $M$ samples from the biasing density in the local FIS $\widetilde{\ve{\theta}}_r\sim {\pi}_{\text{bias}}^{(r)}(\cdot; \widehat{\ve{\upsilon}}_r)$\;
			
			Generate $M$ samples from the prior density in the local CS $\widetilde{\ve{\theta}}_\perp\sim {\pi}_{\text{pr}}^{(\perp)}(\cdot)$\;
			
			Contruct the full set of parameter samples $\ve{\theta} \leftarrow \mat{\Phi}_r \widetilde{\ve{\theta}}_r + \mat{\Phi}_\perp \widetilde{\ve{\theta}}_\perp$\;
			
			Evaluate LSF $\ve{g}_{\text{eval}}\leftarrow g(\ve{\theta})$ and indicator function $\ve{d}_{\text{extra}} \leftarrow \ve{g}_{\text{eval}}\leq 0  = \I{\mathcal{F}}{\ve{\theta}}$\;
			
			Compute the weights $\ve{w}_{\mathrm{extra}}\leftarrow\exp\left( \ln\pi_{\text{pr}}^{(r)}(\widetilde{\ve{\theta}}_r) - \ln {\pi}_{\text{bias}}^{(r)}(\widetilde{\ve{\theta}}_r;\widehat{\ve{\upsilon}}_r)\right)$\;
			
			Append the extra indicator function and weight values $\ve{d}_{\text{eval}} \leftarrow [\ve{d}_{\text{eval}}, \ve{d}_{\text{extra}}]$ and $\ve{w} \leftarrow [\ve{w}, \ve{w}_{\mathrm{extra}}]$\;	
			
			$k\leftarrow k+1$ and $N\leftarrow N+M$ 
		}
	}
	\Output{$\widehat{p}_{\mathcal{F}}$}
\end{algorithm}
\DecMargin{1em}

\section{Numerical experiments}\label{sec:numexp}
We test the proposed method on three examples. For the first two experiments, the gradient of the LSF is available analytically and the failure probabilities are independent of the dimension of the input parameter space. This allows us to perform several  parameter studies to validate the applicability and accuracy of the iCEred method. In the final example, we consider a structural mechanics application involving spatially variable parameters modeled as random fields. In this case, the LSF gradient is computed using the adjoint method, which is derived in \Cref{sec:adjoint} for this particular problem.

Moreover, we consider the gradients of the natural logarithm of the smooth indicators \cref{eq:ind} 
\begin{equation}\label{eq:grad_logind1}
\nabla \ln f^{\mathrm{log}}(\ve{\theta};s) = 
-\dfrac{\nabla g(\ve{\theta})}{s} \left[1+\text{tanh} \left(\dfrac{g(\ve{\theta})}{s}\right) \right], \qquad
\nabla \ln  f^{\mathrm{erf}}(\ve{\theta};s) = 
-\dfrac{\nabla g(\ve{\theta})}{s} \cdot \dfrac{ \phi\left(-\dfrac{g(\ve{\theta})}{s}\right)}{\Phi\left(-\dfrac{g(\ve{\theta})}{s}\right)},
\end{equation}
for the computation of the estimator $\widehat{\mat{H}}$ in \cref{eq:H_MCS}.

In all experiments, the target coefficient of variation for the iCEred method is set to $\delta=1.5$ and the tolerance of the certified approximation is $\varepsilon=0.01$. For the refinement step, the target coefficient of variation for the failure probability estimate is $\overline{\delta}=0.05$, and the increment in the sample size is $M=50$. The parametric family of biasing densities is chosen as single Gaussian densities, i.e., $\Pi = \{{\pi}_{\text{bias}}(\ve{\theta}; \ve{\upsilon})= \mathcal{N}(\ve{\theta};\ve{\upsilon})\given\allowbreak \ve{\upsilon}=[\ve{\mu},\mat{\Sigma}]\in\Upsilon\}$.


\subsection{Linear LSF in varying dimensions}\label{subsec:ex1}
We consider a LSF expressed as a linear combination of $d$ independent standard Gaussian random variables \cite{engelund_and_rackwitz_1993}:
\begin{equation}\label{eq:bm_lsf_example1}
g(\ve{\theta}) = \beta - \frac{1}{\sqrt{d}}\sum_{i=1}^{d} \theta_i \qquad \text{with gradient} \quad \nabla g(\ve{\theta}) = \left[-\frac{1}{\sqrt{d}}\cdot \mathbf{1}_d\right]^\tran,
\end{equation}
where $\beta$ is a given maximum threshold, and $\mathbf{1}_d$ denotes a $d$-dimensional vector containing ones. The probability of failure is independent of the dimension $d$ and it can be computed analytically as $\pf = \Phi(-\beta)$, with $\Phi(\cdot)$ denoting the standard Gaussian CDF.

We initially test the two smooth approximations of the log-indicator gradient \cref{eq:grad_logind1} used within the iCEred method. In this case, the refinement step of the algorithm is not implemented. This is to evaluate the standalone performance of the method. Ten different values of decreasing target failure probabilities are fixed as $\pf=[1\times 10^{-1}, 1\times 10^{-2}, \ldots,\allowbreak 1\times 10^{-10}]$, which have an associated list of thresholds $\beta=-\Phi^{-1}(\pf)\approx[1.282, 2.326, \ldots, 6.361]$. Moreover, for each threshold $\beta$, three different dimensions $d\in\{2,358,\allowbreak 1000\}$ are employed to define the LSF \cref{eq:bm_lsf_example1}. These allows us to test the method for low- and high-dimensional settings. A parameter study on the number of samples per level is also performed and we choose this value from $N\in\{100,250,500,1000\}$. In this example, there exists a clear low-dimensional structure of the parameter space and the rank of the projector $\mat{P}_r$ is $r=1$ for all parameter cases.

\begin{figure}[!ht]
	\centering
	\hspace{1.7cm}\raisebox{-0.5cm}{\small \rotatebox{0}{\textbf{$d=2$}}}\hspace{4.cm} \raisebox{-0.5cm}{\small \rotatebox{0}{\textbf{$d=358$}}}\hspace{4cm} \raisebox{-0.5cm}{\small \rotatebox{0}{\textbf{$d=1000$}}}\\
	\raisebox{1.7cm}{\small \rotatebox{90}{$f^{\text{erf}}$}}\hspace{0.05cm}
	\includegraphics[width=0.315\textwidth]{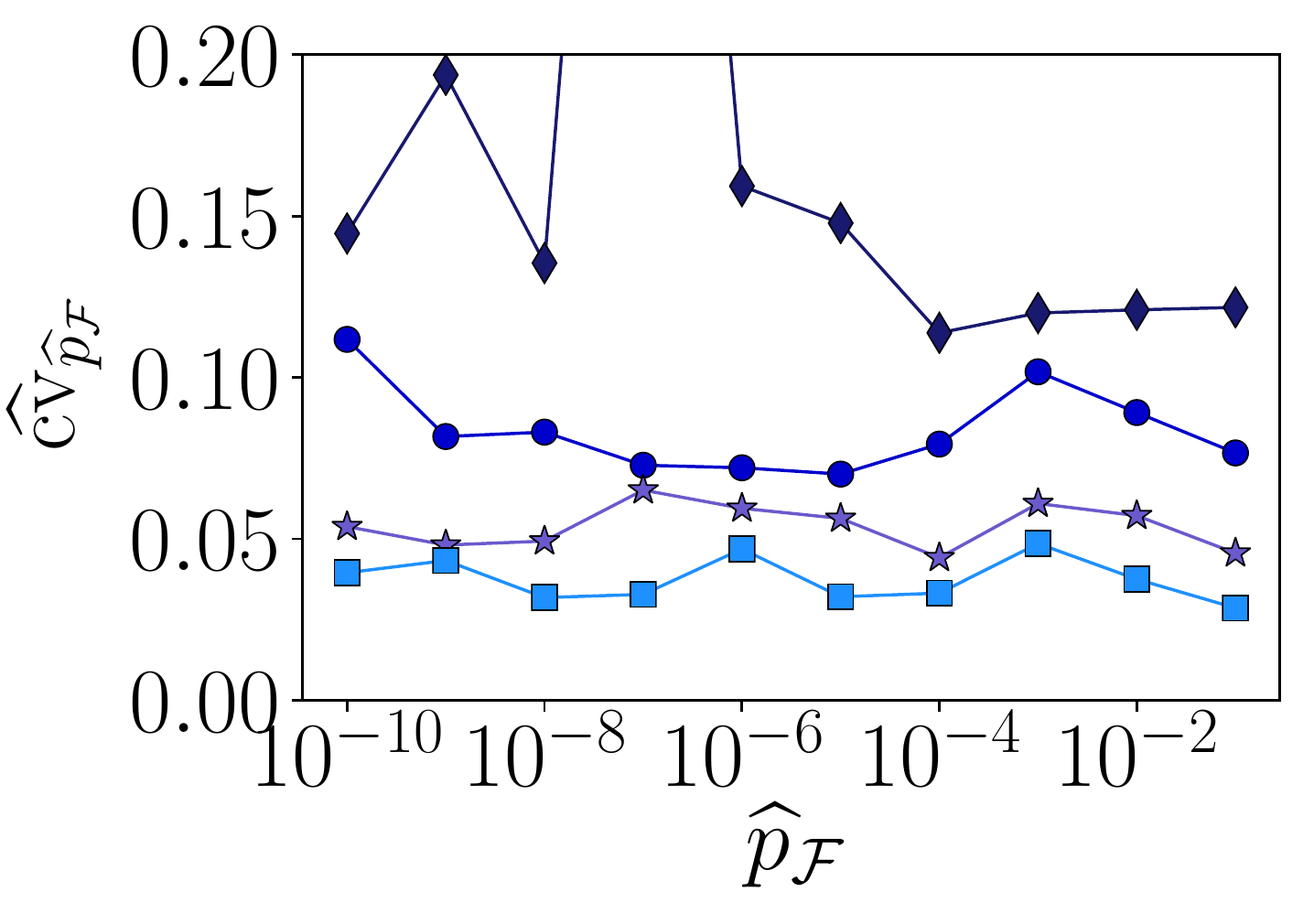}
	\includegraphics[width=0.315\textwidth]{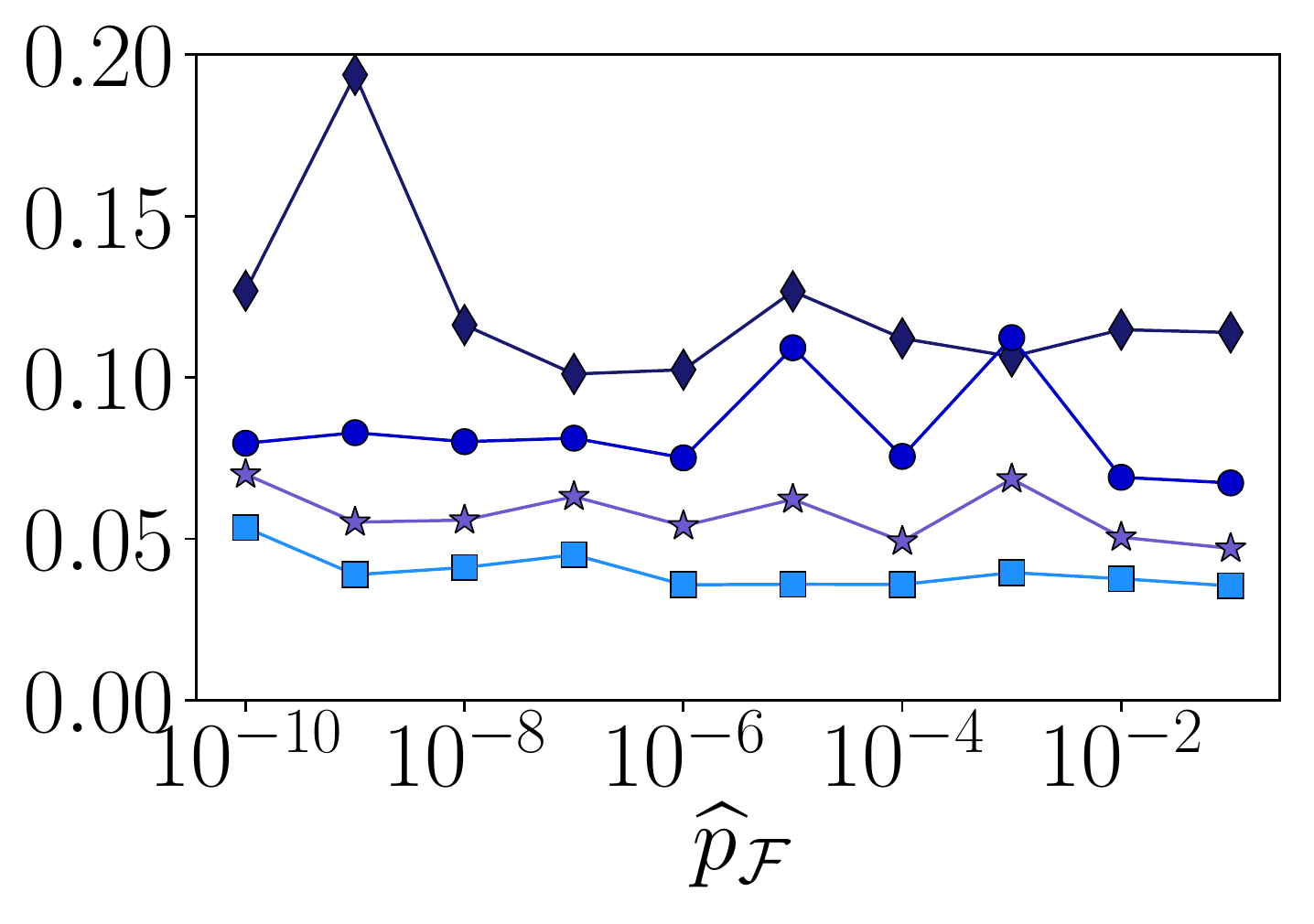}
	\includegraphics[width=0.315\textwidth]{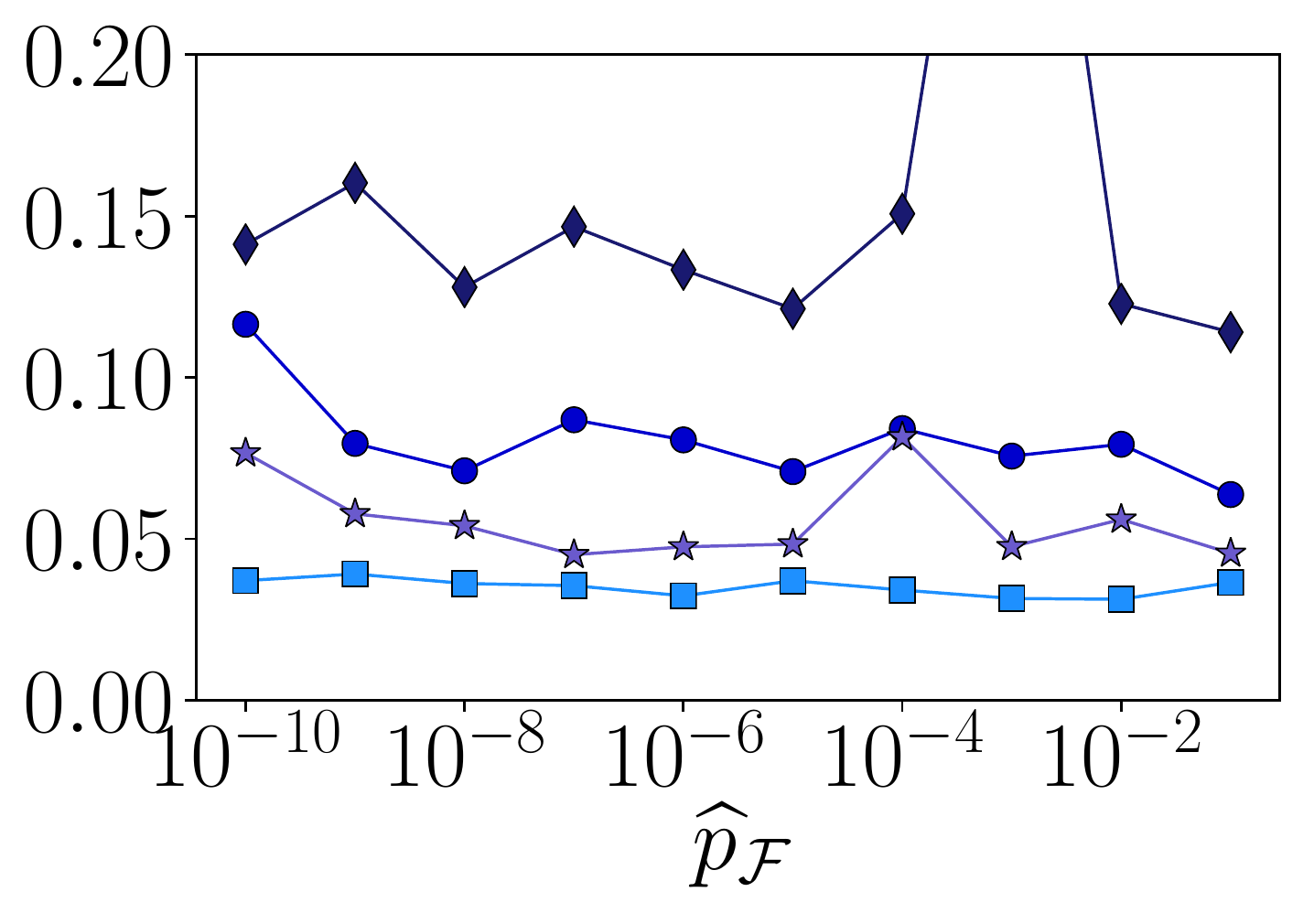}\\
	\raisebox{1.7cm}{\small \rotatebox{90}{$f^{\text{log}}$}}\hspace{0.05cm}
	\includegraphics[width=0.315\textwidth]{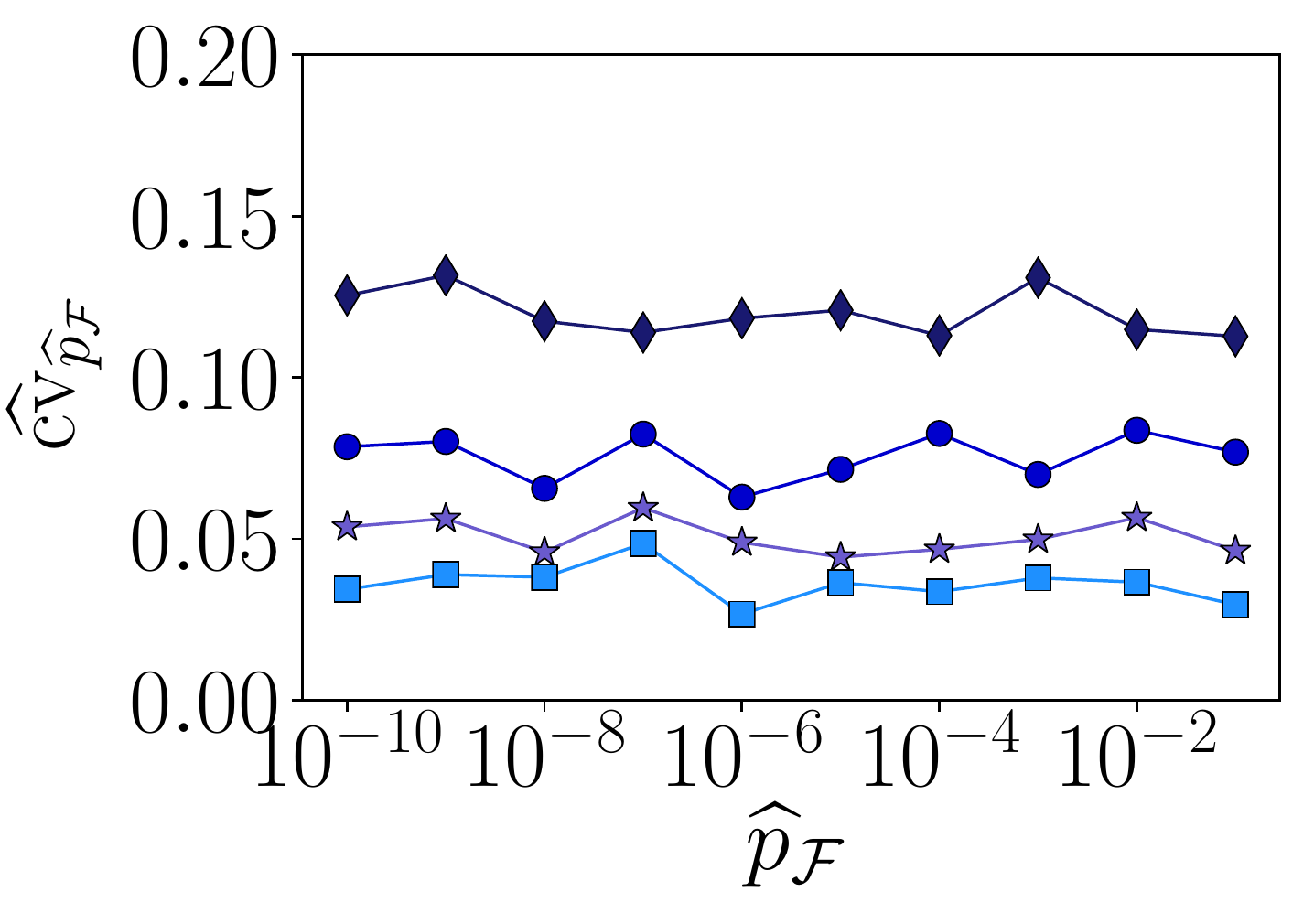}
	\includegraphics[width=0.315\textwidth]{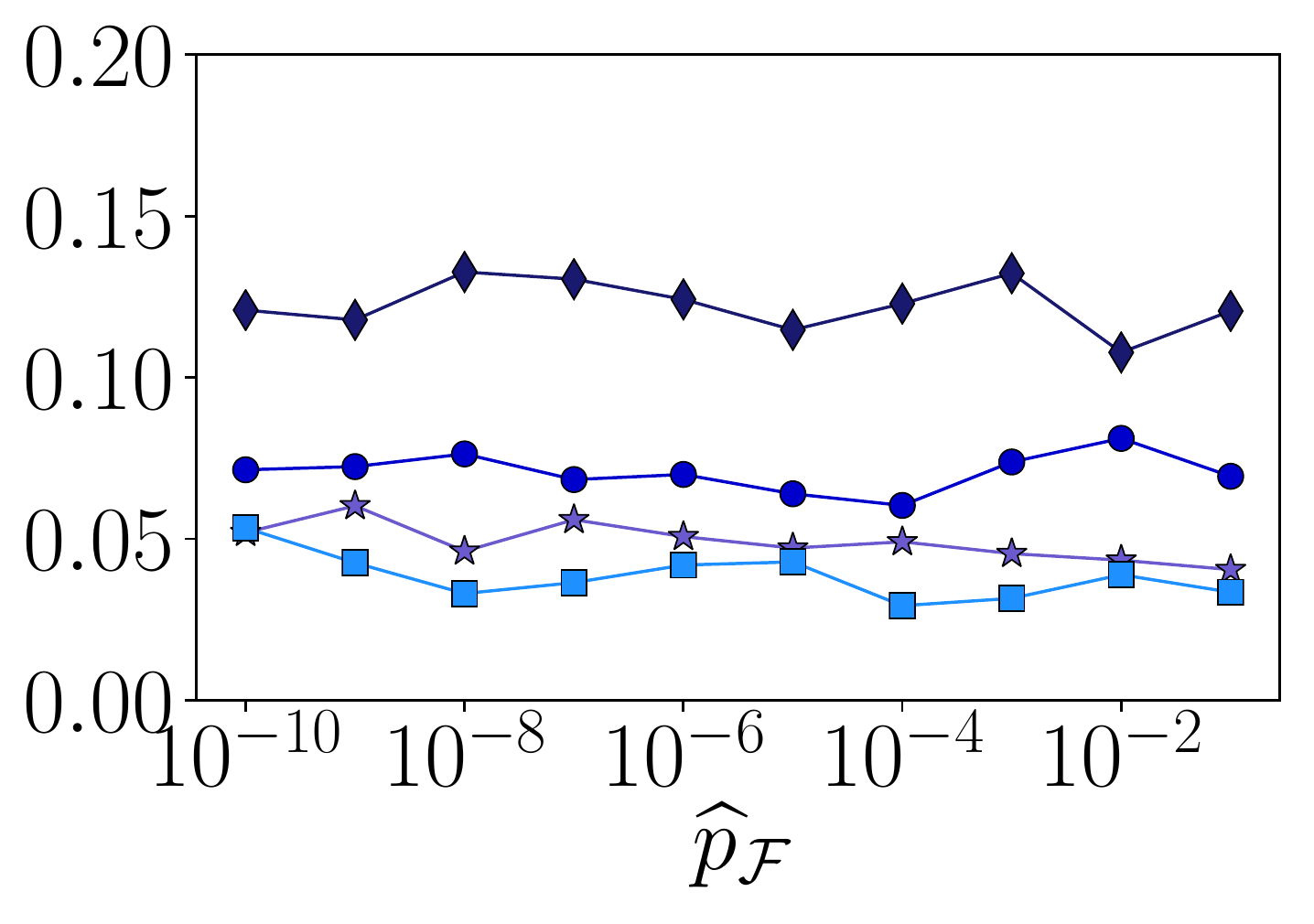}
	\includegraphics[width=0.315\textwidth]{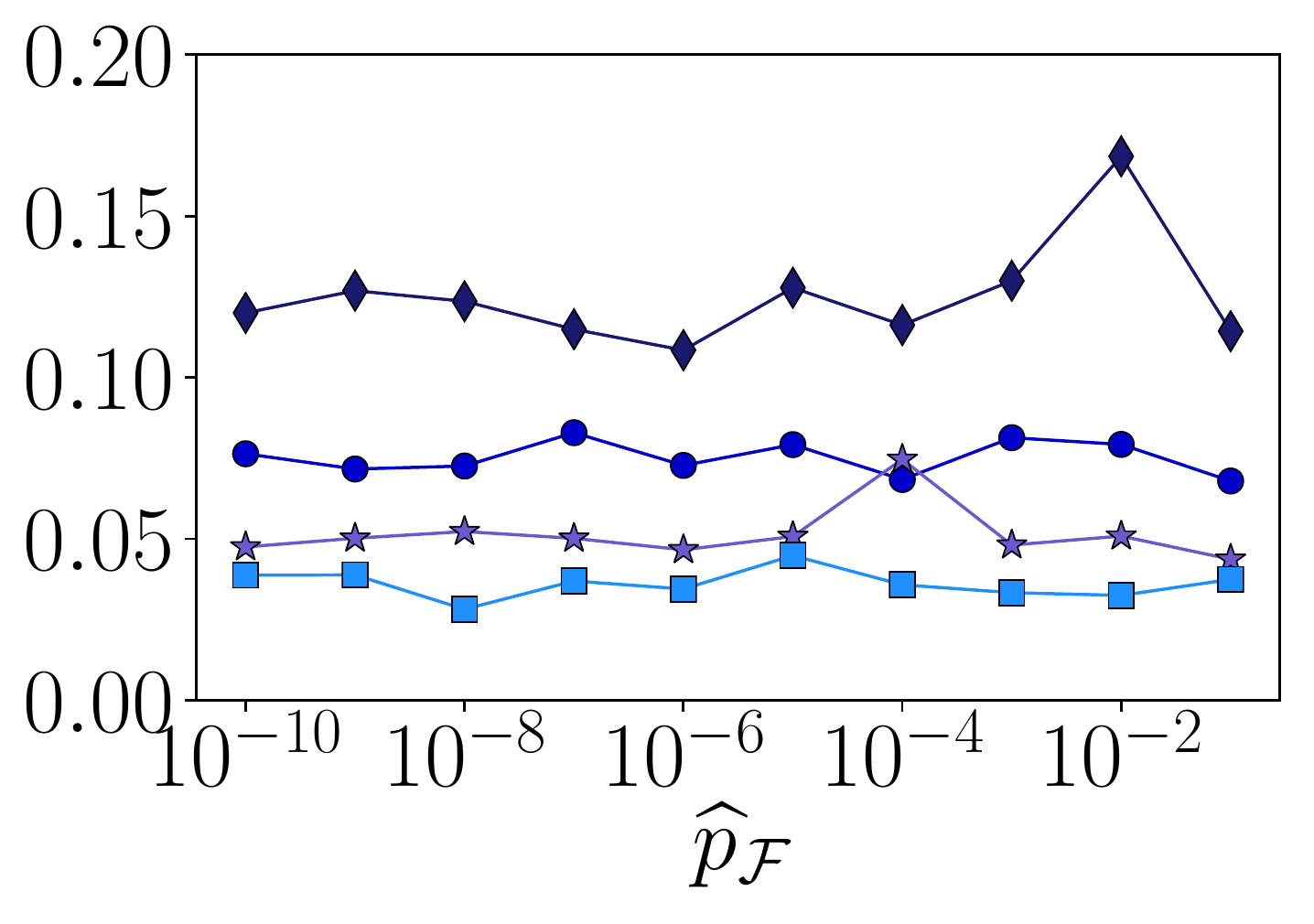}
	\adjincludegraphics[trim={1cm 0 0 {10.6cm} },clip=true]{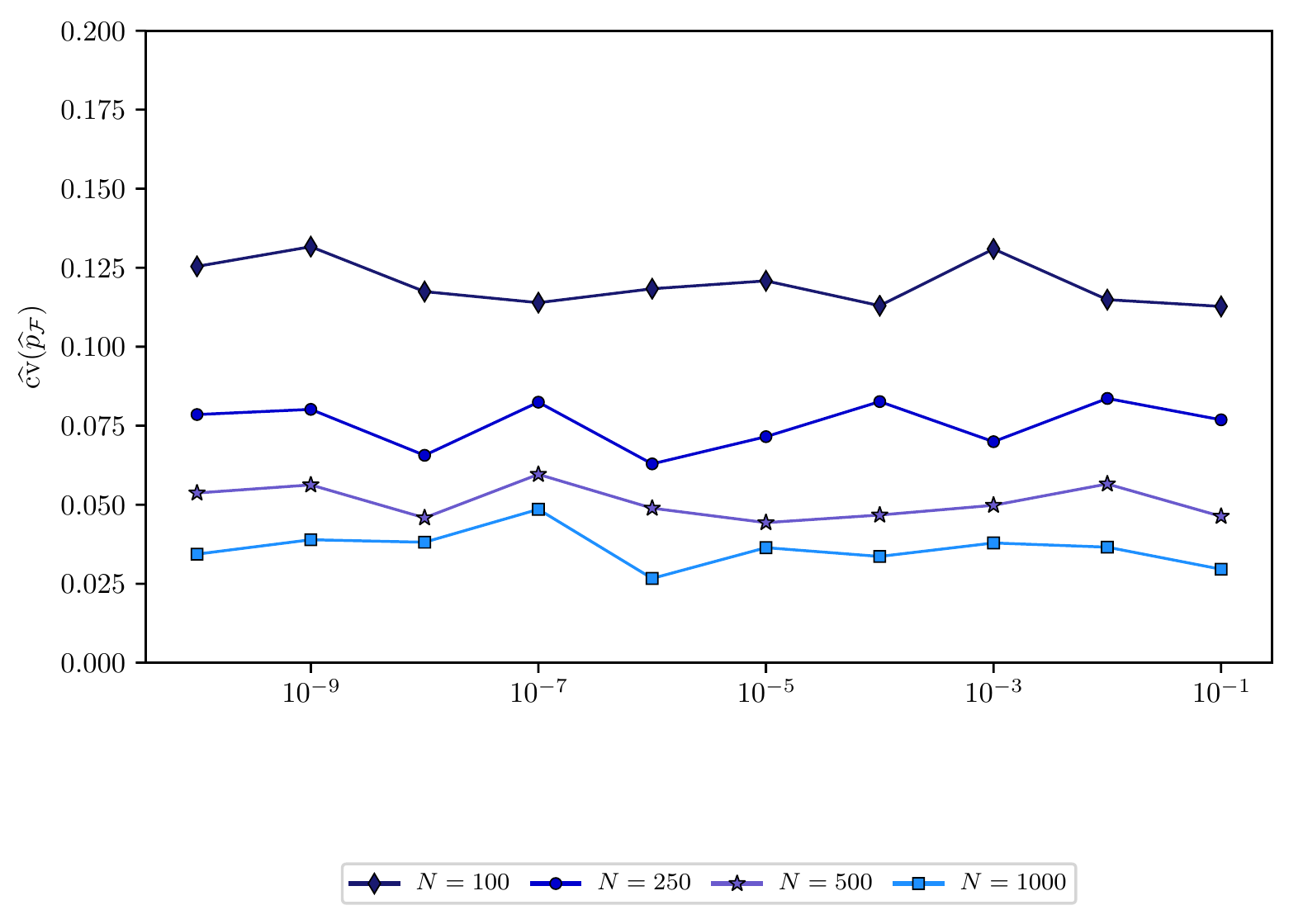}\vspace*{-0.5cm}
	\caption{Coefficient of variation of the failure probability estimate for example \ref{subsec:ex1} using different target probabilities of failure, number of samples, smooth indicators (rows), and increasing dimension (columns).}
	\label{fig:EX1_results_smooth_dimvsPf}
\end{figure}

\Cref{fig:EX1_results_smooth_dimvsPf} shows the coefficient of variation of the failure probability estimates, computed as an average of $100$ independent iCEred runs. One observes that under the LSF \cref{eq:bm_lsf_example1}, the performance of the iCEred method is independent of the dimension and magnitude of the failure probability. As the number of samples increases, the smooth approximations yield similar results. However, when the number of samples per level is small, the smooth approximation $f^{\mathrm{erf}}$ which is based on the standard Gaussian CDF produces larger $\cv{\pfe}$ values. Note in \cref{eq:grad_logind1} that the standard Gaussian CDF is evaluated in the denominator. As the number of iCEred levels increases, the smoothing parameter $s\to 0$. For very small values of $s$, $\Phi(-g(\ve{\theta})/s)$ takes values close to zero, and hence the value of $\nabla \ln f(\ve{\theta};s)$ is not defined. When this occurs, the matrix $\widehat{\mat{H}}$ cannot be estimated and the basis of the previous level is used. Although this problem is inherent to any approximation of the indicator function, it appears that $f^{\mathrm{log}}$ based on the logistic function is more robust than $f^{\mathrm{erf}}$. This is probably because $f^{\mathrm{erf}}$ involves numerical approximation of the error function. As a result, $f^{\mathrm{log}}$ yields more stable results for sample size $N=100$. 

\begin{figure}[!ht]
	\centering
	\hspace{1.2cm}\raisebox{-0.5cm}{\small \rotatebox{0}{\textbf{$N=250$}}}\hspace{3.8cm} \raisebox{-0.5cm}{\small \rotatebox{0}{\textbf{$N=1000$}}}\hspace{3.8cm} \raisebox{-0.5cm}{\small \rotatebox{0}{\textbf{$N=2500$}}}\\
	\includegraphics[width=0.32\textwidth]{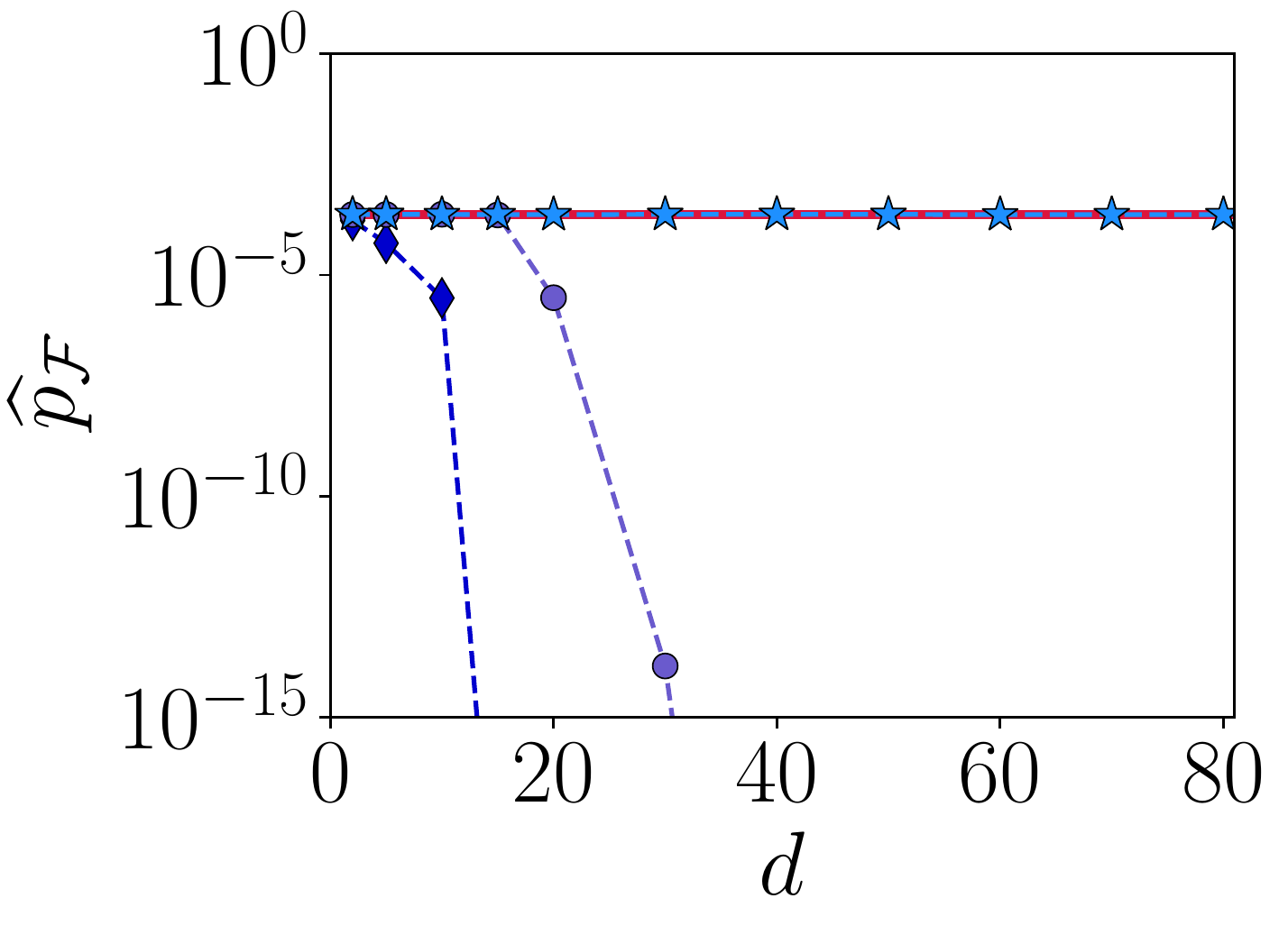}
	\includegraphics[width=0.32\textwidth]{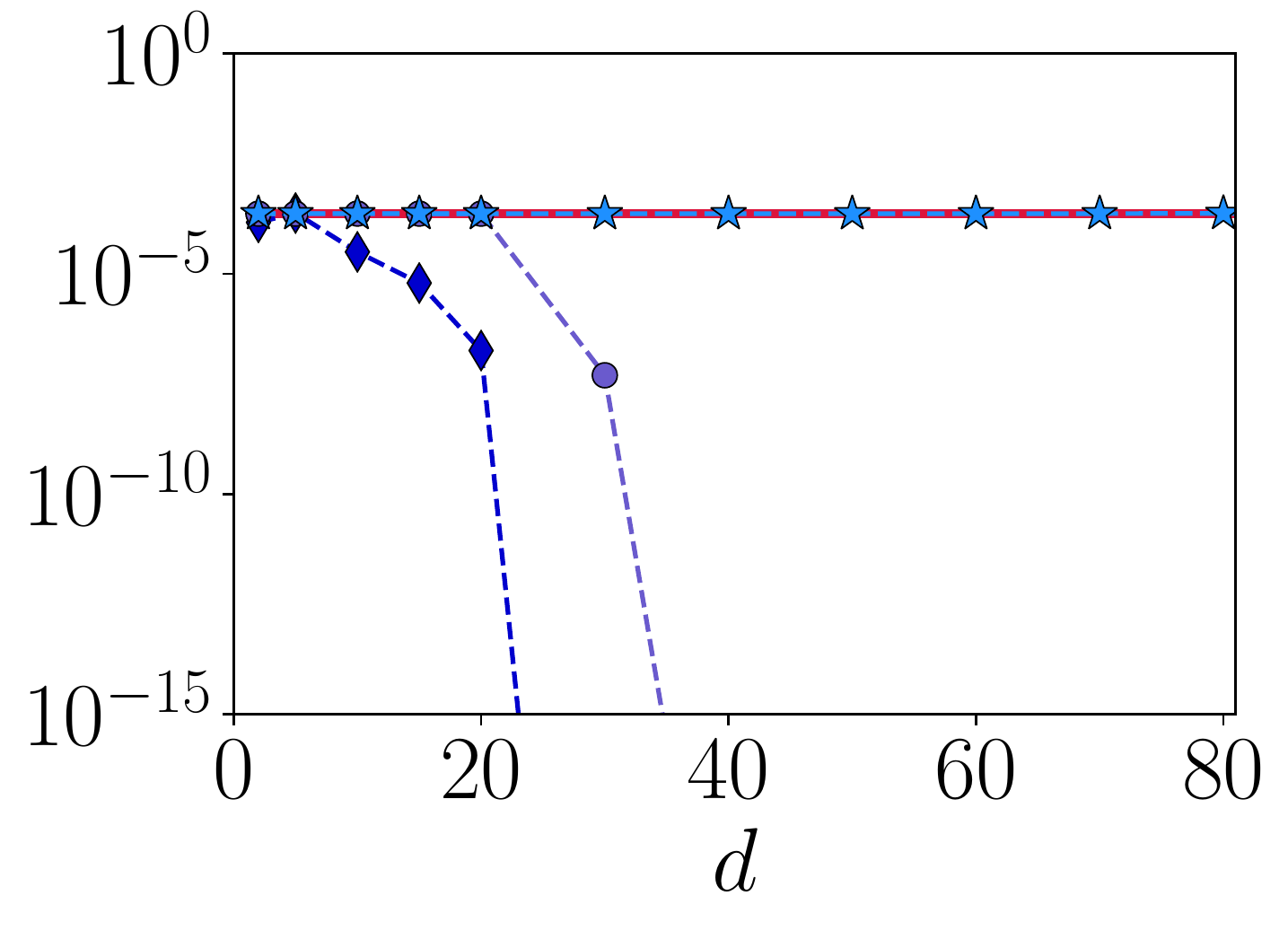}
	\includegraphics[width=0.32\textwidth]{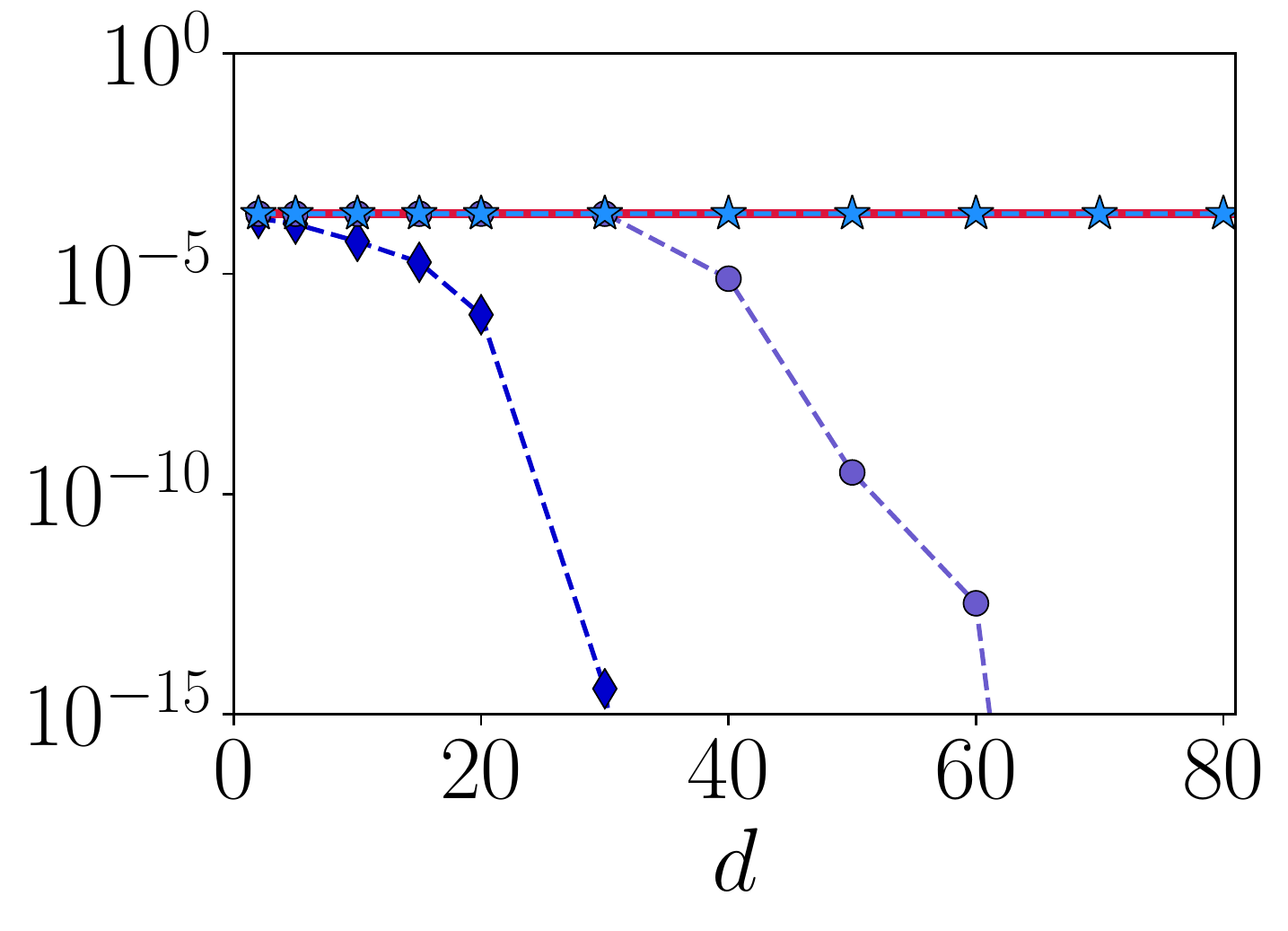}
	\adjincludegraphics[trim={1cm 0 0 {10.7cm} },clip=true]{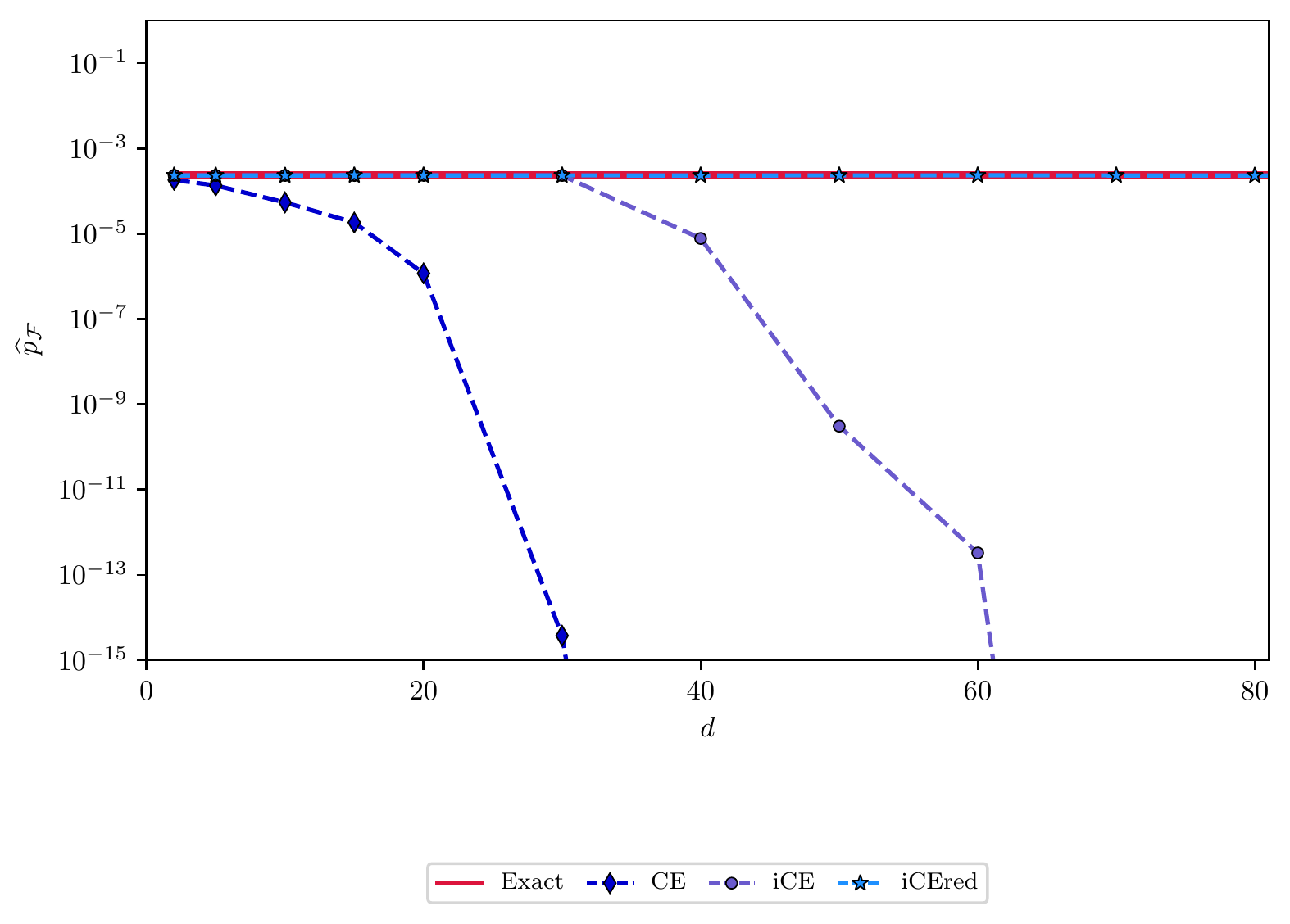}\vspace*{-0.5cm}
	\caption{Failure probability estimate for example \ref{subsec:ex1} using the CE, iCE, and iCEred methods (with single Gaussian parametric family) for increasing dimension of the parameter space and different number of samples per level (columns).}
	\label{fig:EX1_CEcomp}
\end{figure}

We now fix the smooth approximation of the indicator function to $f^{\mathrm{log}}$ and the threshold to $\beta=3.5$ ($\pf\approx 2.33\times 10^{-4}$). In \Cref{fig:EX1_CEcomp}, we plot a comparison between the iCEred method (without refinement) with the standard CE and iCE methods for increasing dimension of the parameter space and number of samples per level $N\in\{250,1000,2500\}$. The CE method is implemented with $\rho=0.1$ (cf., \cref{subsec:CEmethod}) and the iCE method with $\delta=1.5$ (cf., \cref{subsec:iCEmethod}); for both approaches the biasing distribution family consist of single Gaussian densities. We observe that the performance of standard CE and iCE methods deteriorates with increasing dimension of the parameter space, and that augmenting the sample size improves the accuracy in the estimation of the target failure probability. This is related to the solution of the stochastic optimization problem required in both approaches, which amounts to fitting high-dimensional Gaussian densities. The number of samples required to properly perform this step depends on the rank of the covariance matrix of the Gaussian biasing density, and it roughly scales quadratically with the dimension. Although the results can improve if other parametric families of biasing densities are employed (as those proposed in \cite{wang_and_song_2016,papaioannou_et_al_2019}), under the Gaussian parametric family the standard methods are inadequate to perform reliability analysis in high dimensions. The iCEred method estimates consistently the target failure probability for all the dimensions and even at small sample size. Since the rank of the projector is $r=1$, the updating of reference parameters in iCEred amounts to fitting one-dimensional Gaussian densities for all the dimension cases.

\begin{figure}[!ht]
	\centering
	\hspace{0.1cm}\raisebox{-0.5cm}{\small \rotatebox{0}{\textbf{$d=2$}}}\hspace{4.3cm} \raisebox{-0.5cm}{\small \rotatebox{0}{\textbf{$d=358$}}}\hspace{4cm} \raisebox{-0.5cm}{\small \rotatebox{0}{\textbf{$d=1000$}}}\\
	\includegraphics[width=0.32\textwidth]{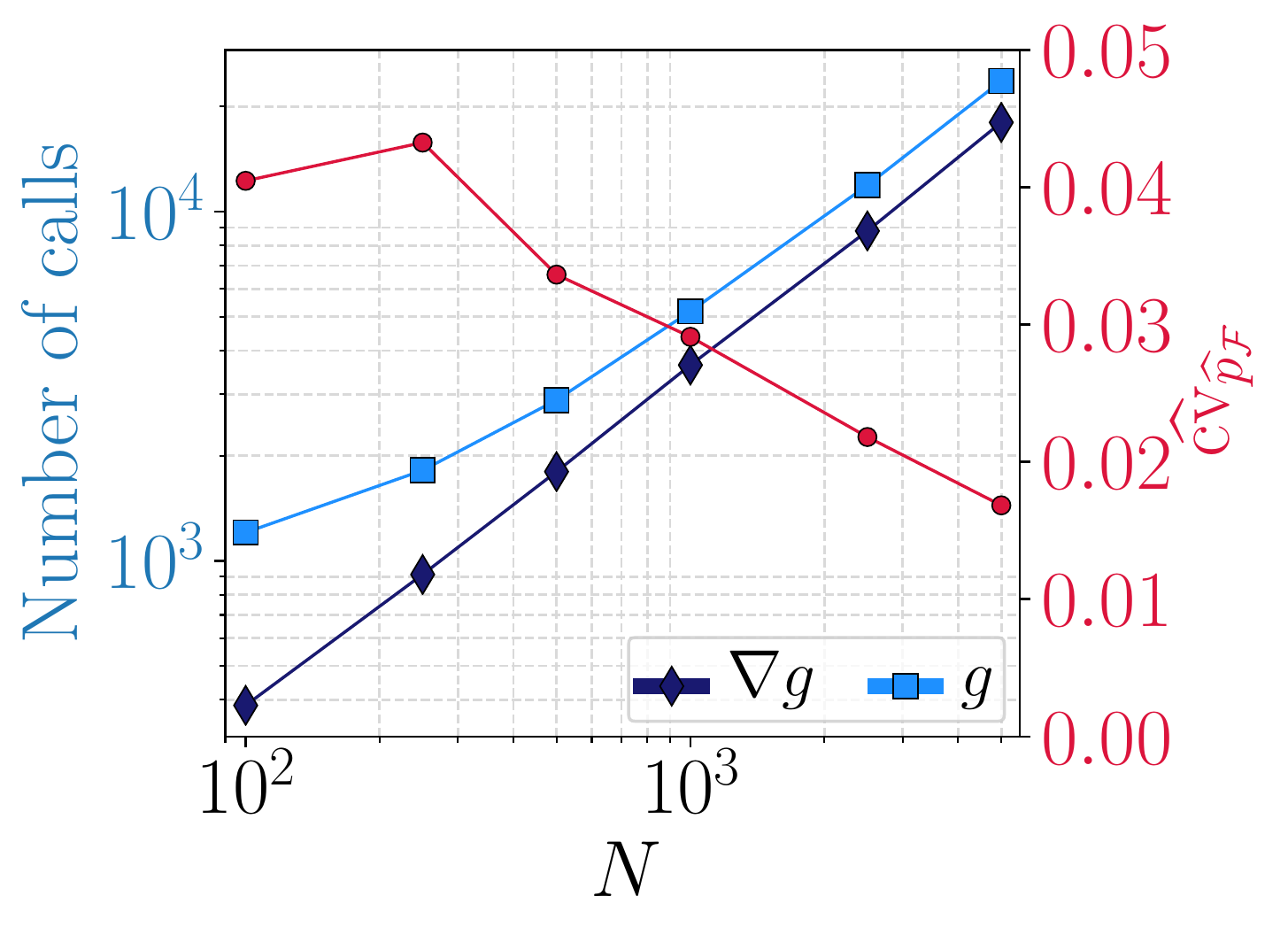}
	\includegraphics[width=0.32\textwidth]{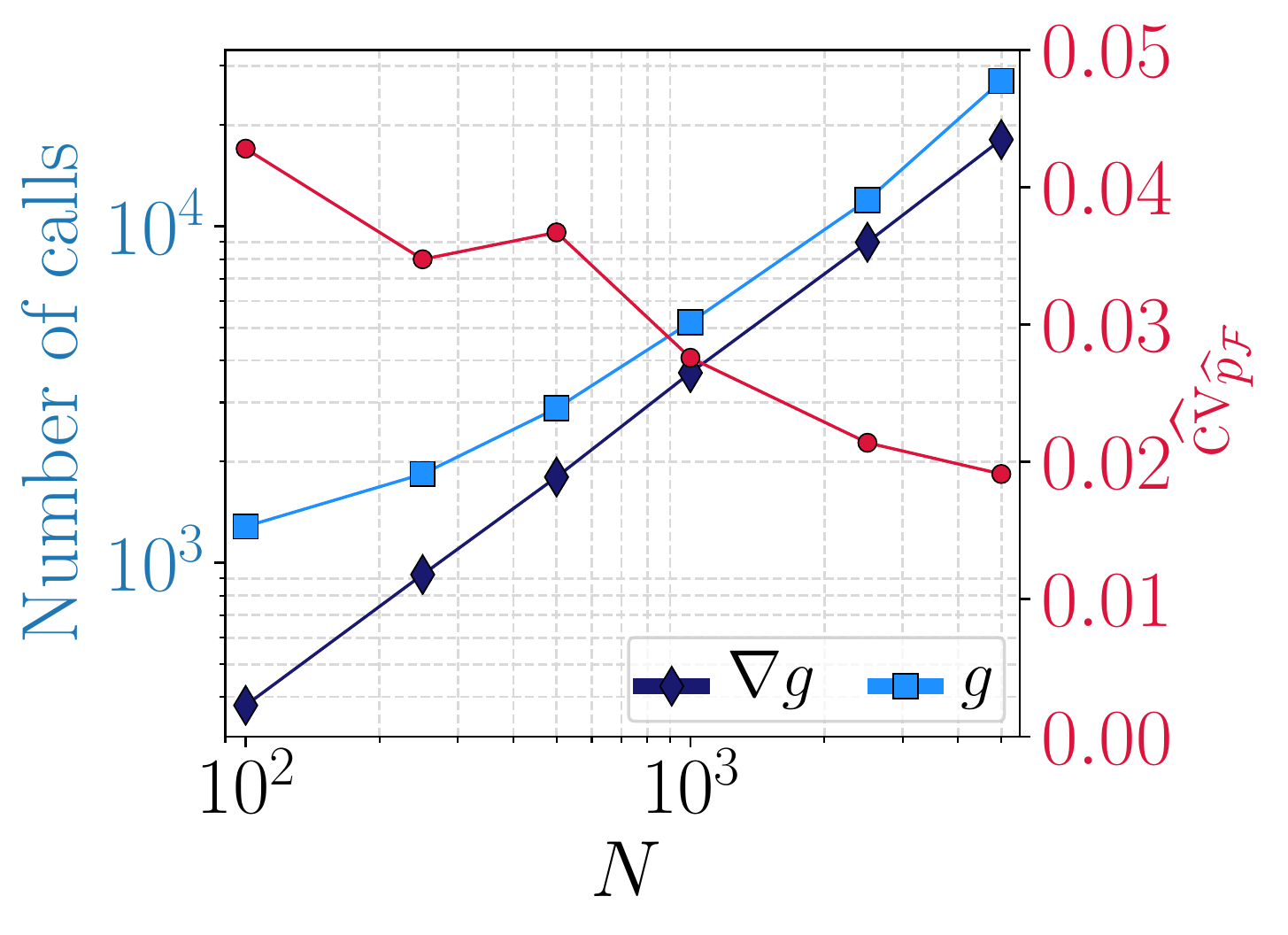}
	\includegraphics[width=0.32\textwidth]{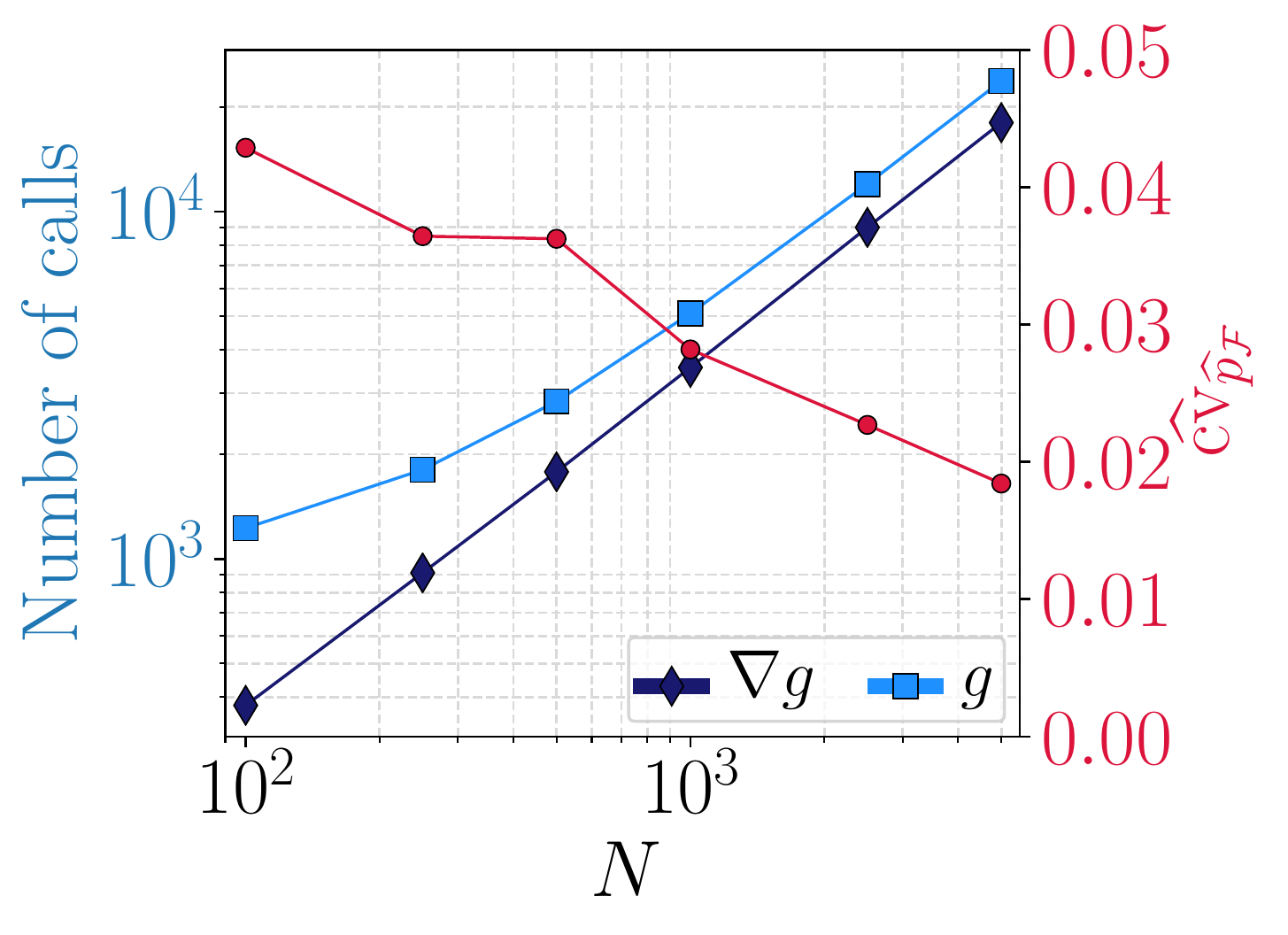}
	\caption{iCEred with refinement for example \ref{subsec:ex1}: number of LSF and gradient calls for different number of samples, and dimensions (columns). The $\widehat{\mathrm{cv}}$ of the failure probability estimate is also shown (in red).}
	\label{fig:EX1_adapt_dim}
\end{figure}

We conclude this example by showing the advantage of the refinement step for iCEred. In this case, the number of samples per level is selected from the set $N\in\{100,250,500,1000,2500,5000\}$. \Cref{fig:EX1_adapt_dim} shows the number of LSF and gradient calls, together with the coefficient of variation of the failure probability estimate (again, as an average of 100 independent simulations). Note that for sample sizes 100, 250 and 500, there is an increase in the number of LSF evaluations, due to the refinement. This is because small values of $N$ are not sufficient to compute a probability of failure estimate that has $\cv{\pfe}$ smaller than the value of $\overline{\delta}$. Therefore, extra LSF computations are performed in the refinement algorithm to reduce the $\cv{\pfe}$ to $\overline{\delta}$. For large sample sizes, this step has no effect since the $\cv{\pfe}$ obtained from iCEred is already below the predefined $\overline{\delta}$. 

\subsection{Quadratic LSF in varying dimensions}\label{subsec:ex2}
We add a quadratic term to the LSF \cref{eq:bm_lsf_example1}:
\begin{subequations}\label{eq:bm_lsf_example2_all}
\begin{align}
g(\ve{\theta}) &= \beta + \frac{\kappa}{4}\left(\theta_1-\theta_2\right)^2 - \frac{1}{\sqrt{d}}\sum_{i=1}^{d} \theta_i \qquad \text{with gradient} \label{eq:bm_lsf_example2}  \\ 
\nabla g(\ve{\theta}) &= \left[\frac{\kappa}{2}(\theta_1-\theta_2)- \frac{1}{\sqrt{d}}, ~\frac{\kappa}{2}(\theta_2-\theta_1)- \frac{1}{\sqrt{d}}, ~-\frac{1}{\sqrt{d}}\cdot \mathbf{1}_{d-2}\right]^\tran,
\end{align}
\end{subequations}
where $\beta=4$ and the parameter $\kappa$ defines the curvature of the LSF at the point in the parameter space with the largest probability density (the larger the curvature, the smaller $\pf$). The reference probability of failure is independent of the dimension and it can be computed as $\pf = \int_{-\infty}^{\infty}\int_{-\infty}^{\infty} \phi(-u+v^2(\kappa/2) )\phi(v)~\dd u\dd v$, where $\phi$ denotes the standard Gaussian density \cite{papaioannou_et_al_2015}. We employ the iCEred method with adaptation to estimate the probability of failure associated to the LSF \cref{eq:bm_lsf_example2}. The target failure probabilities are $\pf \approx [6.62\times 10^{-6}, 4.73\times 10^{-6}]$ for curvature values $\kappa\in\{5,10\}$. Three different dimensions $d\in\{2,334,1000\}$ are employed to define the LSF. 

\begin{figure}[!ht]
	\centering
	\hspace{1cm}\raisebox{-0.5cm}{\small \rotatebox{0}{\textbf{$d=2$}}}\hspace{4.5cm} \raisebox{-0.5cm}{\small \rotatebox{0}{\textbf{$d=334$}}}\hspace{4cm} \raisebox{-0.5cm}{\small \rotatebox{0}{\textbf{$d=1000$}}}\\
	\includegraphics[width=0.32\textwidth]{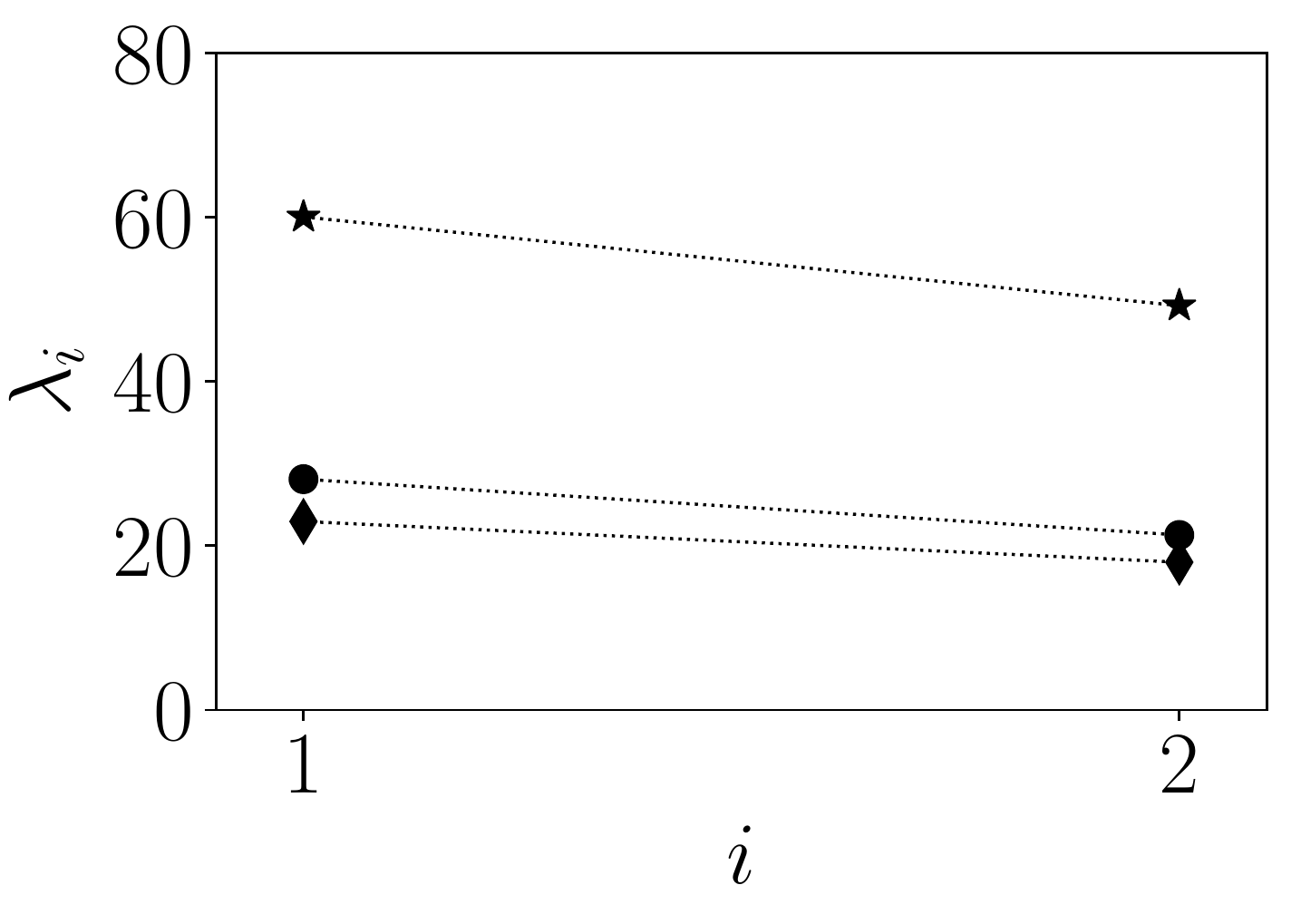}
	\includegraphics[width=0.32\textwidth]{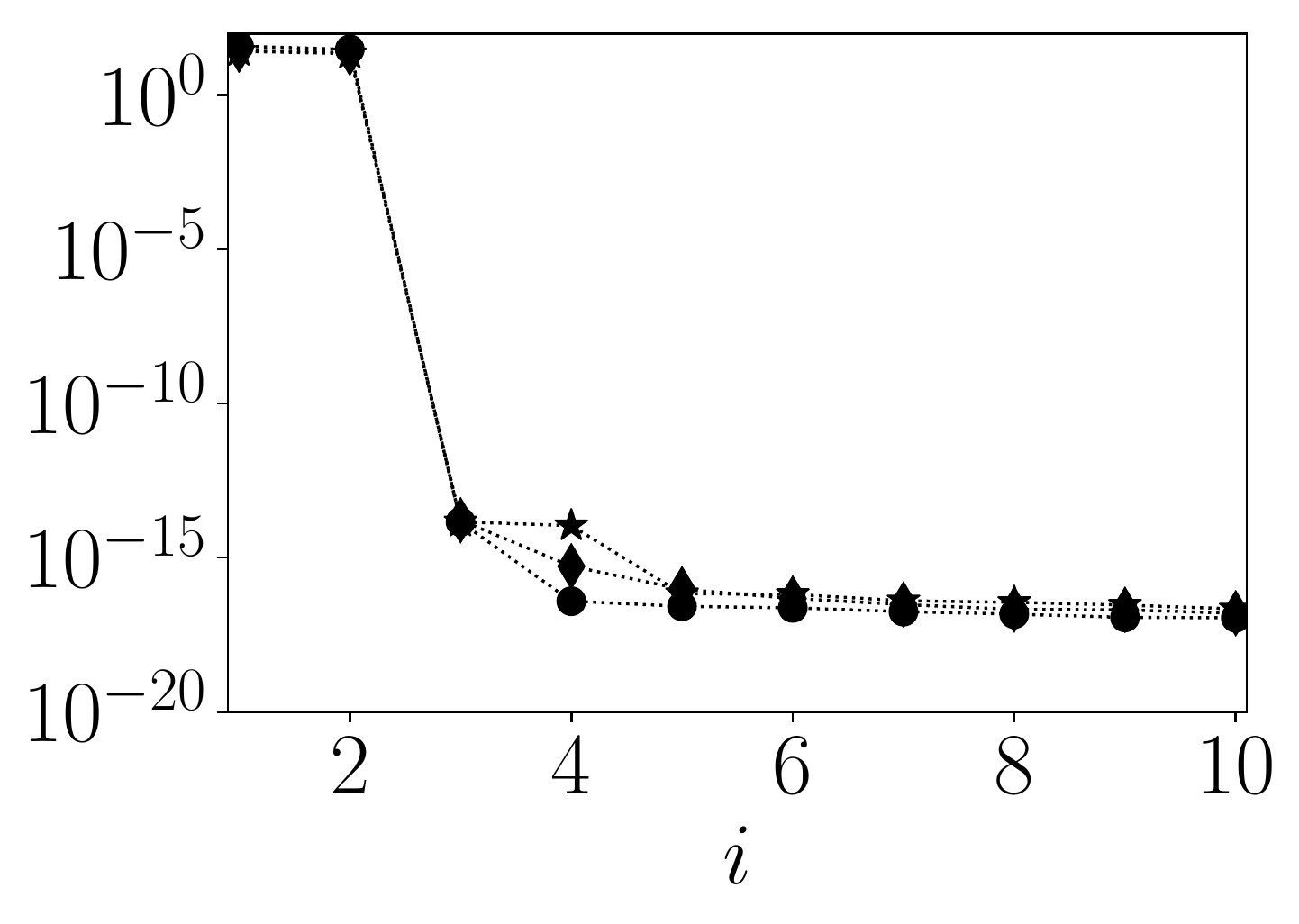}
	\includegraphics[width=0.32\textwidth]{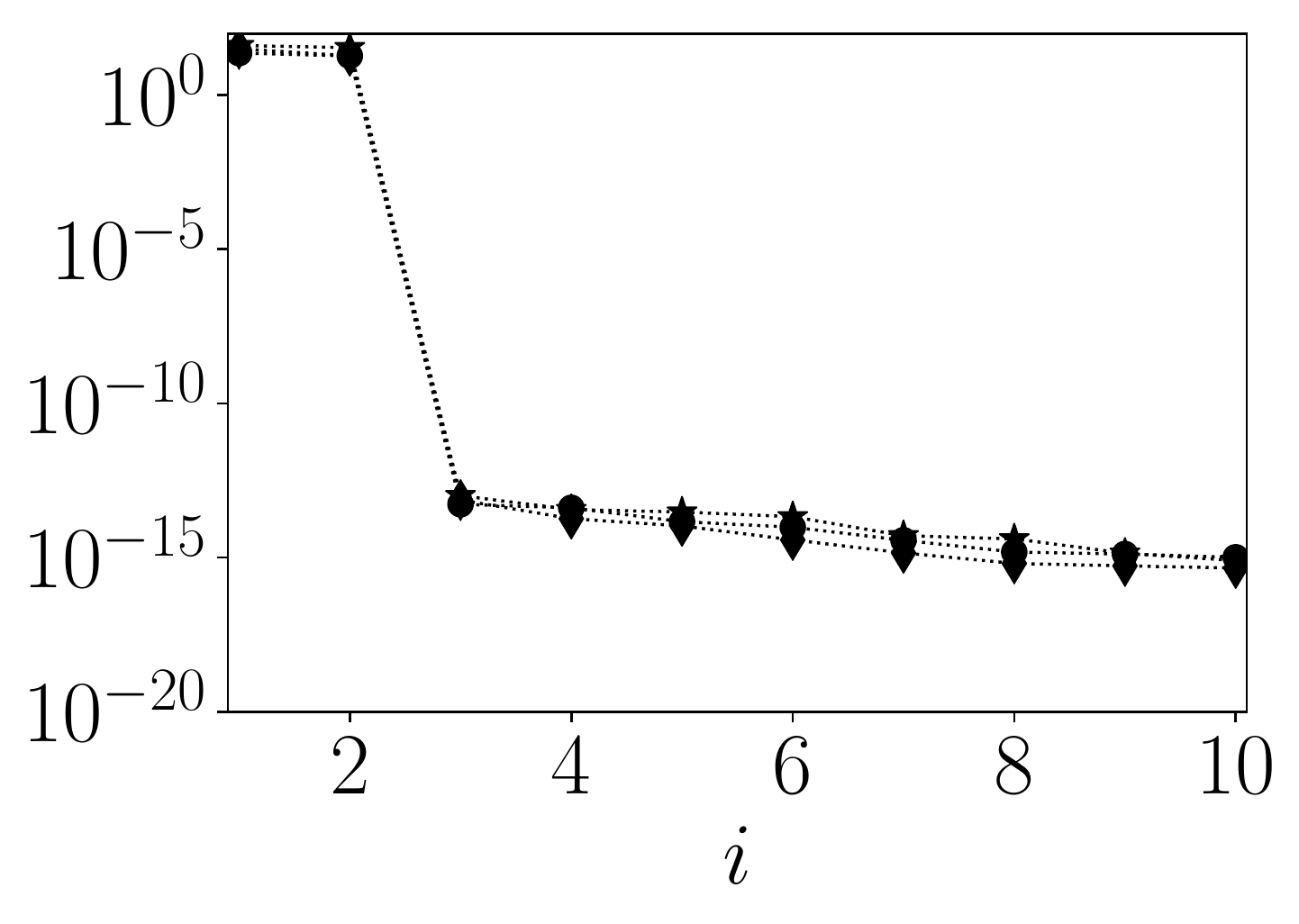}\\
	\includegraphics[width=0.32\textwidth]{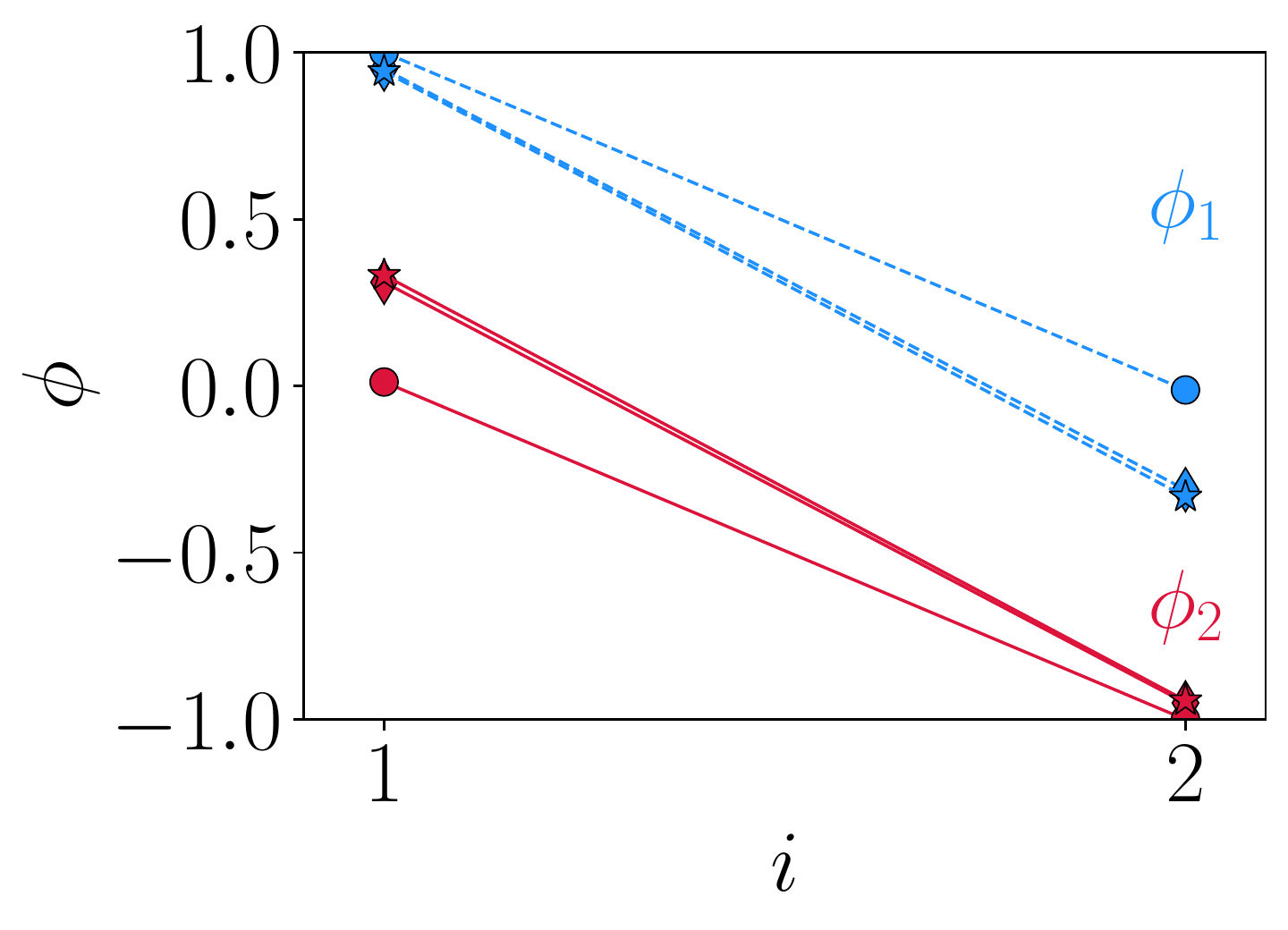}
	\includegraphics[width=0.32\textwidth]{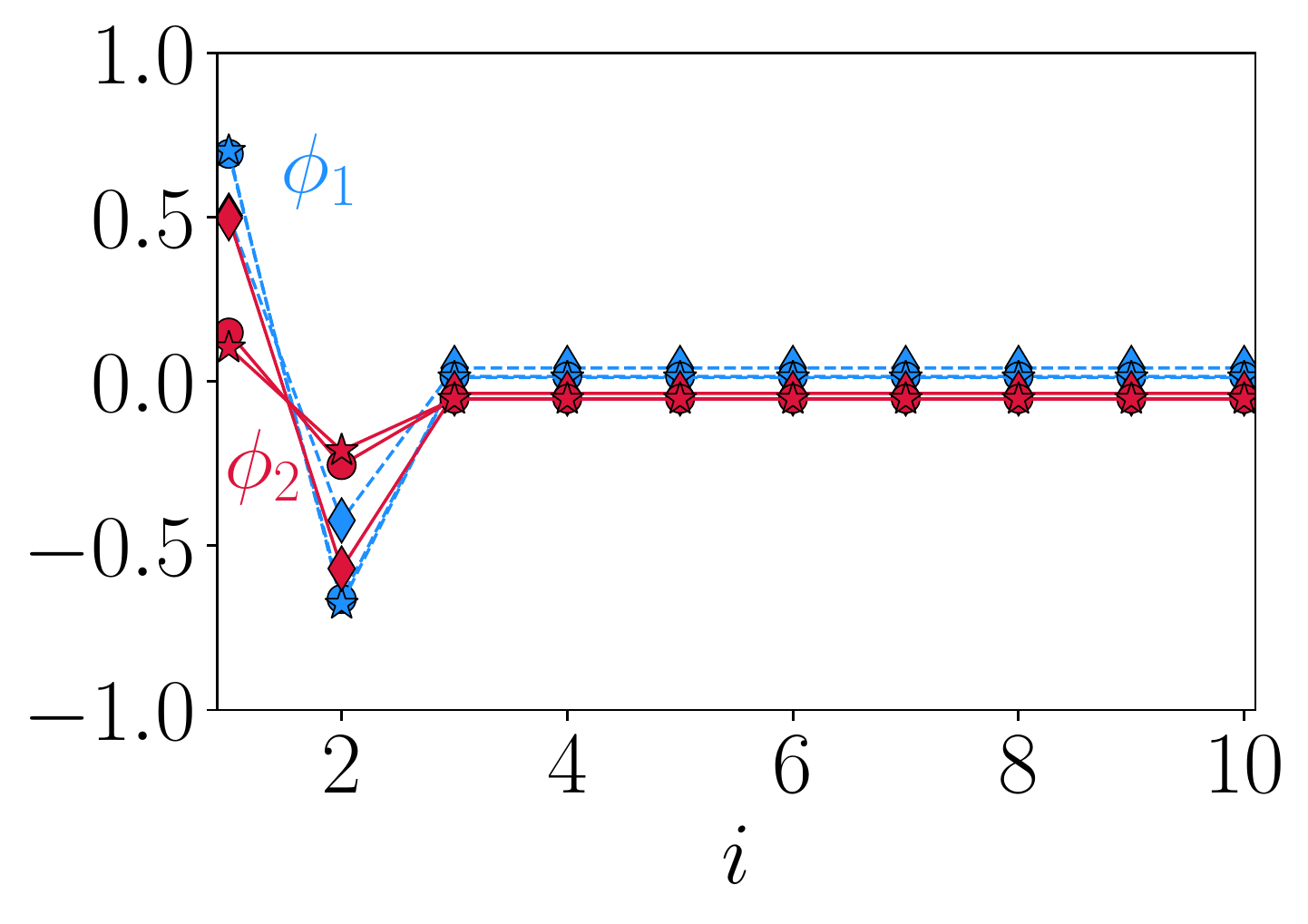}
	\includegraphics[width=0.32\textwidth]{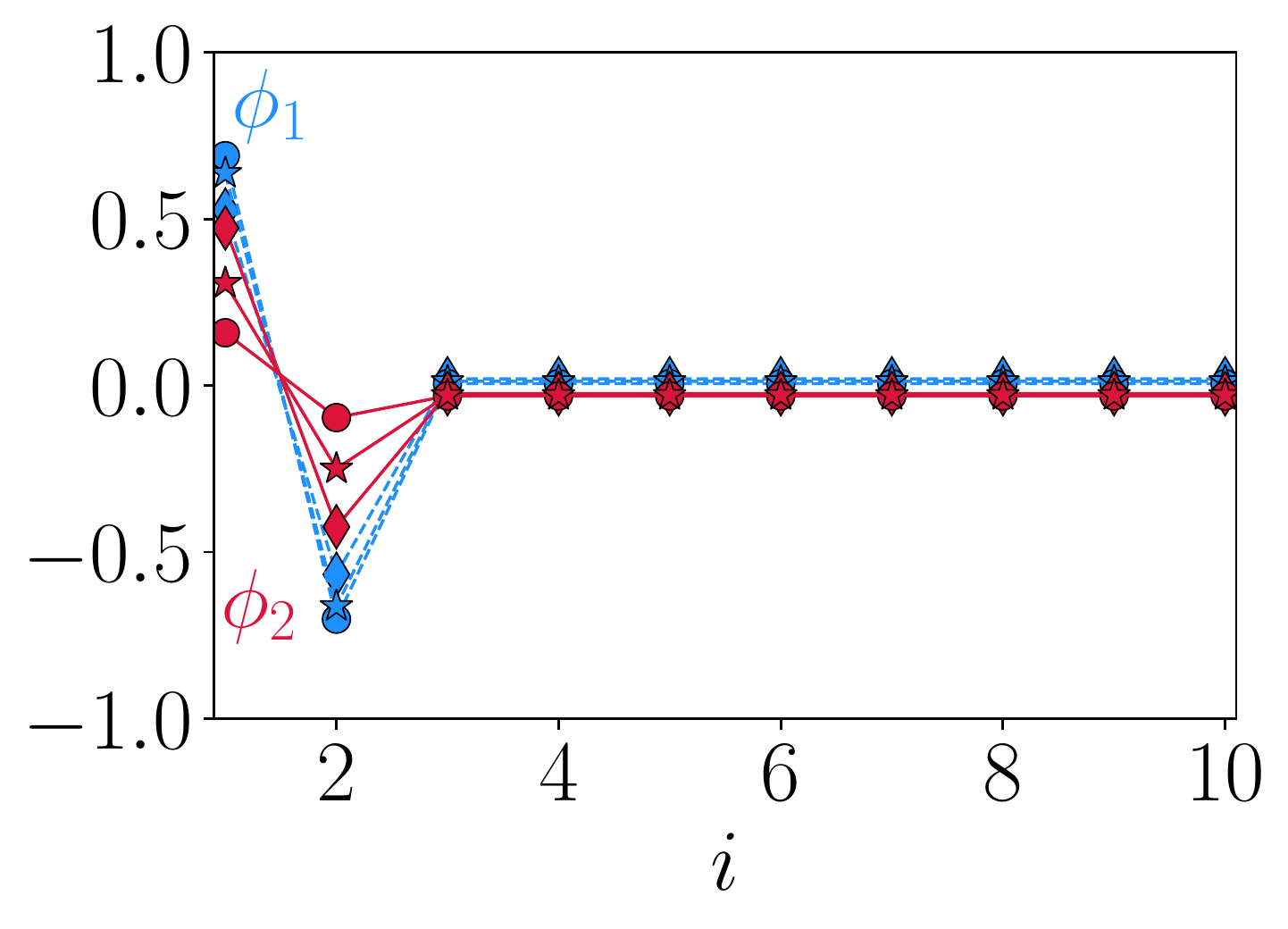}
	\adjincludegraphics[trim={1cm 0 0 {10.7cm} },clip=true]{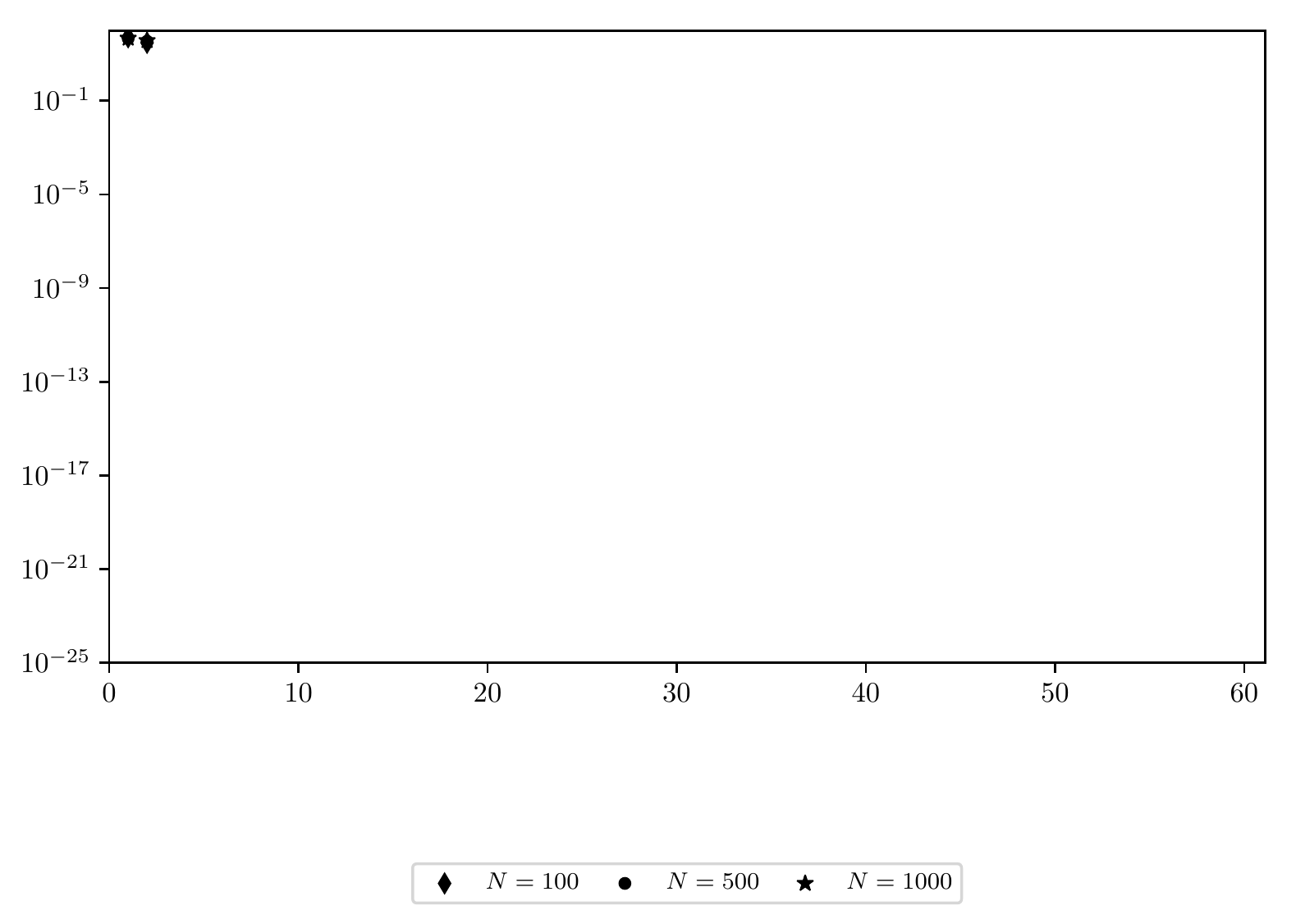}\vspace*{-0.5cm}
	\caption{Eigenvalue decay (1st row) and first two eigenvectors (2nd row) of the matrix $\widehat{\mat{H}}$ corresponding to example \ref{subsec:ex2} with $\kappa=5$, for different number of samples and increasing dimension (columns).}
	\label{fig:EX2_results_eigs}
\end{figure}

\begin{figure}[!ht]
\centering
\hspace{0.6cm}\raisebox{-0.5cm}{\small \rotatebox{0}{\textbf{$d=2$}}}\hspace{4.2cm} \raisebox{-0.5cm}{\small \rotatebox{0}{\textbf{$d=334$}}}\hspace{3.8cm} \raisebox{-0.5cm}{\small \rotatebox{0}{\textbf{$d=1000$}}}\\
\raisebox{1.7cm}{\small \rotatebox{90}{$\kappa=5$}}\hspace{0.05cm}
\includegraphics[width=0.315\textwidth]{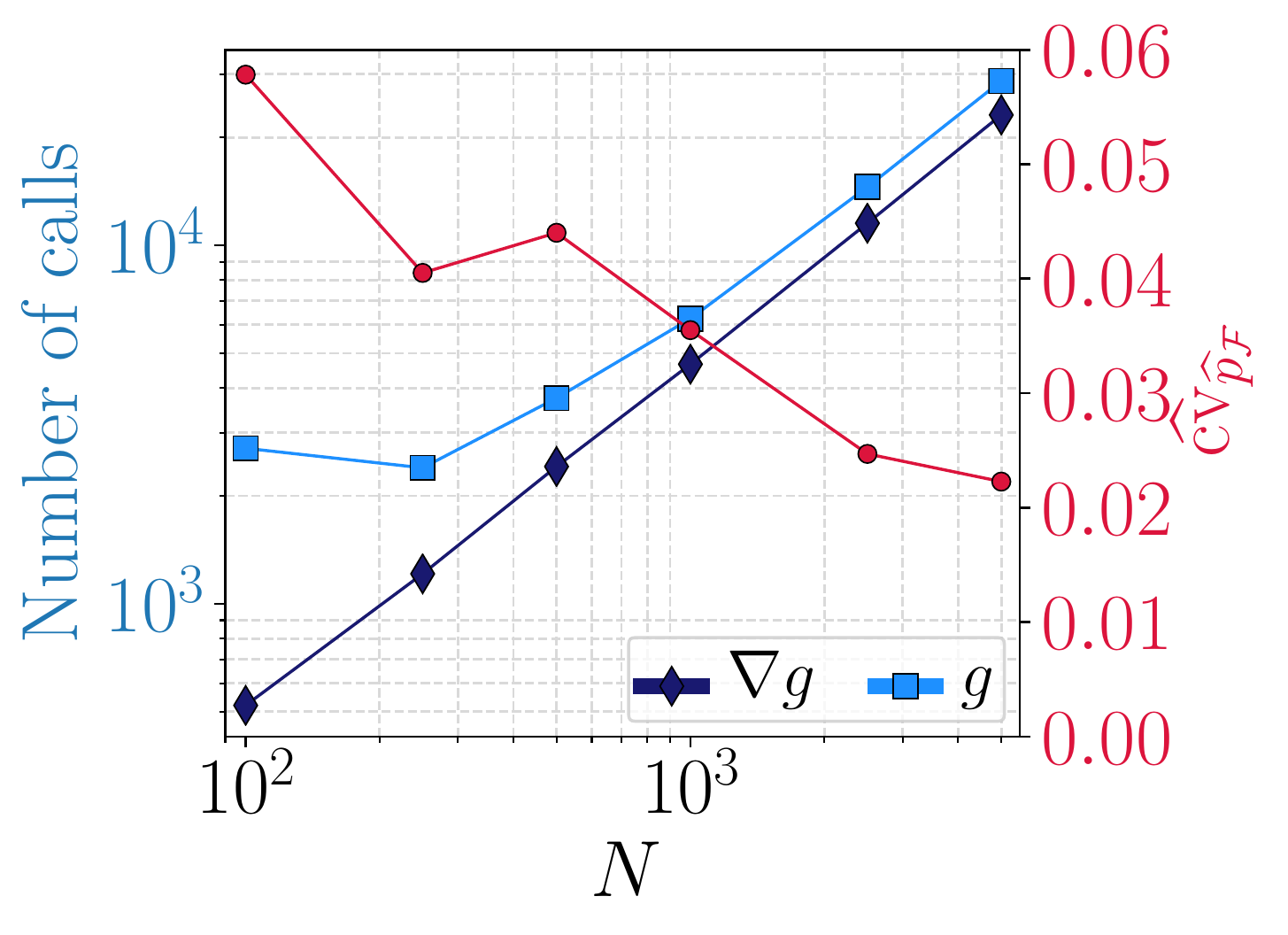}
\includegraphics[width=0.315\textwidth]{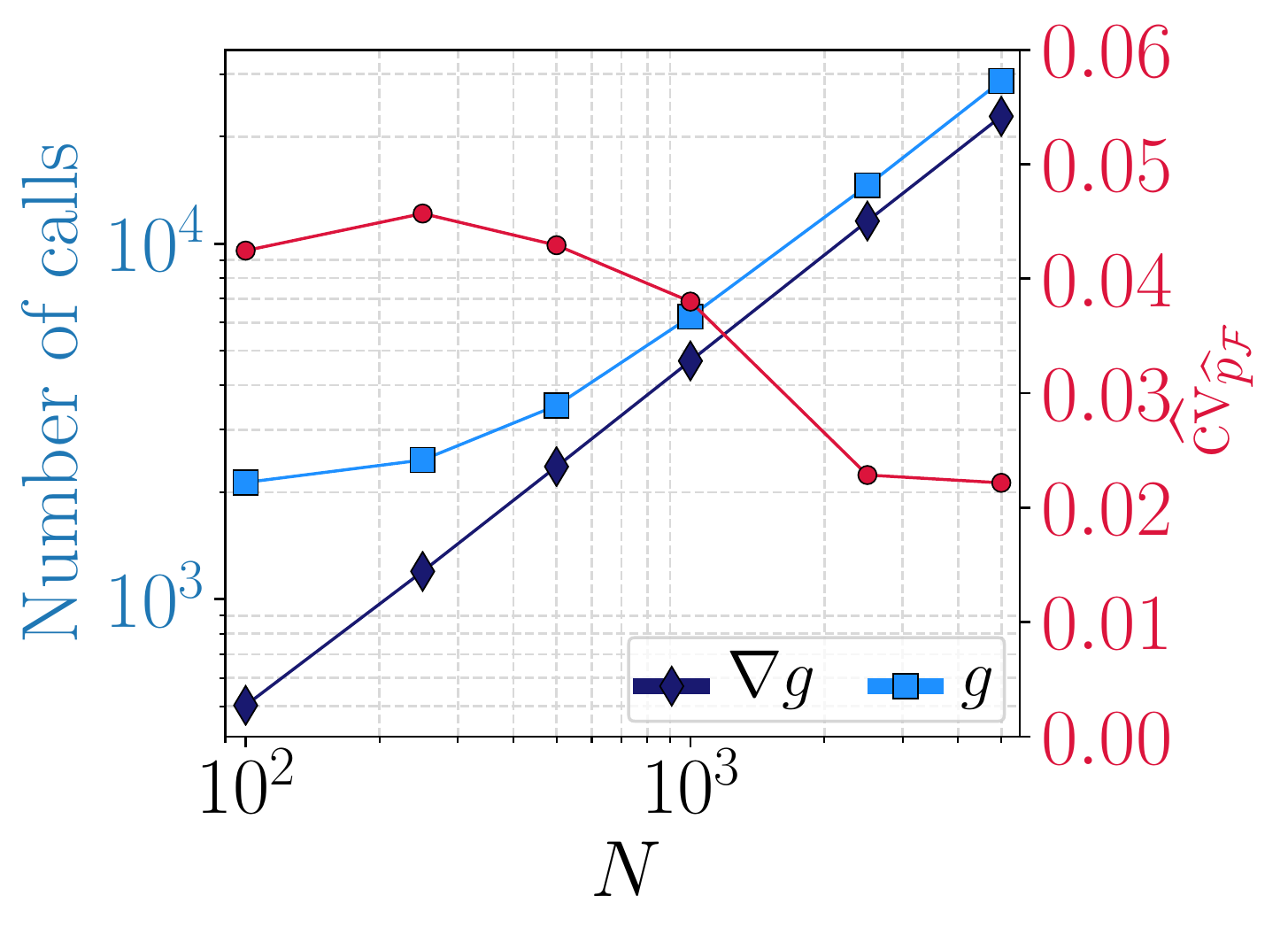}
\includegraphics[width=0.315\textwidth]{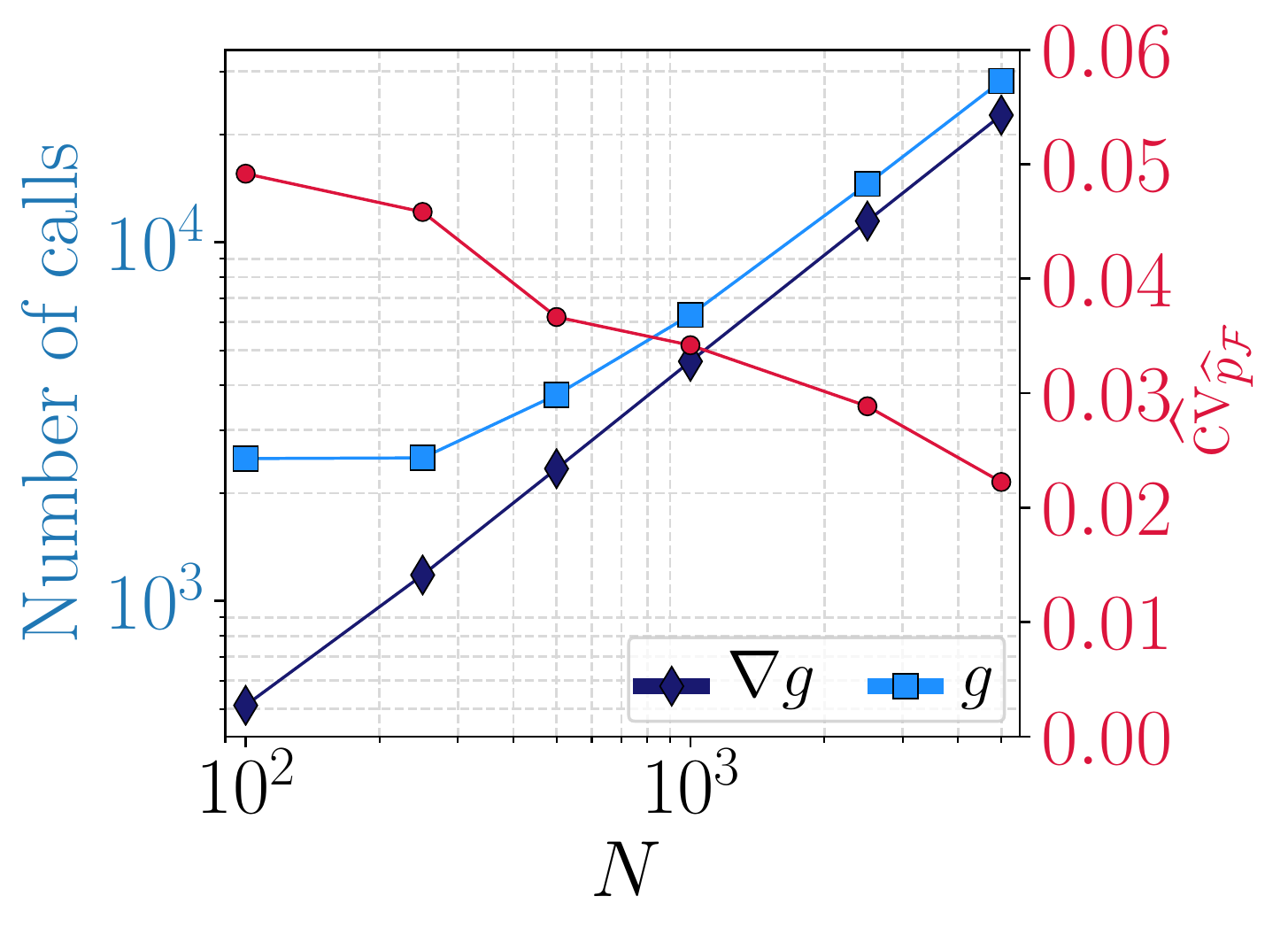}\\
\raisebox{1.7cm}{\small \rotatebox{90}{$\kappa=10$}}\hspace{0.05cm}
\includegraphics[width=0.315\textwidth]{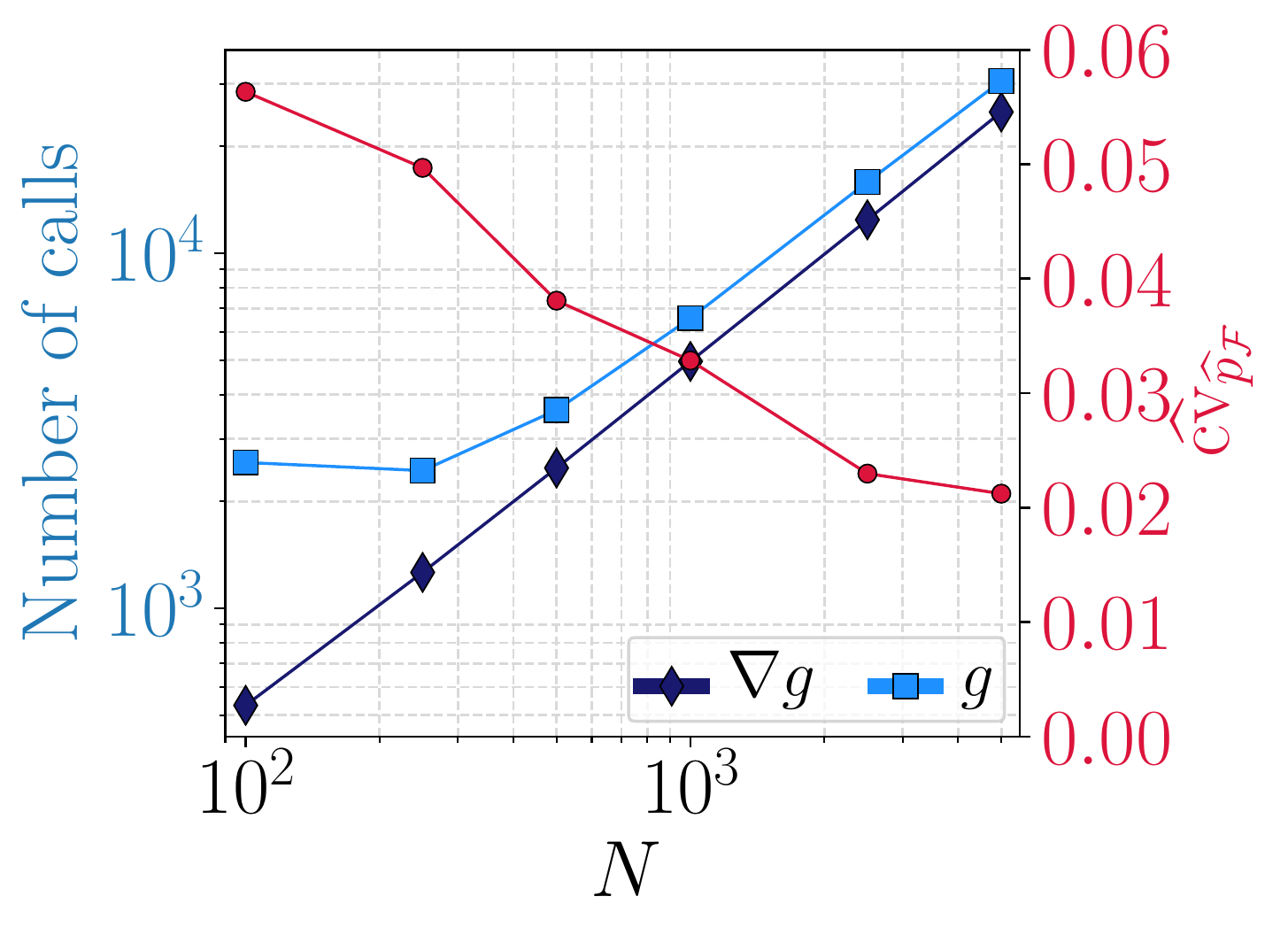}
\includegraphics[width=0.315\textwidth]{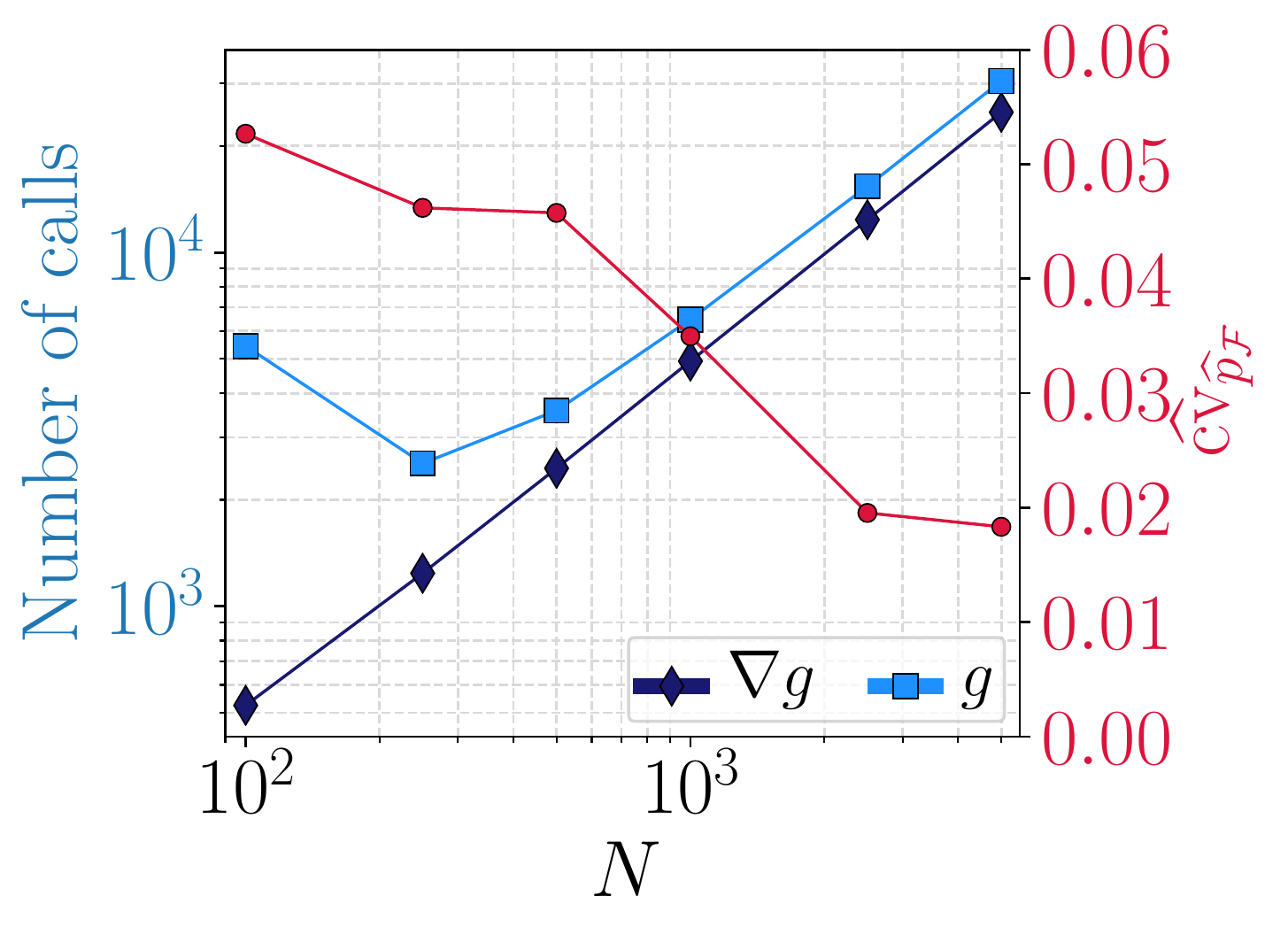}
\includegraphics[width=0.315\textwidth]{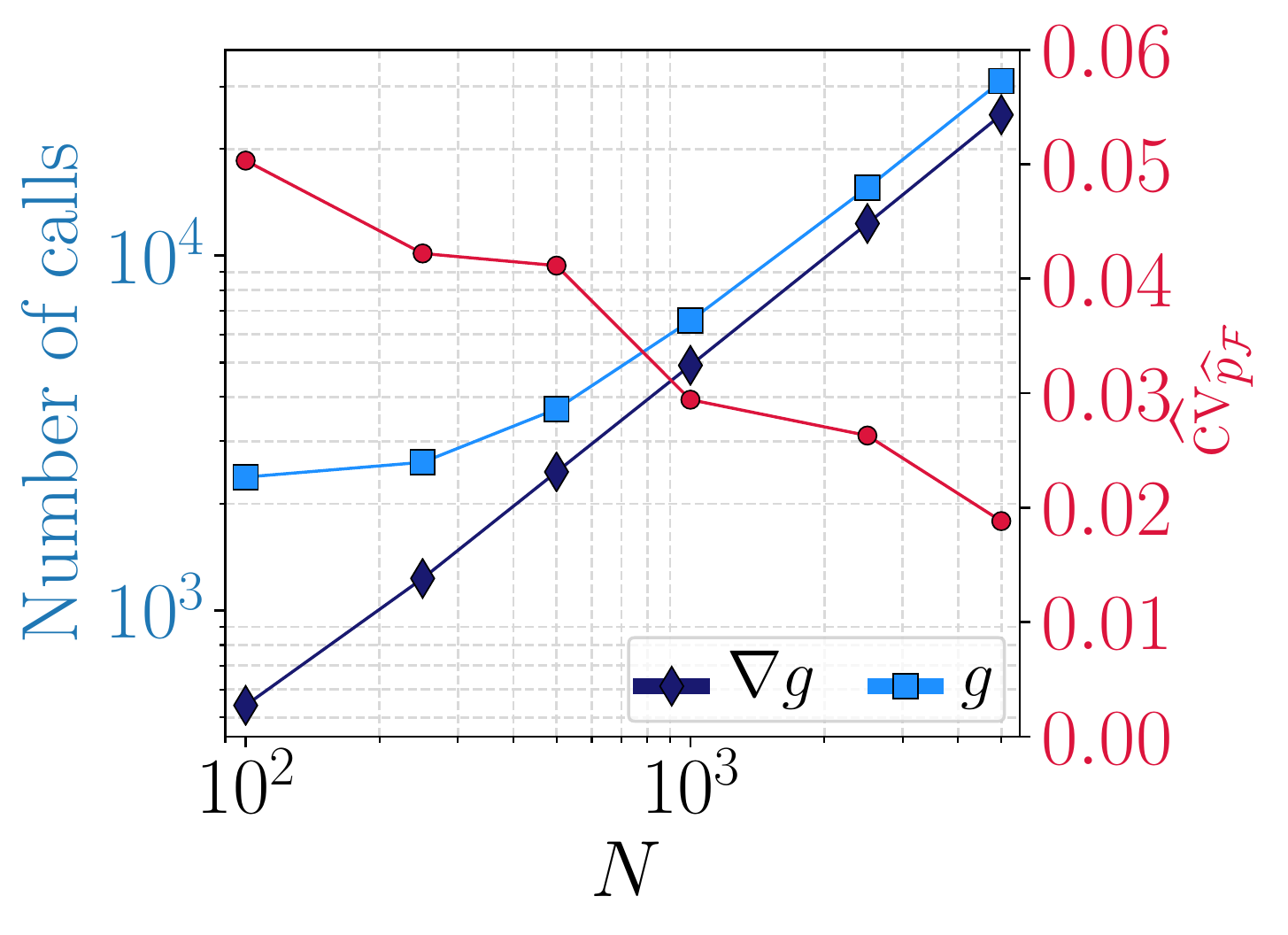}
\caption{iCEred with refinement for example~\ref{subsec:ex2}: number of LSF and gradient calls for varying sample size, curvature values (rows) and dimensions (columns). The $\widehat{\mathrm{cv}}$ of the failure probability estimate is also shown (in red).}
\label{fig:EX2_adapt_dim}
\end{figure}

In \Cref{fig:EX2_results_eigs}, we plot the eigenvalue decay and first two eigenvectors of the estimator $\widehat{\mat{H}}$, for different number of samples per level $N\in\{100,500,1000\}$ and a fixed curvature $\kappa=5$. The results are for illustration purposes only since they correspond to a single iCEred run. We note that these eigenpairs correspond to the eigendecomposition of $\widehat{\mat{H}}$ at the last iCEred iteration. The rank of the projector is $r=2$ and it is constant through all the iCEred levels. As in example \ref{subsec:ex1}, this is a problem with a clear low-dimensional structure since only the parameter components $\theta_1$ and $\theta_2$ define the geometry of the LSF; the remaining components represent a linear extrusion in the complementary directions. Thus, the rank remains unchanged even when using small number of samples per level. Note also that for $d=2$ there is no associated dimension reduction and the iCEred method reduces to standard iCE (as seen in \cref{fig:EX1EX2_gradplot}).  

\Cref{fig:EX2_adapt_dim} shows the number of calls of the LSF and its gradient, together with the coefficient of variation of the failure probability estimates; the number of samples per level is now chosen as $N\in\{100,250,500,1000,2500,5000\}$. The results are computed as an average of 100 independent runs. Since we perform refinement, all $\cv{\pfe}$ values are close to the predefined $\overline{\delta}$ for sample size smaller than $N=1000$. This behavior is similar for all studied dimension cases. Although the curvature parameter increases the nonlinearity of the problem, the performance of iCEred remains insensitive to this value. For instance, when using $N=1000$ samples per level, the failure probability estimates are $\pfe \approx [6.61\times 10^{-6}, 4.74\times 10^{-6}]$.

\subsection{2D plate in plane stress: LSF with random fields}\label{subsec:ex3}
We consider a steel plate model defined by a square domain $D$ with length $0.32$~m, thickness $t = 0.01$~m, and a hole of radius $0.02$~m located at its center. Spatial coordinates are denoted by $\ve{x}=[x_1,x_2]\in D$. The displacement field $\ve{u}(\ve{x}):=[u_{x_1}(\ve{x}),u_{x_2}(\ve{x})]^\tran$ is computed using elasticity theory through a set of elliptic PDEs (Cauchy--Navier equations) \cite{johnson_2009}. Due to the geometry of the plate, the PDEs can be simplified under the plane stress hypothesis to
\begin{equation}\label{eq:plate}
G(\ve{x})\nabla^2\ve{u}(\ve{x}) + \frac{E(\ve{x})}{2(1-\nu)} \nabla(\nabla\cdot\ve{u}(\ve{x})) + \mat{b} =0,
\end{equation}
where $G(\ve{x}):= E(\ve{x})/(2(1+\nu))$ is the shear modulus, $\nu = 0.29$ is the Poisson ratio of the steel, and $\mat{b}$ is the vector of body forces acting on the plate, assumed to be negligible.

\begin{figure}[!ht]
\centering
\includegraphics[width=0.46\textwidth]{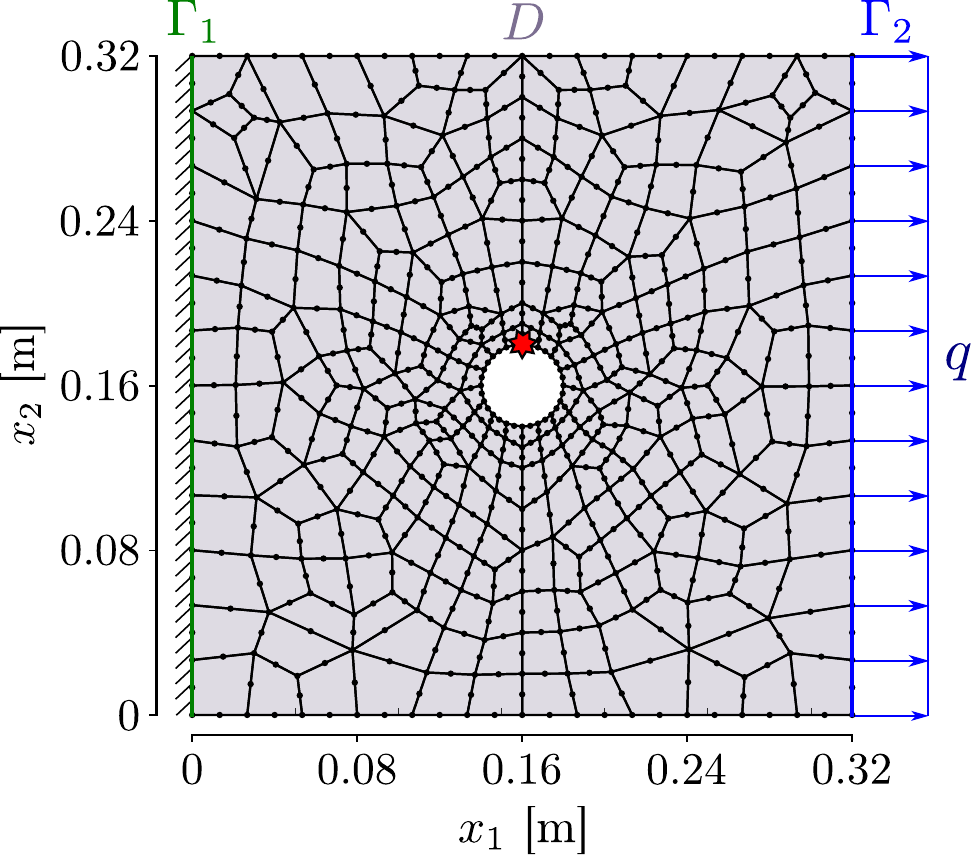}
\caption{Plate configuration (star marks the location of the control node).}
\label{fig:plate}
\end{figure}

A Dirichlet boundary condition is imposed at the left edge of the plate, $\ve{u}(\ve{x})=\ve{0}$ for $\ve{x}\in \Gamma_1$. Moreover, a random surface load $q$ is applied at the right boundary $\Gamma_2$. This action is modeled as a Gaussian random variable with mean $\mu_q = 60$~MPa and standard deviation $\sigma_q = 12$~MPa. \Cref{eq:plate} is solved with the finite element method using $282$ eight-node serendipity quadrilateral elements \cite{onate_2009}, as shown in \Cref{fig:plate}.

The Young's modulus $E(\ve{x})$ is random and spatially variable. A log-normal random field with mean value $\mu_{E} = 2\times 10^{5}$~MPa and standard deviation $\sigma_{E} = 3\times 10^{4}$~MPa is used for its representation. We select the isotropic exponential kernel $k(\ve{x},\ve{x}')= \exp(-\ell^{-1}\norm{\ve{x}-\ve{x}'}_2)$ as the autocorrelation function of the underlying Gaussian field; the correlation length is set to $\ell=0.04$ m. The random field is approximated with the Karhunen--Loève (K-L) expansion as \cite{ghanem_and_spanos_2012}
\begin{equation}\label{eq:K-Llog}
E(\ve{x})\approx \widehat{E}\big(\ve{x};\ve{\theta}^{(\text{K-L})}\big) := \exp\left[ \mu_{E'} + \sum_{k=1}^{K} \sqrt{\alpha_{k}} \varphi_{k}(\ve{x}) \theta_{k}^{(\text{K-L})}\right],
\end{equation}
where $\mu_{E'}$ and $\sigma_{E'}$ are the mean and standard deviation of the associated Gaussian field, $\alpha_{k}\in[0,\infty)$ ($\alpha_{k}\geq \alpha_{k+1}$ and $\lim_{k\rightarrow \infty}\alpha_k = 0$), $\varphi_{k}(\ve{x})$ are the eigenvalues and eigenfunctions of the covariance operator $C(\ve{x},\ve{x}')=\sigma_{E'}^2\cdot k(\ve{x},\ve{x}')$, and $\ve{\theta}^{(\text{K-L})}\in\mathbbm{R}^K$ is the standard Gaussian random vector of K-L coefficients. The number of terms in the expansion is fixed to $K=868$, which accounts for $92.5\%$ of the spatial average of the variance of the Gaussian random field ($\ln E$). The eigenpairs are estimated with the Nyström method using $100$ Gauss--Legendre (GL) points in each direction. 

We select the principal stress ${\sigma}_1(\ve{\sigma}) =\nicefrac{1}{2}\left(\sigma_{x_1}+\sigma_{x_2}\right) + [\left(\nicefrac{1}{2}\left(\sigma_{x_1}-\sigma_{x_2}\right)\right)^2 + \tau_{x_1x_2}^2]^{\nicefrac{1}{2}}$ as the target QoI. This quantity depends on the stress field of the plate $\ve{\sigma}(\ve{x}):= [\sigma_{x_1}(\ve{x}), \sigma_{x_2}(\ve{x}),\allowbreak \tau_{x_1x_2}(\ve{x})]^\tran$, which is computed after obtaining the displacement field via \cref{eq:plate} and applying Hooke's law for continuous media (see \cite{solecki_and_conant_2003} for details). The failure of the plate occurs when the value of $\sigma_1$ at a control point $\ve{x}_{\mathrm{ctr}}=[0.16,0.18]$ exceeds a yield tensile strength of $320$ MPa. The LSF is defined as
\begin{equation}\label{eq:LSF_plate}
g(\ve{\theta}) = 320 - \sigma_1(\ve{\sigma}(\ve{x}_{\mathrm{ctr}};\ve{\theta})).
\end{equation}

The $\sigma_1$ stress defining the LSF is evaluated at the GL point of the element closest to the control node $\ve{x}_{\mathrm{ctr}}$. The uncertain parameter vector $\ve{\theta}=[{\theta}^{(q)},\ve{\theta}^{(\text{K-L})}]$ includes the load random variable and the Young's modulus random field. The random variable $\theta^{(q)}$ is also standard Gaussian since we apply the transformation $\theta^{(q)}=(q-\mu_q)/\sigma_q$. The stochastic dimension of the problem is $d=K+1=869$. 

This example cannot be solved efficiently using standard CE or iCE, since a large number of effective samples per level are required for fitting high-dimensional parametric densities. Therefore, to compare the results of iCEred, we estimate the failure probability by an average of 100 independent runs of subset simulation (SuS) using $N=3000$ samples per level \cite{au_and_beck_2001}. Although it tends to produce failure probability estimates with a relatively high coefficient of variation, SuS is the standard algorithm for solving efficiently reliability problems in high dimensions. 
\begin{table}[!ht]
	\centering 
	\caption{Failure probability estimates: the iCEred results are shown as an average of 40 simulations; the SuS estimate is computed as an average of 100 simulations.}
	\label{tab:pf}
	\begin{tabular}{ccccc|ccccc}
		\toprule
		\multicolumn{5}{c|}{iCEred} & \multicolumn{5}{c}{SuS} \\
		$N$ &  $g_{\mathrm{call}}$ & $(\nabla g)_{\mathrm{call}}$ & $\widehat{p}_\mathcal{F}$ & $\cv{\pf}$  & $N$  & $n_{\mathrm{lv}}$& $g_{\mathrm{call}}$ &$\widehat{p}_\mathcal{F}$ & $\cv{\pf}$  \\
		\midrule
		100 &  2396 & 675  & $3.57\times 10^{-6}$& 0.050  & \multirow{3}{*}{3000} & \multirow{3}{*}{6} & \multirow{3}{*}{$1.62\times 10^4$} &\multirow{3}{*}{$3.75\times 10^{-6}$} &  \multirow{3}{*}{0.215}    \\
		250 &  2416 & 1612 & $3.60\times 10^{-6}$  & 0.046   &  &  &    &  &      \\
		500 &  3685  & 3037 & $3.62\times 10^{-6}$ & 0.043  &   &    & & &      \\
		\bottomrule
	\end{tabular}
\end{table}

We apply the iCEred method with number of samples per level selected from the set $N\in\{100,250,500\}$. \Cref{tab:pf} shows the results as an average of 40 independent simulations. Note that the coefficient of variation of the failure probabilities are close to the target $\overline{\delta}$. This is also reflected in the averaged number of LSF and gradient calls, denoted respectively as $g_{\mathrm{call}}$ and $(\nabla g)_{\mathrm{call}}$. Both number of evaluations are the same when refinement is not required, and the values are closer to each other when number of required iterations in the refinement step is small. As seen also in the previous examples, the value of $g_{\mathrm{call}}$ increases considerably at a small value of $N$ in order to match $\overline{\delta}$. However, there is a trade-off between the computational cost of the extra LSF and the LSF gradient calls. If the gradient is expensive to evaluate, a small number of samples is recommended since the value of the probability estimate is later improved by the refinement. On the contrary, if an efficient way to compute the LSF gradient is available, the extra LSF evaluations in the refinement might exceed the cost of the gradient computations; in this case using a large sample size is recommended. Note also in \Cref{tab:pf} that the mean value of the probability of failure computed by repeated runs of SuS is close to the values estimated by repeated runs of iCEred, whereas the coefficient of variation of the SuS estimate is significantly higher than the one of iCEred. Moreover, the iCEred method requires considerably fewer LSF evaluations, and thus can be much more efficient than SuS provided an effective gradient computation is feasible.
\begin{figure}[!ht]
	\centering
	\hspace*{0.3cm}
	\raisebox{-0.5cm}{\small\rotatebox{0}{{$N=100$}}}\hspace{3.5cm} 
	\raisebox{-0.5cm}{\small\rotatebox{0}{{$N=250$}}}\hspace{3.6cm} \raisebox{-0.5cm}{\small\rotatebox{0}{{$N=500$}}}\\
	\includegraphics[width=0.32\textwidth]{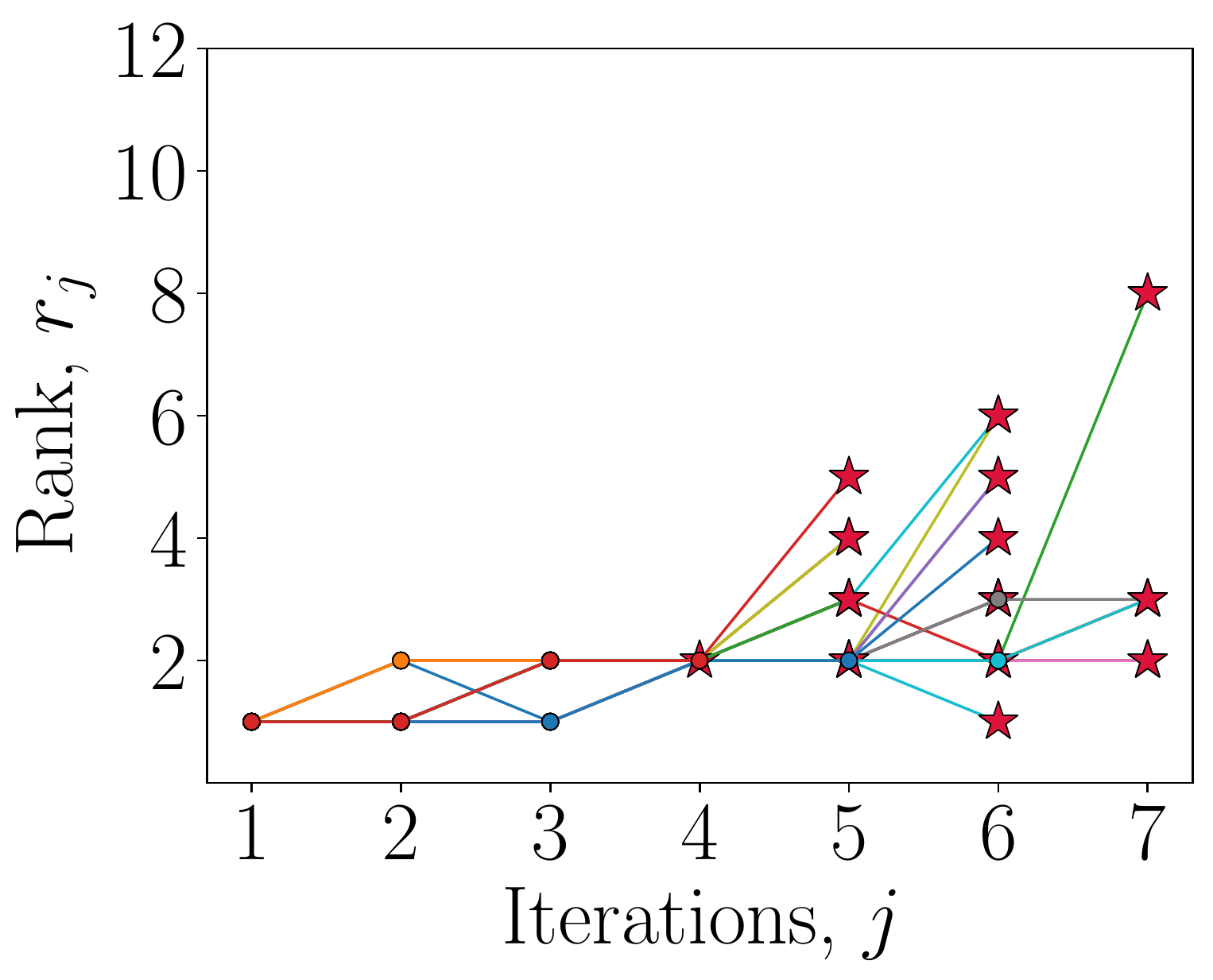}
	\includegraphics[width=0.3\textwidth]{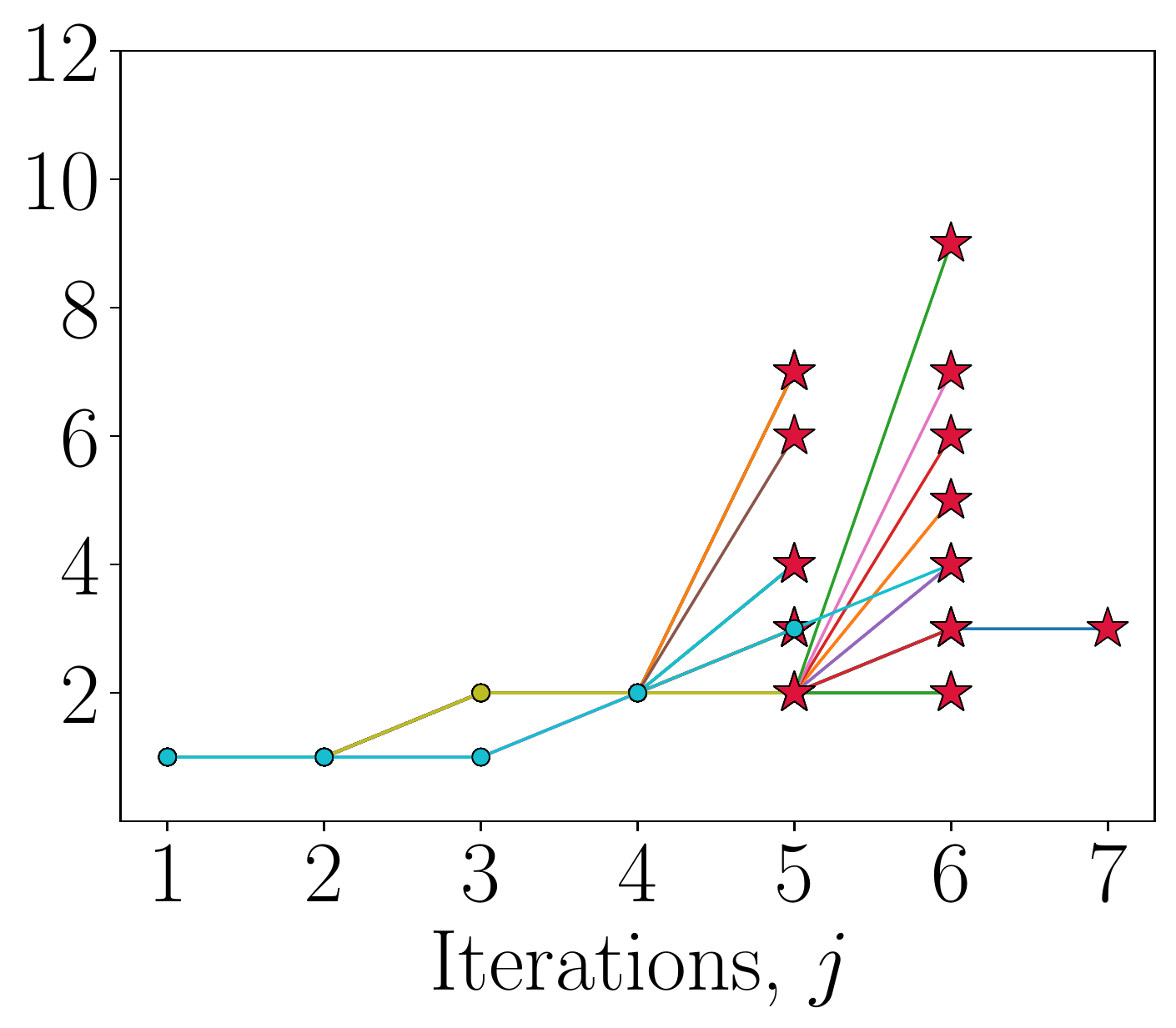}
	\includegraphics[width=0.3\textwidth]{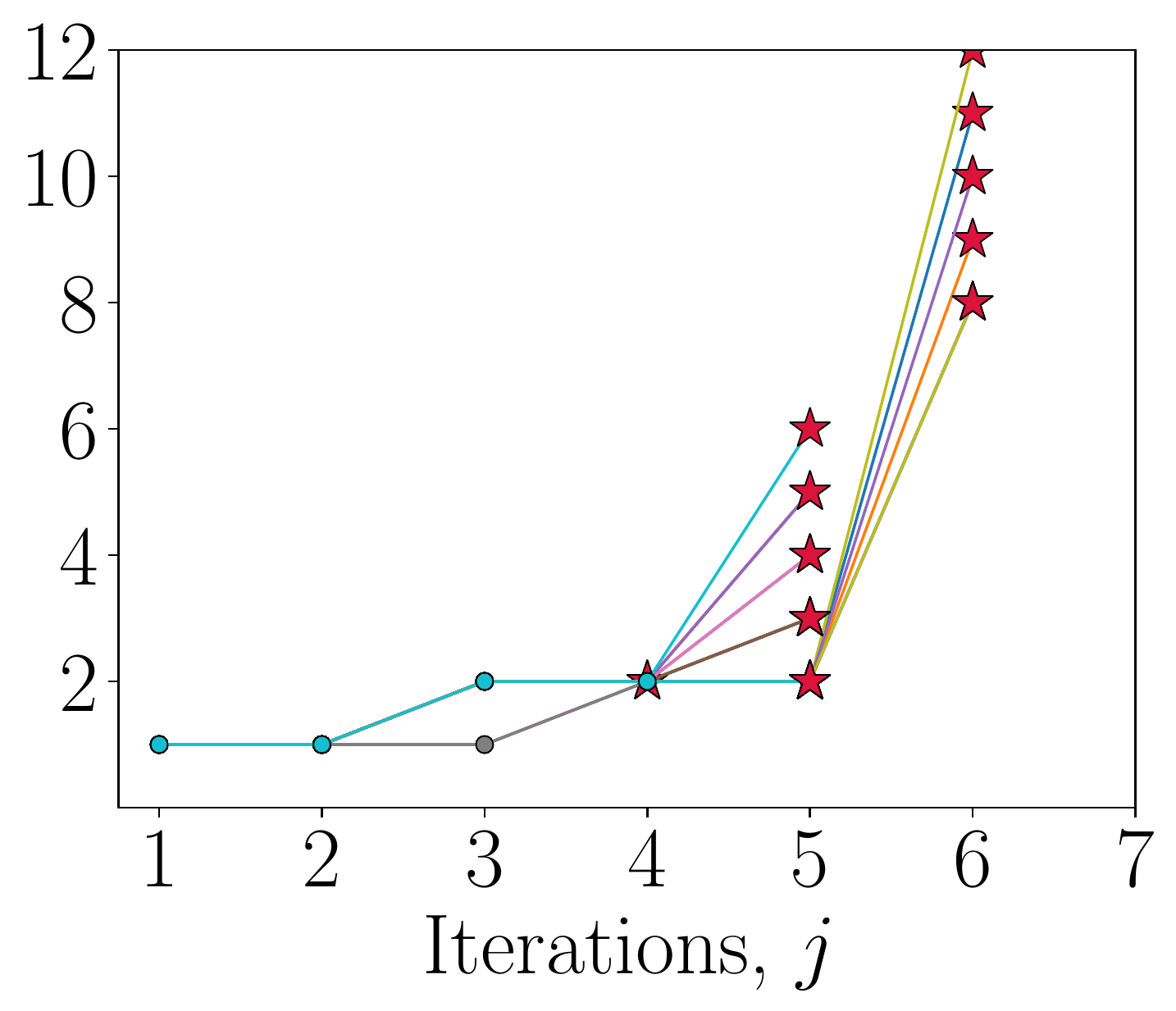}
	\caption{Example~\ref{subsec:ex3}: evolution of the rank with the iCEred iterations for several simulation runs and $N\in\{100,250,500\}$ (columns). The star marks the final value of the rank at the given simulation run.}
	\label{fig:EX3_j_vs_r}
\end{figure}

We plot in \Cref{fig:EX3_j_vs_r} the evolution of the rank with the iCEred iterations (intermediate levels). The results are shown for different sample sizes and independent simulation runs. As $N$ increases the number of iterations becomes smaller; on average, for each of the employed sample sizes, the number of levels are $n_{\mathrm{lv}}=[5.75,5.45,5.08]$. For the investigated number of samples per level, the maximum observed ranks across the 40 independent simulations are $r=[8,9,12]$; however, the final rank is on average $r=3$ for $N=100$, and $r=4$ for $N=[250,500]$. We also observe that the value of the smoothing parameters at the final iteration are on average $s_{n_{\mathrm{lv}}}=[0.098,0.084,0.051]$, which shows that $s$ reaches values closer to zero for larger sample sizes. Note in \cref{eq:grad_logind1} that the gradient of the log-smooth indicator is essentially driven by the gradient of the LSF; the smoothing parameter indirectly determines the locations at which the gradient is evaluated. In this example, the increment in the rank with the number of intermediate levels seems to be related to the LSF gradient discovering a larger FIS as the samples move towards the failure domain where the nonlinearity of the LSF increases.
\begin{figure}[!ht]
	\centering
	\includegraphics[width=0.55\textwidth]{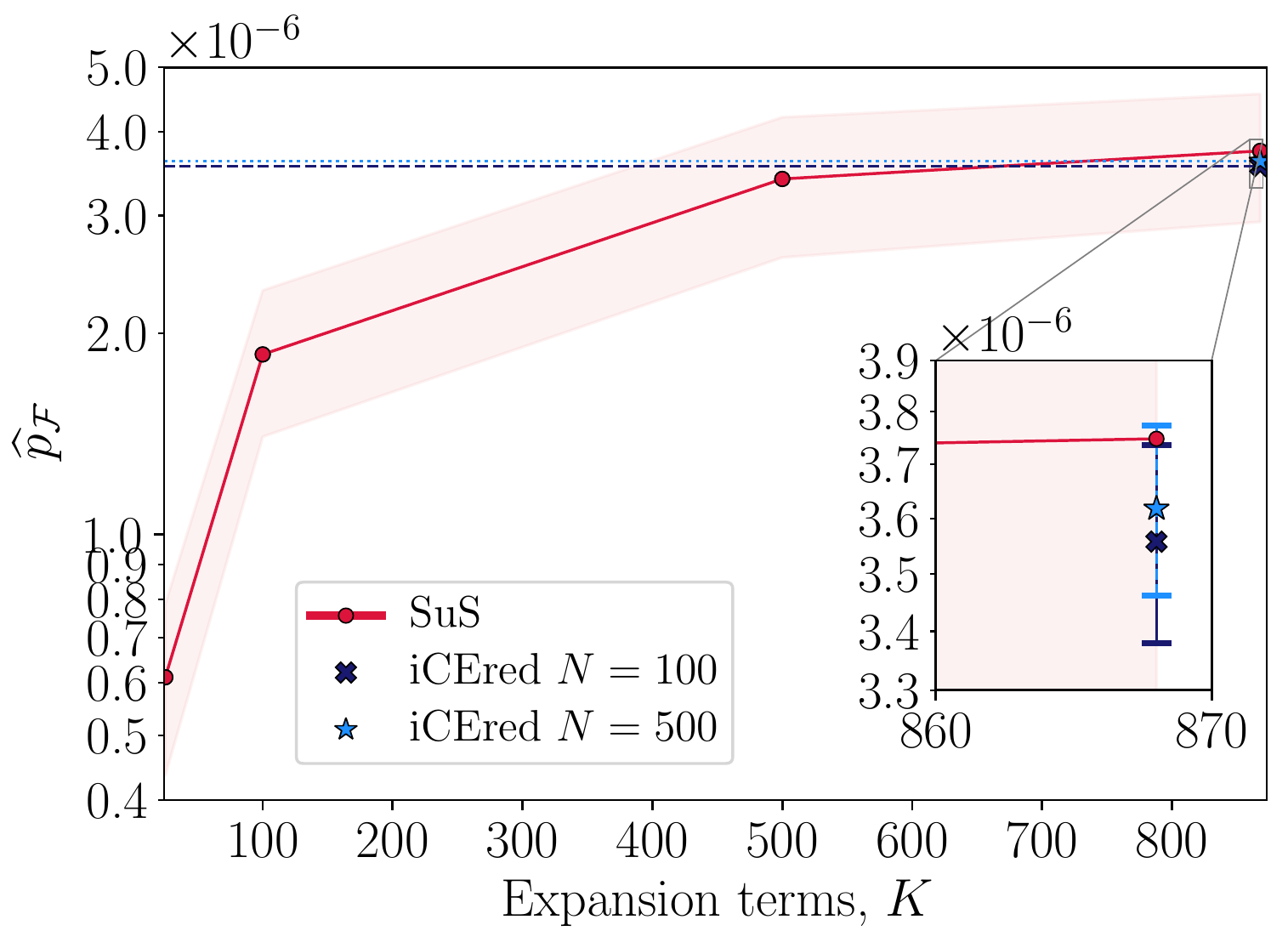}
	\caption{Example~\ref{subsec:ex3}: probability of failure estimated by SuS at selected truncation orders in the K-L expansion (mean and standard deviation bounds), and probability of failure estimated by iCEred at the $K=868$ term (zoomed area: mean and standard deviation bounds).}
	\label{fig:EX3_Pf_MSEa}
\end{figure}

Finally, \Cref{fig:EX3_Pf_MSEa} shows the evolution of the probability of failure for increasing terms in the \mbox{K-L} expansion \cref{eq:K-Llog}. Different failure probabilities are computed by SuS for truncation orders $K\in\{25, 100, 500, \allowbreak 868\}$; the iCEred method is only evaluated for $K=868$. In both approaches, the estimators are given as average of 40 independent simulations. Note the difference between the failure probabilities computed by reducing the number of input dimensions (K-L terms) and those estimated by identifying the FIS. Solving the rare event simulation problem on the FIS is \emph{not} equivalent to truncating the parameter space, as in the latter case, there is an associated loss of information (e.g., random field variability). The effective dimension of the local FIS in this example is on average $r=4$ (when using $N=500$ samples), and thus the iCEred method can compute the failure probability associated with an input dimension of $K=868$ efficiently and with a very small coefficient of variation.

\section{Summary and conclusions}\label{sec:conclusions}
We developed a computational framework for solving high-dimensional rare event simulation problems. This includes settings where the underlying system properties are spatially inhomogeneous and random field models are required for their representation. The approach (called iCEred) is an extension of importance sampling with the cross-entropy method that enhances its efficiency in high dimensions. The main idea is to adapt dimension reduction techniques developed for Bayesian inference through expressing rare event simulation as a Bayesian inversion problem. This enables the identification of the potential low-dimensional structure of the rare event simulation problem. As a result, we construct a so-called failure-informed subspace, where efficient low-dimensional biasing distributions for importance sampling can be defined. This process ultimately requires the computation of the limit-state function gradient. 

When the limit-state function is highly nonlinear, the rank of the projectors that map the parameters onto the failure-informed subspace increases as one approaches the final iterations of the algorithm. This indicates that when the smooth approximation gravitates to the indicator function, the gradient samples are discovering a larger subspace due to the increase in nonlinearity at the failure region. 

The numerical experiments show that the iCEred is able to effectively compute rare event probability estimates in high dimensions with considerably smaller variability and required sample size compared to standard simulation approaches that typically rely on Markov chain Monte Carlo. We also remark that when dealing with problems that have multiple failure points (multi-modal failure hypersurface), mixture distributions can also be adapted within the iCEred framework.

A possible improvement of the method is to define a measure to assess whether or not the second moment matrix of the gradient of the log-smooth indicator function needs to be computed at a given level. Since the main interest is the estimation of the failure probability and not the construction of the optimal biasing distribution itself, it might be sufficient to build the failure-informed space at the initial iterations and continue with the standard methodology in the remainder levels. Furthermore, rather than using the same sample size for the limit-state function evaluations and gradient computations, one can define two different sample sizes. The number of samples per level required to evaluate the gradient can be adapted based on the value of the rank at the given intermediate level. These suggestions could potentially save significant computational demands related to gradient evaluations in iCEred. Moreover, the low-dimensional structure of the failure-informed subspace can be exploited in other rare event simulation algorithms.

\appendix
\section{Adjoint solution}\label{sec:adjoint}
In example~\ref{subsec:ex3}, the eigenfunctions of the covariance operator are interpolated at the GL points of the finite element mesh using Nystr\"{o}m formula \cite{press_et_al_2007}. Thus, for a $n_{\mathrm{GP}}$-point spatial discretization, the K-L representation of the Young's modulus truncated at the $K$-th term is expressed in matrix form as $\widehat{\ve{E}} = \exp\left[\mu_{E'} + \mat{A}\ve{\theta}^{(\mathrm{K-L})}\right]$, where $\ve{\Phi} \ve{\Lambda}=\mat{A}\in \mathbbm{R}^{n_{\mathrm{GP}}\times K}$, $\ve{\Lambda} = \text{diag}(\sqrt{\ve{\lambda}}) \in\mathbbm{R}^{K\times K}$ is a diagonal matrix with the square root of the eigenvalues of the covariance operator, and $\ve{\Phi}\in\mathbbm{R}^{n_{\mathrm{GP}}\times K}$ is a matrix containing the associated eigenfunctions evaluated at the GL points. 

For a given realization of the uncertain parameters $\ve{\theta}=[{\theta}^{(q)}, \ve{\theta}^{(\text{K-L})}]$, the finite element formulation yields the global matrix equilibrium equation $\mat{K}(\ve{\theta})\ve{u}(\ve{\theta}) = \ve{f}(\ve{\theta})$, where $\mat{K}\in\mathbbm{R}^{n_{\text{dof}}\times n_{\text{dof}}}$ is the stiffness matrix, $\ve{f}\in\mathbbm{R}^{n_{\text{dof}}}$ is the force vector, and $\ve{u}\in\mathbbm{R}^{n_{\text{dof}}}$ is the vector of displacements; $n_{\text{dof}}$ denotes the total number of degrees of freedom. Furthermore, the stress field can be computed at a specific GL point as,
\begin{equation}\label{eq:target_stress}
\ve{\sigma}^\star(\ve{\theta}) = \left( \mat{D}^{\star}(\ve{\theta})~\mat{B}^{\star}~ \mat{M}\right) \ve{u}(\ve{\theta})
\end{equation}
where the constitutive matrix  $\mat{D}^{\star}\in\mathbbm{R}^{3\times 3}$ and the deformation matrix $\mat{B}^{\star}\in\mathbbm{R}^{3\times n_{\text{eq}}}$ are those evaluated at the GL point closest to the control point (cf., \Cref{fig:plate}). In our case, the number of element equations is $n_{\text{eq}}=16$, since there are two degrees of freedom for each element node. Note that in \cref{eq:target_stress}, we employ a matrix $\mat{M}\in \mathbbm{R}^{n_{\text{eq}}\times n_{\text{dof}}}$ to `activate' the degrees of freedom corresponding to the nodes of the element that has the control point as one of its nodes. 

Our aim is to compute directly the gradient of the principal stress ${\sigma}_1(\ve{\sigma}^\star(\mat{D}(\ve{\theta}), \ve{u}(\ve{\theta})))$ with respect to $\ve{\theta}$. We consider the total derivative 
\begin{equation}\label{eq:adjoint}
\frac{\dd {\sigma}_1}{\dd \ve{\theta}} = \frac{\dd {\sigma}_1}{\dd \ve{\sigma}^\star}\Big( \frac{\dd \ve{\sigma}^\star}{\dd \mat{D}^\star} \frac{\dd \mat{D}^\star}{\dd \ve{\theta}} + \frac{\dd \ve{\sigma}^\star}{\dd \ve{u}}\frac{\dd \ve{u}}{\dd \ve{\theta}} \Big). 
\end{equation}

The multiplicative term in \cref{eq:adjoint} is obtained from the definition of the $\sigma_1$ stress
\begin{equation}
\frac{\dd {\sigma}_1}{\dd \ve{\sigma}^\star} =  \left[ \frac{1}{2} \left(1+ \frac{{\sigma}^\star_x-{\sigma}^\star_y}{2e}\right), \frac{1}{2} \left(1 - \frac{{\sigma}^\star_x-{\sigma}^\star_y}{2e}\right),  \frac{{\tau}^\star_{xy}}{e}\right], \quad e=\sqrt{{\left(\frac{{\sigma}^\star_x-{\sigma}^\star_y}{2}\right)}^2+{{\tau}^\star_{xy}}^2}.
\end{equation}

To compute the first term inside the parenthesis \cref{eq:adjoint}, we use the fact that the element constitutive matrix can be factored as $\widehat{\ve{E}}\cdot\mat{D}_0= \mat{D}$ (under plane stress assumption \cite[p.122]{onate_2009}), and that the derivative of the K-L expansion of a lognormal field is $\widehat{\ve{E}} \mat{A} = \widehat{\ve{E}}'\in \mathbbm{R}^{n_{\mathrm{GP}}\times K}$. Hence,
\begin{equation}
\frac{\dd \ve{\sigma}^\star}{\dd \mat{D}^\star} = (\mat{B}^{\star}\mat{M})\ve{u} \qquad \text{and} \qquad
\frac{\dd \mat{D}^\star}{\dd \ve{\theta}}  =\frac{\dd \mat{D}^\star}{\dd \widehat{\ve{E}}} \frac{\dd \widehat{\ve{E}}}{\dd \ve{\theta}^{(\mathrm{K-L})}} = \mat{D}_0\widehat{\ve{E}}'^\star,
\end{equation}
where $\widehat{\ve{E}}'^\star\in \mathbbm{R}^{K}$ is the K-L expansion derivative at the GL point closest to the control node. Moreover, the second term inside the parenthesis in \cref{eq:adjoint} is
\begin{equation}
\frac{\dd \ve{\sigma}^\star}{\dd \ve{u}} = \mat{D}^{\star}\mat{B}^{\star}\mat{M} \qquad \text{and} \qquad \frac{\dd \ve{u}}{\dd \ve{\theta}}  = \mat{K}^{-1}\left(\dfrac{\dd \ve{f}}{\dd {\theta}^{(q)}} - \dfrac{\dd \mat{K}}{\dd \ve{\theta}^{(\mathrm{K-L})}}\ve{u}\right),
\end{equation}
and the components of the derivative $\frac{\dd \ve{u}}{\dd \ve{\theta}}$ are obtained from the finite element formulation as
\begin{equation}\nonumber
\dfrac{\dd \ve{f}}{{\theta}^{(q)}}= \dfrac{\dd \ve{f}}{\dd {q}}\dfrac{\dd {q}}{\dd \theta^{(q)}} = \ve{c}\cdot\sigma_q \qquad \text{and} \qquad \dfrac{\dd \mat{K}}{\dd \ve{\theta}^{(\mathrm{K-L})}} = \bigcup_{e} \int_{A^{(e)}} \mat{B}^{(e),\tran}\left(\mat{D}_0\widehat{\ve{E}}'^{(e)}\right)\mat{B}^{(e)} ~t~ \dd A^{(e)},
\end{equation}
where $\ve{c}\in\mathbbm{R}^{n_{\mathrm{dof}}}$ is a constant vector that maps the surface load $q$ to equivalent nodal forces, $\bigcup_e$ denotes assembly procedure over all elements, $t$ is the thickness of the plate, and the integration is performed over the area of each element $A^{(e)}$. The term  $\frac{\dd \mat{K}}{\dd \ve{\theta}^{(\mathrm{K-L})}}$ is a rank-3 tensor with size $K\times n_{\mathrm{dof}}\times n_{\mathrm{dof}}$, and its assembly is in general computationally intensive. 

The idea of the adjoint method \cite{arora_and_haug_1979} is to expand and re-organize the terms in \cref{eq:adjoint} to avoid the computation of expensive matrix operations. Specifically consider the term
\begin{equation}
\frac{\dd {\sigma}_1}{\dd \ve{\sigma}^\star}\frac{\dd \ve{\sigma}^\star}{\dd \ve{u}} \frac{\dd \ve{u}}{\dd \ve{\theta}}  = \frac{\dd {\sigma}_1}{\dd \ve{\sigma}^\star}\frac{\dd \ve{\sigma}^\star}{\dd \ve{u}}  \left(\mat{K}^{-1}\left(\dfrac{\dd \ve{f}}{\dd {\theta}^{(q)}} - \dfrac{\dd \mat{K}}{\dd \ve{\theta}^{(\mathrm{K-L})}}\ve{u}\right)\right) = \underbrace{\frac{\dd {\sigma}_1}{\dd \ve{\sigma}^\star}\frac{\dd \ve{\sigma}^\star}{\dd \ve{u}} \mat{K}^{-1}}_{\ve{\lambda}} \underbrace{\left(\dfrac{\dd \ve{f}}{\dd {\theta}^{(q)}} - \dfrac{\dd \mat{K}}{\dd \ve{\theta}^{(\mathrm{K-L})}}\ve{u}\right)}_{\frac{\dd \ve{p}}{\dd \ve{\theta}} },
\end{equation}
where $\ve{p}=\mat{K}\ve{u}-\ve{f}$, and the adjoint multiplier $\ve{\lambda}\in\mathbbm{R}^{n_{\mathrm{dof}}}$ is computed by solving the adjoint equation
\begin{equation}\label{eq:adjoint0}
\mat{K}^{\tran}\ve{\lambda} = \left[\frac{\dd {\sigma}_1}{\dd \ve{\sigma}^\star}\frac{\dd \ve{\sigma}^\star}{\dd \ve{u}} \right]^{\tran},
\end{equation}
we remark that the boundary conditions defining the finite element problem also need to be imposed on \cref{eq:adjoint0}. Therefore, finding the gradient
$\frac{\dd {\sigma}_1}{\dd \ve{\theta}} = \left[\frac{\dd {\sigma}_1}{\dd {\theta}^{(q)}},\frac{\dd {\sigma}_1}{\dd\ve{\theta}^{(\mathrm{K-L})}}\right]$ amounts to compute
\begin{equation}\label{eq:adjointfinal_comp}
\frac{\dd {\sigma}_1}{\dd {\theta}^{(q)}} = \ve{\lambda}^\tran\dfrac{\dd \ve{f}}{\dd {\theta}^{(q)}} \qquad \text{and} \qquad \frac{\dd {\sigma}_1}{\dd \ve{\theta}^{(\mathrm{K-L})}} =
 \frac{\dd {\sigma}_1}{\dd \ve{\sigma}^\star}\frac{\dd \ve{\sigma}^\star}{\dd \mat{D}^\star}\frac{\dd \mat{D}^\star}{\dd \ve{\theta}} + \ve{\lambda}^\tran
\left( - \dfrac{\dd \mat{K}}{\dd \ve{\theta}^{(\mathrm{K-L})}}\ve{u}\right).
\end{equation}

\bibliographystyle{siamplain}
\bibliography{bibfile}

\begin{thebibliography}{10}

\bibitem{arora_and_haug_1979}
{\sc J.~S. Arora and E.~J. Haug}, {\em Methods of design sensitivity analysis
  in structural optimization}, AIAA Journal, 17 (1979), pp.~970--974.

\bibitem{ash_and_doleansdade_2000}
{\sc R.~B. Ash and C.~Doléans-Dade}, {\em Probability and measure theory},
  Harcourt/Academic Press, 2~ed., 2000.

\bibitem{asmussen_2002}
{\sc S.~Asmussen}, {\em {Large deviations in rare events simulation: examples,
  counterexamples and alternatives}}, in Monte Carlo and Quasi-Monte Carlo
  Methods 2000, K.-T. Fang, H.~Niederreiter, and F.~J. Hickernell, eds.,
  Springer, 2002, pp.~1--9.

\bibitem{au_and_beck_2001}
{\sc S.-K. Au and J.~L. Beck}, {\em Estimation of small failure probabilities
  in high dimensions by subset simulation}, Probabilistic Engineering
  Mechanics, 16 (2001), pp.~263--277.

\bibitem{botev_and_kroese_2012}
{\sc Z.~I. Botev and D.~P. Kroese}, {\em {Efficient Monte Carlo simulation via
  the generalized splitting method}}, Statistics and Computing, 22 (2012),
  pp.~1--16.

\bibitem{botev_et_al_2007}
{\sc Z.~I. Botev, D.~P. Kroese, and T.~Taimre}, {\em Generalized cross-entropy
  methods with applications to rare-event simulation and optimization},
  SIMULATION, 83 (2007), pp.~785--806.

\bibitem{bucklew_2004}
{\sc J.~A. Bucklew}, {\em An introduction to rare event simulation}, Springer,
  2004.

\bibitem{cerou_et_al_2012}
{\sc F.~C{\'e}rou, P.~Del~Moral, T.~Furon, and A.~Guyader}, {\em {Sequential
  Monte Carlo for rare event estimation}}, Statistics and Computing, 22 (2012),
  pp.~795--808.

\bibitem{constantine_2015}
{\sc P.~G. Constantine}, {\em {Active subspaces: emerging ideas for dimension
  reduction in parameter studies}}, Society for Industrial and Applied
  Mathematics (SIAM), 2015.

\bibitem{constantine_et_al_2014}
{\sc P.~G. Constantine, E.~Dow, and Q.~Wang}, {\em {Active subspace methods in
  theory and practice: applications to kringing surfaces}}, SIAM Journal on
  Scientific Computing, 36 (2014), pp.~A1500--A1524.

\bibitem{cui_et_al_2014}
{\sc T.~Cui, J.~Martin, Y.~M. Marzouk, A.~Solonen, and A.~Spantini}, {\em
  {Likelihood-informed dimension reduction for nonlinear inverse problems}},
  Inverse Problems, 30 (2014), p.~0114015.

\bibitem{delmoral_et_al_2006}
{\sc P.~Del~Moral, A.~Doucet, and A.~Jasra}, {\em {Sequential Monte Carlo
  samplers}}, Journal of the Royal Statistical Society: Series B, 68 (2006),
  pp.~411--436.

\bibitem{ditlevsen_et_al_1990}
{\sc O.~Ditlevsen, R.~E. Melchers, and H.~Gluver}, {\em General
  multi-dimensional probability integration by directional simulation},
  Computers \& Structures, 36 (1990), pp.~355--368.

\bibitem{engelund_and_rackwitz_1993}
{\sc S.~Engelund and R.~Rackwitz}, {\em A benchmark study on importance
  sampling techniques in structural reliability}, Structural Safety, 12 (1993),
  pp.~255--276.

\bibitem{fiessler_et_al_1979}
{\sc B.~Fiessler, H.-J. Neumann, and R.~Rackwitz}, {\em Quadratic limit states
  in structural reliability}, Journal of the Engineering Mechanics Division
  (ASCE), 105 (1979), pp.~661--676.

\bibitem{geyer_et_al_2019}
{\sc S.~Geyer, I.~Papaioannou, and D.~Straub}, {\em {Cross entropy-based
  importance sampling using Gaussian densities revisited}}, Structural Safety,
  76 (2019), pp.~15--27.

\bibitem{ghanem_and_spanos_2012}
{\sc R.~G. Ghanem and P.~D. Spanos}, {\em {Stochastic finite elements: a
  spectral approach}}, Dover Publications, {Revised}~ed., 2012.

\bibitem{hohenbichler_and_rackwitz_1988}
{\sc M.~Hohenbichler and R.~Rackwitz}, {\em Improvement of second-order
  reliability estimates by importance sampling}, Journal of Engineering
  Mechanics, 114 (1988), pp.~2195--2199.

\bibitem{johnson_2009}
{\sc C.~Johnson}, {\em Numerical solution of partial differential equations by
  the finite element method}, Dover Publications, 2009.

\bibitem{kahn_and_marshall_1953}
{\sc H.~Kahn and A.~W. Marshall}, {\em {Methods of reducing sample size in
  Monte Carlo computations}}, Journal of the Operations Research Society of
  America, 1 (1953), pp.~263--278.

\bibitem{koutsourelakis_et_al_2004}
{\sc P.~S. Koutsourelakis, H.~J. Pradlwarter, and G.~I. Schu\"eller}, {\em
  {Reliability of structures in high dimensions, part I: algorithms and
  applications}}, Probabilistic Engineering Mechanics, 19 (2004), pp.~409--417.

\bibitem{lacaze_et_al_2015}
{\sc S.~Lacaze, L.~Brevault, S.~Missoum, and M.~Balesdent}, {\em Probability of
  failure sensitivity with respect to decision variables}, Structural and
  Multidisciplinary Optimization, 52 (2015), pp.~375--381.

\bibitem{lemaire_et_al_2009}
{\sc M.~Lemaire, A.~Chateauneuf, and J.~Mitteau}, {\em Structural reliability},
  Wiley-ISTE, 2009.

\bibitem{margossian_2019}
{\sc C.~C. Margossian}, {\em {A review of Automatic Differentiation and its
  efficient implementation}}, \href{https://arxiv.org/pdf/1811.05031.pdf}{\tt
  arXiv:1811.05031v2} eprint,  (2019), pp.~1--32.

\bibitem{owen_2013}
{\sc A.~B. Owen}, {\em Monte Carlo theory, methods and examples},
  \href{http://statweb.stanford.edu/~owen/mc/}{\texttt{statweb.stanford.edu/$\sim$owen/mc/}},
  2013.

\bibitem{onate_2009}
{\sc E.~Oñate}, {\em Structural analysis with the finite element method.
  Linear statics. Volume 1: Basis and Solids}, Springer, 2009.

\bibitem{papaioannou_et_al_2015}
{\sc I.~Papaioannou, W.~Betz, K.~Zwirglmaier, and D.~Straub}, {\em {MCMC
  algorithms for subset simulation}}, Probabilistic Engineering Mechanics, 41
  (2015), pp.~89--103.

\bibitem{papaioannou_et_al_2019}
{\sc I.~Papaioannou, S.~Geyer, and D.~Straub}, {\em Improved cross
  entropy-based importance sampling with a flexible mixture model}, Reliability
  Engineering \& System Safety, 191 (2019), p.~106564.

\bibitem{papaioannou_et_al_2016}
{\sc I.~Papaioannou, C.~Papadimitriou, and D.~Straub}, {\em {Sequential
  importance sampling for structural reliability analysis}}, Structural Safety,
  62 (2016), pp.~66--75.

\bibitem{peherstorfer_et_al_2018}
{\sc B.~Peherstorfer, B.~Kramer, and K.~Willcox}, {\em {Multifidelity
  preconditioning of the cross-entropy method for rare event simulation and
  failure probability estimation}}, SIAM/ASA Journal on Uncertainty
  Quantification, 6 (2018), pp.~737--761.

\bibitem{press_et_al_2007}
{\sc W.~H. Press, S.~A. Teukolsky, W.~T. Vetterling, and B.~P. Flannery}, {\em
  Numerical recipes in C: the art of scientific computing}, Cambridge
  University Press, 3~ed., 2007.

\bibitem{rackwitz_and_fiessler_1978}
{\sc R.~Rackwitz and B.~Fiessler}, {\em Structural reliability under combined
  random load sequences}, Computers \& Structures, 9 (1978), pp.~489--494.

\bibitem{rosenthal_2006}
{\sc J.~S. Rosenthal}, {\em {A first look at rigorous probability theory}},
  World Scientific Publishing Company, 2~ed., 2006.

\bibitem{rubinstein_1997}
{\sc R.~Y. Rubinstein}, {\em Optimization of computer simulation models with
  rare events}, European Journal of Operational Research, 99 (1997),
  pp.~89--112.

\bibitem{rubinstein_and_glynn_2009}
{\sc R.~Y. Rubinstein and P.~W. Glynn}, {\em {How to deal with the curse of
  dimensionality of likelihood ratios in Monte Carlo simulation}}, Stochastic
  Models, 25 (2009), pp.~547--568.

\bibitem{rubinstein_and_kroese_2017}
{\sc R.~Y. Rubinstein and D.~P. Kroese}, {\em {Simulation and the Monte Carlo
  method}}, John Wiley \& Sons, 3~ed., 2017.

\bibitem{schilling_2005}
{\sc R.~L. Schilling}, {\em Measures, integrals and martingales}, Cambridge
  University Press, 2005.

\bibitem{shinozuka_1983}
{\sc M.~Shinozuka}, {\em {Basic analysis of structural safety}}, Journal of
  Structural Engineering, 109 (1983), pp.~721--740.

\bibitem{solecki_and_conant_2003}
{\sc R.~Solecki and R.~J. Conant}, {\em Advanced mechanics of materials},
  Oxford University Press, 2003.

\bibitem{spantini_et_al_2015}
{\sc A.~Spantini, A.~Solonen, T.~Cui, J.~Martin, L.~Tenorio, and Y.~Marzouk},
  {\em {Optimal low-rank approximations of Bayesian linear inverse problems}},
  SIAM Journal of Scientific Computing, 37 (2015), pp.~A2451--A2487.

\bibitem{stuart_2010}
{\sc A.~M. Stuart}, {\em {Inverse problems: a Bayesian perspective}}, Acta
  Numerica, 19 (2010), pp.~451--559.

\bibitem{ullmann_and_papaioannou_2015}
{\sc E.~Ullmann and I.~Papaioannou}, {\em {Multilevel estimation of rare
  events}}, SIAM/ASA Journal on Uncertainty Quantification, 3 (2015),
  pp.~922--953.

\bibitem{valdebenito_et_al_2010}
{\sc M.~A. Valdebenito, H.~J. Pradlwarter, and G.~I. Schuëller}, {\em {The
  role of the design point for calculating failure probabilities in view of
  dimensionality and structural nonlinearities}}, Structural Safety, 32 (2010),
  pp.~101--111.

\bibitem{wahal_and_biros_2019a}
{\sc S.~Wahal and G.~Biros}, {\em {BIMC: the Bayesian inverse Monte Carlo
  method for goal-oriented uncertainty quantification. Part I.}},
  \href{https://arxiv.org/pdf/1911.00619.pdf}{\tt arXiv:1911.00619} eprint,
  (2019), pp.~1--38.

\bibitem{wahal_and_biros_2019b}
{\sc S.~Wahal and G.~Biros}, {\em {BIMC: the Bayesian inverse Monte Carlo
  method for goal-oriented uncertainty quantification. Part II.}},
  \href{https://arxiv.org/pdf/1911.01268.pdf}{\tt arXiv:1911.01268} eprint,
  (2019), pp.~1--35.

\bibitem{wang_and_song_2016}
{\sc Z.~Wang and J.~Song}, {\em {Cross-entropy-based adaptive importance
  sampling using von Mises-Fisher mixture for high dimensional reliability
  analysis}}, Structural Safety, 59 (2016), pp.~42--52.

\bibitem{zahm_et_al_2018a}
{\sc O.~Zahm, T.~Cui, K.~Law, A.~Spantini, and Y.~Marzouk}, {\em {Certified
  dimension reduction in nonlinear Bayesian inverse problems}},
  \href{https://arxiv.org/pdf/1807.03712.pdf}{\tt arXiv:1807.03712v2} eprint,
  (2018), pp.~1--41.

\end{thebibliography}

\end{document}